\documentclass[twocolumn]{aastex701}
\usepackage{natbib}

\usepackage{subfig}

\usepackage{amsmath}

\usepackage{listings}

\usepackage{xcolor}

\definecolor{codegreen}{rgb}{0,0.6,0}
\definecolor{codegray}{rgb}{0.5,0.5,0.5}
\definecolor{codepurple}{rgb}{0.58,0,0.82}
\definecolor{backcolour}{rgb}{0.95,0.95,0.92}

\lstdefinestyle{mystyle}{
    backgroundcolor=\color{white},   
    commentstyle=\color{black},
    keywordstyle=\color{black},
    numberstyle=\tiny\color{black},
    stringstyle=\color{black},
    basicstyle=\ttfamily\footnotesize,
    breakatwhitespace=false,         
    breaklines=true,                 
    captionpos=b,                    
    keepspaces=true,                                   
    numbersep=5pt,                  
    showspaces=false,                
    showstringspaces=false,
    showtabs=false,                  
    tabsize=2,
    basicstyle=\ttfamily,
  upquote=true,
  columns=fullflexible,
  keepspaces=true,
  showstringspaces=false,
  breaklines=true, 
}

\usepackage{rotating}

\received{June 29, 2026}
\revised{September 2, 2026}
\accepted{September 8, 2026}
\submitjournal{ApJS} 

\begin{document}

\title{BOSS-CLAM: Utilizing a Constrained Linear Absorption Model to Infer Stellar Parameters from BOSS Spectra}

\author[0000-0003-3410-5794]{Ilija Medan}
\affiliation{Department of Physics and Astronomy,
	Vanderbilt University,
	Nashville, TN 37235, USA}
\affiliation{Canadian Institute for Theoretical Astrophysics, University of Toronto, Toronto, ON M5S-98H, Canada}
\email{ilija.medan@vanderbilt.edu}

\author[0000-0003-0174-0564]{Andrew R. Casey}
\affiliation{School of Physics and Astronomy, Monash University, Clayton VIC 3800, Australia}
\affiliation{Center for Computational Astrophysics, Flatiron Institute, 162 Fifth Avenue, New York, NY 10010, USA}
\email{andrew.casey@monash.edu}

\author[0000-0002-4863-8842]{Alexander~P.~Ji}
\affiliation{Department of Astronomy \& Astrophysics, University of Chicago, 5640 S Ellis Avenue, Chicago, IL 60637, USA}
\affiliation{Kavli Institute for Cosmological Physics, University of Chicago, Chicago, IL 60637, USA}
\affiliation{NSF-Simons AI Institute for the Sky (SkAI), 172 E. Chestnut St., Chicago, IL 60611, USA}
\email{alexji@uchicago.edu}

\author[0000-0003-2602-4302]{Jonah M.~Otto}
\affiliation{Department of Physics and Astronomy, Texas Christian University, TCU Box 298840 
Fort Worth, TX 76129, USA }
\email{j.otto@tcu.edu}

\author[0009-0006-3626-7585]{Kayvon Sharifi}
\affiliation{Department of Physics and Astronomy, Georgia State University, Atlanta, GA 30302, USA}
\email{ksharifi1@gsu.edu}

\author[0000-0003-0179-9662]{Zachary Way}
\affiliation{Department of Physics and Astronomy, Georgia State University, Atlanta, GA 30302, USA}
\email{zway1@gsu.edu}

\author[0000-0002-1715-1257]{Madeleine McKenzie}
\affiliation{Observatories of the Carnegie Institution for Science, 813 Santa Barbara St., Pasadena, CA 91101}
\email{mmckenzie@carnegiescience.edu}

\author[0000-0001-9738-4829]{Natalie R. Myers}
\affiliation{Department of Physics \&
Astronomy, Johns Hopkins University, 3400 N Charles St, Baltimore, MD 21218, USA}
\email{nrmyers.astro@gmail.com}

\author[0000-0002-3481-9052]{Keivan G. Stassun}
\affiliation{Department of Physics and Astronomy,
	Vanderbilt University,
	Nashville, TN 37235, USA}
\email{keivan.stassun@vanderbilt.edu}

  \author[orcid=0000-0002-7489-5244,sname='Smith']{Peter J. Smith}
  \affiliation{Max Planck Institute for Astronomy, K\"onigstuhl 17, D-69117 Heidelberg, Germany}
  \affiliation{Department of Physics and Astronomy, University of Heidelberg, Im Neuenheimer Feld
  226, D-69120, Heidelberg, Germany}
  \email{pesmith@mpia.de}

\author[0000-0003-0842-2374]{Andrew Tkachenko}
\affiliation{Institute of Astronomy, KU Leuven, Celestijnenlaan 200D, 3001, Leuven, Belgium}
\email{andrew.tkachenko@kuleuven.be}

\author[0000-0002-0572-8012]{Vedant Chandra}
\affiliation{Center for Astrophysics $\mid$ Harvard \& Smithsonian, 60 Garden St, Cambridge, MA 02138, USA}
\email{vedant.chandra@cfa.harvard.edu}

\author[0000-0003-1641-6222]{Michael R.~Blanton}
\email{mblanton@carnegiescience.edu}
\affiliation{
The Observatories of the Carnegie Institution for Science, 813 Santa Barbara Street, \\
Pasadena, CA, 91101, USA}
\affiliation{Center for Cosmology and Particle Physics,
Department of Physics,
New York University, \\
726 Broadway Rm. 1005,
New York, NY 10003, USA}

\author[0000-0002-0740-8346]{Peter M.~Frinchaboy}
\affiliation{Department of Physics and Astronomy, Texas Christian University, TCU Box 298840 
Fort Worth, TX 76129, USA }
\email{p.frinchaboy@tcu.edu}

\author[0000-0003-1479-3059]
{Guy S. Stringfellow}
\affiliation{University of Colorado Boulder, Boulder, Colorado 80309, USA}
\email{Guy.Stringfellow@colorado.edu}

\author[0000-0002-6770-2627]
{Sean Morrison}
\affiliation{Department of Astronomy, University of Illinois at Urbana-Champaign, Urbana, IL 61801, USA}
\email{smorris0@illinois.edu}

\begin{abstract}
Large spectroscopic surveys require robust pipelines capable of inferring stellar parameters over a wide range of the Hertzsprung-Russell (HR) diagram from data of varying quality. SDSS-V is one such survey, where the data from the lower-resolution, optical BOSS spectrograph will provide a large dataset covering a wide range of Galactic stellar populations. To better analyze these data, we present BOSS-CLAM, a generative, forward modeling pipeline for inferring effective temperature ($T_\mathrm{eff}$), surface gravity ($\log g$), metallicity ($[\mathrm{Fe/H}]$), and $\alpha-$abundance ($[\alpha/\mathrm{M}]$) from continuum-normalized BOSS spectra. BOSS-CLAM maps stellar labels to Non-negative Matrix Factorization (NMF) basis vector weights via a polynomial mapping jointly optimized with the spectral decomposition, which provides a more flexible framework for working with the lower-resolution BOSS data. Additionally, training labels are drawn from four complementary sources (ASPCAP, \texttt{BOSS-MINESweeper}, wide binaries, and a hot star validation sample), which enables coverage from cool M dwarfs through hot OB stars, and across a wide range of metallicity. We infer parameters for 1,708,214 BOSS spectra, with a recommended clean catalog of 915,514 sources. Validation against open and globular clusters demonstrates homogeneous, accurate abundances across a wide range of metallicity. Wide binary tests yield abundance uncertainties of $\sigma_{[\mathrm{Fe/H}]} \approx 0.15$ dex and $\sigma_{[\alpha/\mathrm{M}]} \approx 0.06$ dex at SNR = 10. Finally, we demonstrate that the BOSS-CLAM catalog recovers known chemical structure of the Milky Way disk and is well-suited for Galactic archaeology, chemical tagging, and stellar population modeling. The pipeline, trained model, and catalog are publicly released as part of SDSS-V DR20.
\end{abstract}

\keywords{Stellar abundances (1577) --- Sky surveys (1464) --- Surveys (1671)}

\section{Introduction}
\label{sec:intro}

Large spectroscopic surveys have revolutionized many areas of astrophysics. The data from such surveys provide stellar parameters for a large number of stars and/or stars that cover a significant range of the Hertzsprung-Russell (HR) diagram. These data are imperative for studies of the structure and history of the Milky Way \citep[e.g.,][]{Nidever2014, Hayden2015, helmi2018, Ratcliffe2023}, stellar populations in the Galactic disk, halo and clusters \citep[e.g.,][]{Bensby2014, Carrera2019, VasilievBaumgardt2021}, stellar demographics and low-mass star populations relevant to IMF constraints \citep[e.g.,][]{Bochanski2010, Li2023, Li2026}, and binary populations \citep[e.g.,][]{Badenes2018, elb2018}.

The Sloan Digital Sky Survey V \citep[SDSS-V;][]{sdssV} is poised to greatly enhance such work, as it is the first optical plus infrared survey of multi-object spectroscopy for millions of sources across the entire sky. Of particular interest for this work is the data being obtained with the optical BOSS spectrograph \citep[][]{smee13a}. These observations provide some advantages over the higher-resolution, near-infrared ones from APOGEE \citep[][]{wilson19a}. Specifically, observations with BOSS are capable of reaching significantly fainter magnitudes, meaning a much wider range of the HR diagram can be probed. Additionally, there are more BOSS than APOGEE fibers, meaning the number of unique sources observed will be substantial. To fully utilize the data from BOSS under SDSS-V, this means a data analysis pipeline must be constructed that 1) infers parameters for a wide range of the HR diagram and 2) can handle data of varying quality and signal.

\begin{figure*}
	\centering
	\includegraphics[width=\linewidth]{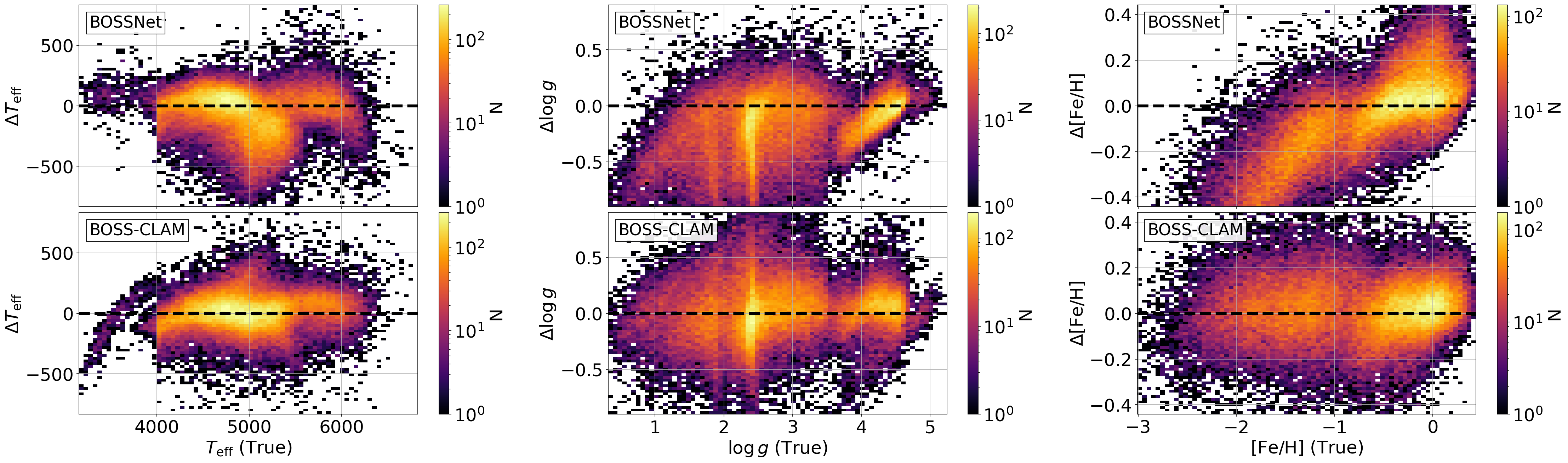}
	\caption{Difference between the stellar parameters of the labels used in this work (True; Section \ref{sec:data}), and the stellar parameters from BOSSNet (top row) and BOSS-CLAM (bottom row).}
	\label{fig:bossnet}
\end{figure*}

Currently, the only pipeline that fits the above criteria is BOSSNet \citep{bossnet}. BOSSNet is a data-driven discriminative model that utilizes a neural network to predict  $T_\mathrm{eff}$, $\log g$, and $[\mathrm{Fe/H}]$ from BOSS spectra. For training, it drew labels from a variety of sources in order to cover a significant range of the HR diagram. On average, BOSSNet achieves these goals with high precision. However, there are some issues with these results. Primarily, there are significant systematic errors across the HR diagram. The top row of Figure \ref{fig:bossnet} shows the difference between the stellar parameters used for training in this work (True; Section \ref{sec:data}) and the BOSSNet parameters. Significant biases are observed, especially for the subgiants and at the most metal-poor end. Additionally, as the model is discriminative, it tends to bias poorly-fitted parameters to the mean of the training distribution. Because of these issues, the goal of this paper is to develop a model that reduces systematic errors and avoids some of the pitfalls of discriminative models.

In this work we focus on generative, forward-modeling approaches where stellar spectra are estimated from labels. Particularly when training a model with higher-precision labels relative to the spectra themselves, as is the case here, this has significant advantages. As discussed in \citet{Ting2025}, when the input features (in this case, lower resolution spectra) have much higher noise than the outputs (stellar labels from higher resolution spectra), this inevitably leads to attenuation bias in discriminative models. In a generative model, as the noisier spectrum is the dependent variable, only the variance of the model estimate is affected, not the bias. Additionally, during inference the uncertainty in the spectra is explicitly incorporated into the likelihood.

There are several methods that employ forward modeling of stellar spectra. The Cannon \citep{Ness2015, Casey2016} predicts continuum-normalized spectra at fixed wavelengths from a set of reference labels. This method has been quite successful in inferring parameters from APOGEE spectra in SDSS-IV and SDSS-V \citep{sdssdr17, Behmard2025}. The Cannon relies on two core assumptions: first, that spectra with identical stellar labels occupy a consistent low-dimensional manifold after continuum normalization, and second, that the training labels are both precise and representative of the underlying stellar population. Such assumptions are difficult to satisfy with the BOSS data. First, BOSS spectra can exhibit flux calibration issues \citep{medan2025} that can be minimized with careful normalization \citep{medan2025,casey2026}, but not completely removed. Second, constructing a homogeneous training set with precise labels across the full range of stellar populations observed by BOSS is challenging, particularly given the broad coverage of the HR diagram. In the method presented here, we address these challenges by using spectral decomposition techniques and explicitly accounting for imperfect labels in the joint optimization.

Additionally, generative models that use synthetic spectra, such as the Payne \citep{Ting2019}, are not suited for these data. The Payne uses neural network interpolation of synthetic spectral grids to infer stellar labels, achieving high precision in APOGEE data. Such methods are sensitive to differences between the observed and synthetic spectra. These differences become increasingly important at lower resolutions, where broad spectral features and continuum variations can dominate the fit. Additionally, it is difficult to cover a wide range of the HR diagram with synthetic models, as many grids are known to produce incompatible spectra for e.g., low-mass stars \citep{jones2005, charams}.

In this work, we will outline the creation and training of a generative model, which we call the ``BOSS-CLAM". BOSS-CLAM predicts continuum normalized spectra from stellar labels by optimizing a smooth, polynomial mapping between stellar parameters and Non-negative Matrix Factorization \citep[NMF;][]{NMF} weights. These weights are multiplied by a series of NMF basis vectors to predict stellar spectra. Using a generative model avoids attenuation bias, the decomposition of the spectra with NMF helps remedy continuum normalization issues remaining in some of the data, and the choice of training data outlined will allow for predictions across the HR diagram, even when there are only partial training labels.

The structure of this paper is as follows: Section \ref{sec:data} describes the training data used, Section \ref{sec:model} describes the BOSS-CLAM model, Section \ref{sec:result} outlines the stellar parameters inferred with BOSS-CLAM with the SDSS-V DR20 spectra, Section \ref{sec:valid} demonstrates validation tests of the BOSS-CLAM parameters with data from external sources, Section \ref{sec:improve} discusses some possible improvements that could be made to the method, Section \ref{sec:applications} provides some examples of applications of these BOSS-CLAM parameters, and Section \ref{sec:summary} provides a summary and some conclusions for this work.

\section{SDSS-V Training Data}\label{sec:data}

In this work, we will utilize optical BOSS spectra from SDSS-V. 
SDSS-V is an all-sky, multi-epoch spectroscopic survey that will observe over six million objects \citep{sdssV}. It consists of three ``Mappers", which are the top-level programs for the survey. The most relevant for our study is the Milky Way Mapper (MWM), which aims to 1) map the stellar populations and chemodynamics
 of the Milky Way to understand its evolution, and 2) probe stellar 
physics and system architecture by observing a variety of stars in the 
Milky Way and Magellanic Clouds. Data for MWM
 is being collected at two observatories: the Sloan Foundation 2.5-m 
telescope at Apache Point Observatory (APO) in New Mexico, USA \citep{APO} and the Irénée du Pont 100-inch telescope at Las Campanas Observatory (LCO) in Chile \citep{LCO}.
 Both observatories are equipped with an optical \citep[BOSS;][]{smee13a} and a near-infrared
\citep[APOGEE;][]{wilson19a} fiber-fed spectrograph. Most importantly, SDSS-V has introduced
 the focal plane system \citep[FPS;][]{FPS}, which utilizes 500 robotic fiber positioners that enable efficient changes between telescope pointings, greatly increasing survey efficiency.
 
 All MWM spectra from SDSS-V are processed through \texttt{astra}, which runs the CLAM to provide a continuum estimate for all MWM stars. For this work, we will use the results from \texttt{astra} in DR20. This corresponds to the \texttt{astra} tag 0.8.1, which is run on the spectra from the BOSS DRP with version \texttt{v6\_2\_1}. Specifically, we use the \texttt{astra} \texttt{mwmStar} data products, which are co-added spectra resampled to the stellar rest frame. These spectra are then normalized using the continuum from the CLAM.

 The main motivation of this work is that SDSS-V DR20 (SDSS Collaboration, et al., {\em in prep}) is largely focused on the release of optical data, with a substantial increase compared to DR19 \citep{sdssVdr19}. Specifically, DR19 included MWM spectra for $\sim$200,000 stars, while DR20 has spectra for $\sim$1,189,000 MWM stars. These stars cover a significant range of the HR diagram, and a pipeline is needed to determine basic stellar parameters across this wide range.

To provide good labels in the form of stellar parameters across such a wide range, we must rely on multiple sources. The following sections discuss the data sources we will use as training labels for our BOSS-CLAM model.

\subsection{Nominal: ASPCAP}\label{sec:data_nom}

The best source of stellar parameters within SDSS-V is from the APOGEE spectra, specifically those from the ASPCAP pipeline. For this dataset, we directly transfer ASPCAP calibrated stellar parameters from SDSS-V DR19 \citep{sdssVdr19, aspcapdr19} to the DR20 BOSS spectra. Here, we filter ASPCAP to include only stars with reliable parameters. This is accomplished by only selecting stars hotter than 4000 K, stars that have SNR higher than 40, removing stars where the spectrum fit and $[\mathrm{Fe/H}]$ were flagged as bad, and ensuring a solution was found for all parameters of interest. Using the ASPCAP data product from \texttt{astra}, these stars are selected like:
\begin{lstlisting}
teff > 4000,
fe_h_flags = 0,
snr > 40,
flag_bad = False,
fe_h is not Null,
raw_alpha_m_atm is not Null,
teff is not Null,
logg is not Null
\end{lstlisting}
We additionally remove sources with $[\mathrm{Fe/H}] < -1.5$ and $\log g < 3.5$, as these metal-poor giants will be better defined by other training set sources. We then transfer the labels to the BOSS spectra with \texttt{snr > 10}, resulting in 33004 spectra for the training/testing set.

Figure \ref{fig:training_set} (top row) shows the Kiel diagram (left column) and $[\alpha/\mathrm{M}]$ vs.~$[\mathrm{Fe/H}]$ (right column) for these data. These data cover a sizable portion of the HR diagram with this single data source. Notably, we are missing data at the low and hot temperature ends, and the low-metallicity end. The following data sources seek to fill in these gaps.

\subsection{BOSS-MINESweeper}\label{sec:data_mine}


To include metal-poor giants, we use the results from the \texttt{BOSS-MINESweeper} value-added catalog \citep{minesweeper}. 
The \texttt{MINESweeper} code utilizes all available information about a star --- the SDSS-V spectrum, broadband photometry from public surveys, and the \textit{Gaia} parallax --- to find the best-matching stellar parameters \citep{Cargile2020}.
The \texttt{BOSS-MINESweeper} VAC reports posterior distributions of the radial velocity $v_\mathrm{r}$, effective temperature $T_\mathrm{eff}$, surface gravity $\log{g}$, metallicity [Fe/H], [$\alpha$/Fe] abundance, and heliocentric distance.

We select stars with the most reliable parameters, defined as stars with no flag during the fitting, within the valid temperature and $\log{g}$ for the method, and are bright enough with a high enough signal for reliable parameters:
\begin{lstlisting}
FLAG = 0,
Teff  < 5600,
logg  < 3.5,
GAIA_G_MAG < 16,
SNR > 20
\end{lstlisting}
This results in 30014 spectra for the training/testing set. Figure \ref{fig:training_set} shows how this source greatly fills in regions for halo-like objects in our parameter space, specifically for the giants and subgiants.

There is an offset in [Fe/H] between the ASPCAP and \texttt{BOSS-MINESweeper} values, particularly in the region of $[\mathrm{Fe/H}] > -1.5$ \citep{minesweeper}. 
In order to train our model on a consistent abundance scale, we modify the \texttt{BOSS-MINESweeper} metallicities to bring them onto the ASPCAP scale. 
For stars with $[\mathrm{Fe/H}] > -1.5$, we fit a second-degree polynomial for the difference between the \texttt{BOSS-MINESweeper} and ASPCAP $[\mathrm{Fe/H}]$ values, as a function of \texttt{BOSS-MINESweeper} $[\mathrm{Fe/H}]$:
\begin{eqnarray}
    \Delta [\mathrm{Fe/H}] &=& -0.0204 [\mathrm{Fe/H}]_{MS}^2 \nonumber \\
                                &&-0.0937 [\mathrm{Fe/H}]_{MS} \nonumber \\
                                &&-0.119
\end{eqnarray}
We correct the \texttt{BOSS-MINESweeper} $[\mathrm{Fe/H}]$ values for $>-1.5$ dex using the above equation, and for $[\mathrm{Fe/H}] \leq -1.5$ we use a constant offset of $\Delta [\mathrm{Fe/H}](-1.5)$. For $[\alpha/\mathrm{M}]$, we similarly see an offset and use a constant value of 
0.106 subtracted from the \texttt{BOSS-MINESweeper} values to put it on a similar scale as ASPCAP.

\subsection{Wide Binaries}\label{sec:data_wb}

To add data for the low-mass dwarfs in the training set, we utilize the wide binaries from \citet{elbadry2021}. We only select binaries where both components are a likely true wide binary according to \citet{elbadry2021}, and pass a series of astrometric and photometric quality cuts, to rule out any unresolved binaries within the system \citep[see][for more details]{way2026}:
\begin{lstlisting}
r_chance_align < 0.1,
ruwe < 1.2,
ipd_frac_multi_peak < 2,
ipd_frac_odd_win < 2,
phot_bp_n_obs > 0,
phot_rp_n_obs > 0,
phot_bp_n_contaminated_transits / phot_bp_n_obs < 10,
phot_bp_n_blended_transits / phot_bp_n_obs < 1,
phot_rp_n_contaminated_transits / phot_rp_n_obs < 10,
phot_rp_n_blended_transits / phot_rp_n_obs < 1,
phot_bp_rp_excess_factor < 2
\end{lstlisting}
Additionally, we are only interested in pairs where the primary has a DR19 ASPCAP measurement and meets the same criteria as the Nominal sample (Section \ref{sec:data_nom}), and the secondary has $M_G > 7.57$, $M_G < 10 (G-RP) + 5$, and a BOSS spectrum with \texttt{snr > 10}. With this, we then transfer the $[\mathrm{Fe/H}]$ and $[\alpha/\mathrm{M}]$ from the ASPCAP parameters of the primary to the BOSS spectrum of the secondary.

For $T_\mathrm{eff}$ and $\log g$, we use the photometric relations from \citet{mann2015, mann2019}. Specifically, for $T_\mathrm{eff}$ we use the relation calibrated with Gaia $BP-RP$ color and $[\mathrm{Fe/H}]$. For $\log g$, we estimate the mass and radius from $M_K$ (2MASS K band absolute magnitude) and $[\mathrm{Fe/H}]$, and then $\log g = \log(M) - 2\ \log(R) + 4.437$. This results in 512 spectra for the training/testing set. Figure \ref{fig:training_set} shows how this fills in the region for low-mass dwarf stars. We should note that the metallicity range of this data source is quite limited, which will influence the final results in this regime.

\subsection{Hot Star}\label{sec:data_hot}

\begin{figure*}
	\centering
    \textbf{Nominal: ASPCAP}\\[0.5em]
	\includegraphics[width=0.8\linewidth]{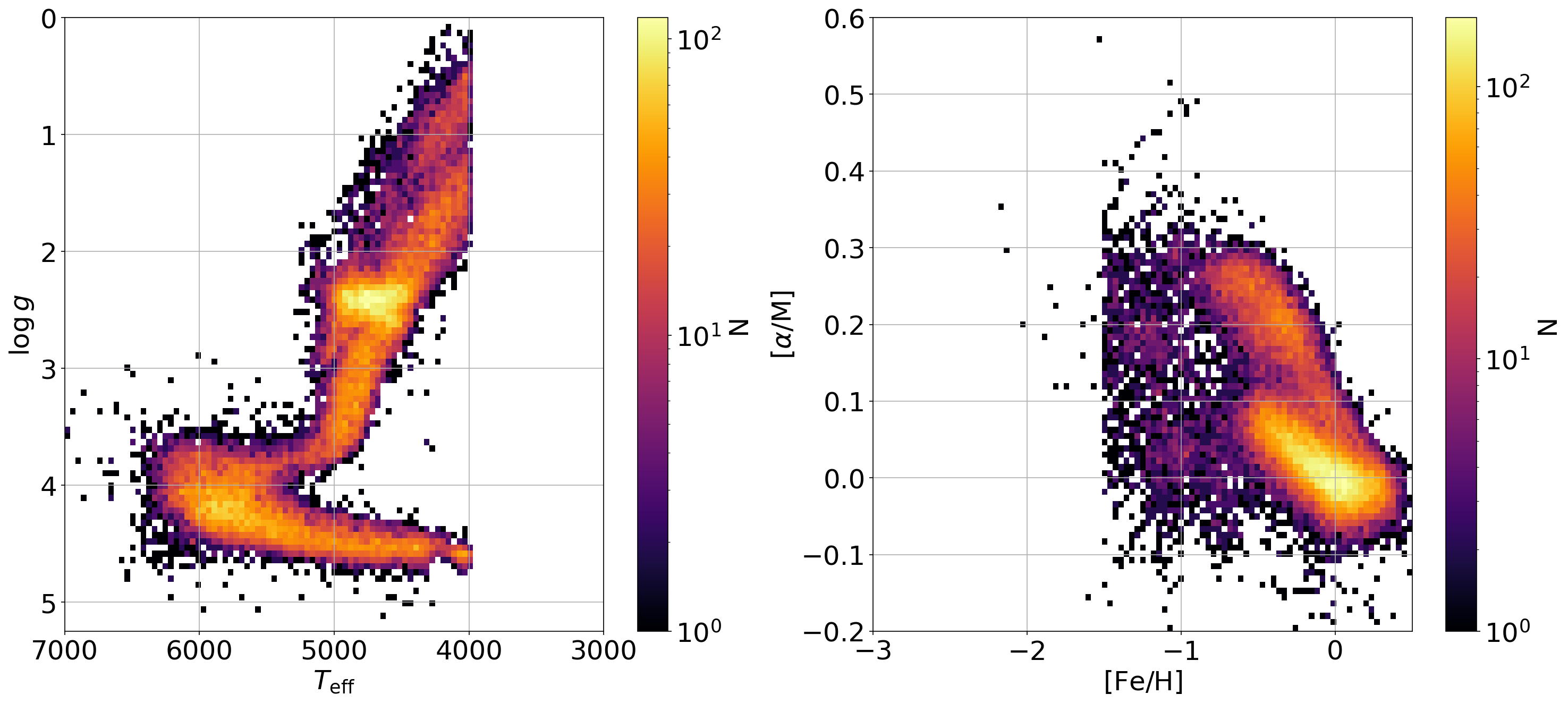}\\[2em]
    
    \textbf{\texttt{BOSS-MINESweeper}}\\[0.5em]
	\includegraphics[width=0.8\linewidth]{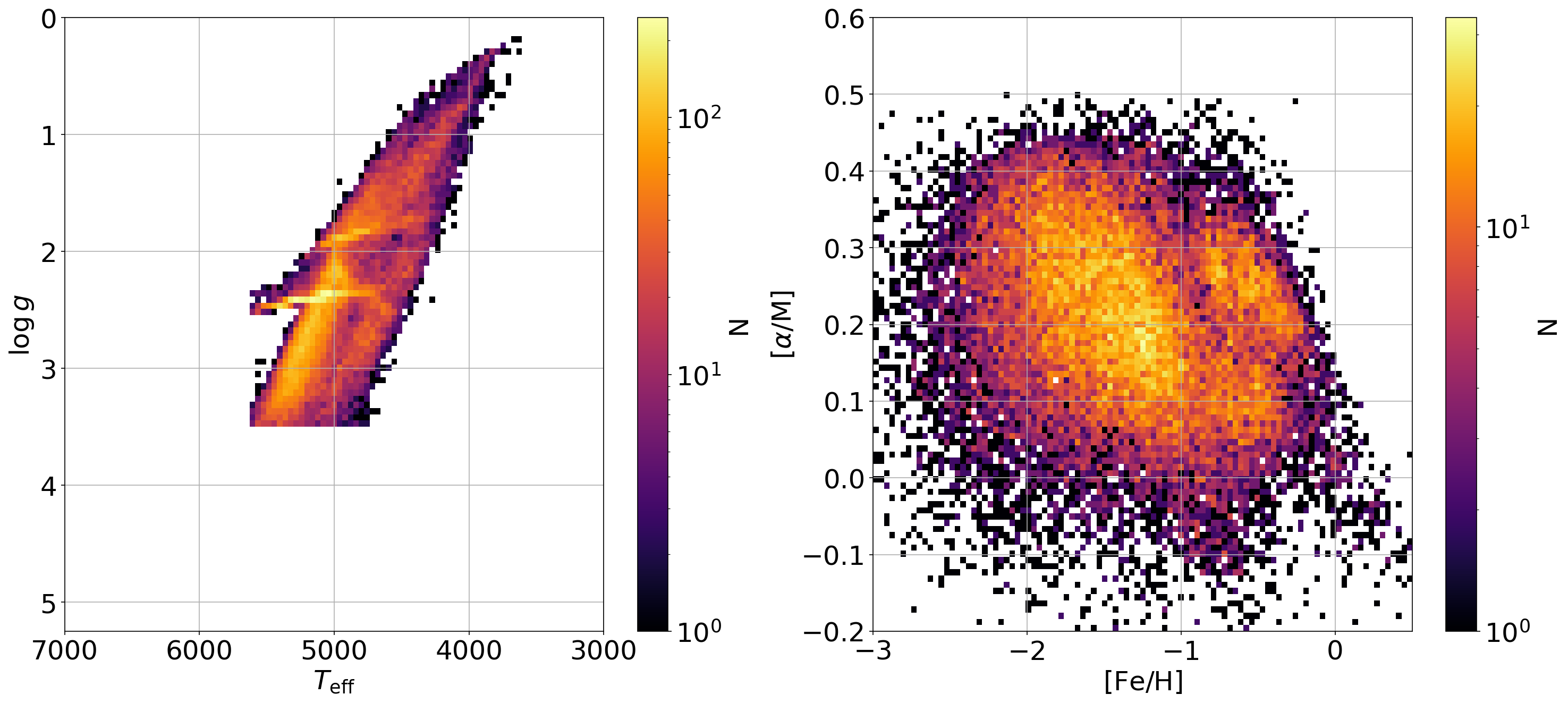}\\[2em]

    \textbf{Wide Binaries}\\[0.5em]
	\includegraphics[width=0.8\linewidth]{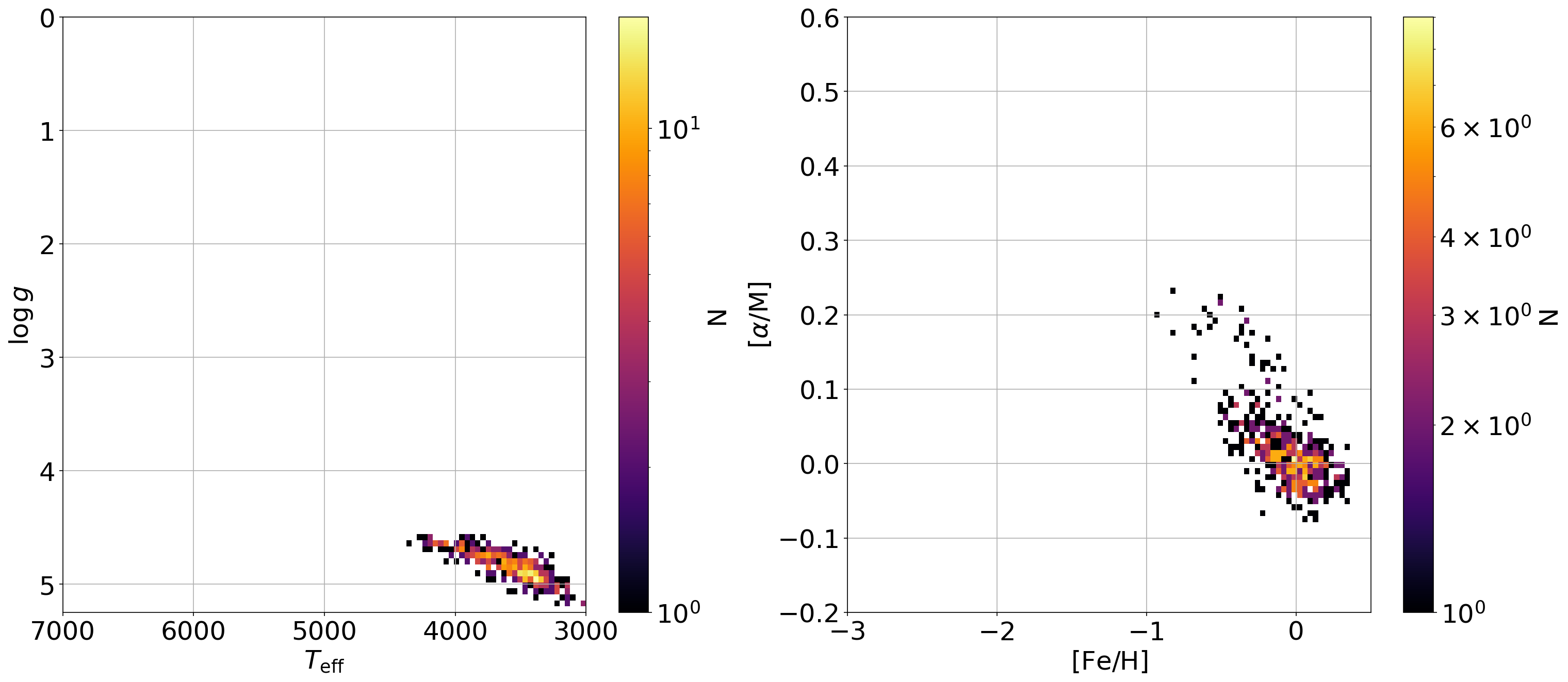}\\
	\caption{Kiel diagram (left column) and $[\alpha/\mathrm{M}]$ vs.~$[\mathrm{Fe/H}]$ (right column) for the data sources used for the training set, as described in Section \ref{sec:data}. The only data source not included in this figure is for the hot stars with $7000 < T_\mathrm{eff} < 53000$ K. These are not included as there are no $[\mathrm{Fe/H}]$ or $[\alpha/\mathrm{M}]$ data for these objects.}
	\label{fig:training_set}
\end{figure*}

To sample stars with $T_\mathrm{eff} > 7000$ K, we use the validation sample from \citet{Tkachenko2026}.
 This is a validation subset consisting of 3535 hot stars with parameters from medium- to
 high-resolution ($R\gtrsim$10\,000) spectra from various studies. These results only 
include $T_\mathrm{eff}$ and $\log g$, so in the training $[\mathrm{Fe/H}]$ and $[\alpha/\mathrm{M}]$ will be left unconstrained. When matched to DR20 BOSS, this results in 626 spectra for the training/testing set. We do not include these data in Figure \ref{fig:training_set}, as it does not include $[\mathrm{Fe/H}]$ or $[\alpha/\mathrm{M}]$ data, but this source provides training data for $7000 < T_\mathrm{eff} < 53000$ K, which is missed by the above data sources.

\section{BOSS-CLAM Model Description}\label{sec:model}


The Constrained Linear Absorption Model (CLAM) simultaneously fits stellar absorption and continuum via non-negative matrix factorization \citep{casey2026}. In this model, line absorption is represented as a linear combination of non-negative (constrained) vectors, where absorption can only be additive. This factorization leads to highly interpretable basis vectors and good performance at the edges of the parameter space (e.g., hot or metal-poor stars) because the model parameters do not need to \emph{subtract} a mean spectrum vector. Here we develop a method we dub ``BOSS-CLAM", which is heavily influenced by this. Similar to the CLAM, BOSS-CLAM decomposes stellar spectra using non-negative matrix factorization (NMF). In this method, though, we will also optimize a mapping between the NMF weights and stellar labels. Once the model is trained, we can infer stellar labels from observed spectra.

This approach could be trained using synthetic labels from a grid or with high-fidelity labels from a training set. In this instance, we use the latter, making BOSS-CLAM a data-driven method that relies on good labels (in the form of stellar parameters) and continuum-normalized spectra for training. Unlike methods such as The Cannon \citep{Ness2015, Casey2016}, however, BOSS-CLAM does not assume that the training labels are exact. As described in Section \ref{sec:model_train}, the labels are optimized jointly with the spectral model during training, allowing the method to accommodate imperfect and partially missing labels. Additionally, spectra are represented through a low-dimensional NMF decomposition, which allows residual variation arising from imperfect continuum normalization, observation systematics, or label uncertainties to be captured within this latent representation. These differences are particularly important for the data used here.

This model will be trained using the labels described in Section \ref{sec:data}. Additionally, all spectra are continuum normalized using the \texttt{astra} results from the CLAM. BOSS-CLAM can then be divided into a training and inference step, both of which are described below.

\subsection{Training}\label{sec:model_train}

The goal of BOSS-CLAM will be to estimate normalized stellar spectra using NMF decomposition, such that
\begin{equation}\label{eq:flux_nmf}
    \hat{\mathbf{F}} = \mathbf{1} - \mathbf{W}\mathbf{H}
\end{equation}
In the above, $\mathbf{H}$ is a series of non-negative basis absorption spectra, which when combined with a set of basis-weights ($\mathbf{W}$) will create an absorption spectrum of a stellar source. For the above, we need to 1) find an optimal set of basis spectra that can used to reconstruct the wide range of objects in our training sample and 2) determine a way to ascribe the correct weights to forward model spectra. This will be accomplished through the following BOSS-CLAM model and training scheme.

First, we will map stellar labels to the  basis-weights. For the labels of our model, let $\boldsymbol{\ell}_n = (\ell_{n,1}, \ldots, \ell_{n,L})^\top$ denote the $L$ stellar labels (in this case $L=4$; $T_\mathrm{eff}$, $\log g$, $[\mathrm{Fe/H}]$, $[\alpha/\mathrm{M}]$) for star $n$, standardized as,
\begin{equation}
    \tilde{\ell}_{n,l} = \frac{\ell_{n,l} - \mu_l}{\sigma_l},
\end{equation}
where $\mu_l$ and $\sigma_l$ are the mean and standard deviation of label $l$ in the training set. Each star's standardized labels are mapped to a quadratic feature vector $\boldsymbol{d}_n \in \mathbb{R}^P$ containing a bias, $L$ linear terms, $L$ quadratic
terms, and $L(L-1)/2$ cross-terms ($P = 15$ for $L = 4$). Stacking over $N$ stars yields the design matrix $\mathbf{D} \in \mathbb{R}^{N \times P}$.

A learned coefficient matrix $\boldsymbol{\Theta} \in \mathbb{R}^{P \times K}$ then maps
$\mathbf{D}$ to our non-negative NMF basis-weights \citep{NMF},
\begin{equation}
    \mathbf{W} = \mathrm{softplus}(\mathbf{D}\,\boldsymbol{\Theta})
\end{equation}
The softplus is a numerical approximation of a ramp function, and is used to enforce positivity and retain numerical stability during the optimization. The above then combines $K$ non-negative basis spectra $\mathbf{H} \in \mathbb{R}^{K \times M}_{\geq 0}$ (parameterized in log-space to enforce positivity) to produce the predicted normalized flux,
\begin{equation}\label{eq:flux_nmf}
    \hat{\mathbf{F}} = \mathbf{1} - \mathbf{W}\mathbf{H}
\end{equation}
In this implementation, we use $K=160$ basis vectors.

During the training stage, the stellar labels ($\boldsymbol{\ell}_n$), polynomial mapping terms ($\boldsymbol{\Theta}$), NMF basis spectra ($\mathbf{H}$), and a per-pixel scatter ($s_\lambda$; see below) are to be optimized, resulting in a total of 
$N \times L + P \times K + K \times M + M$ parameters. We use Adam \citep{kingma2014adam} with a learning rate of 0.001 for 10000 iterations to minimize the loss function,
\begin{eqnarray}
    \mathcal{L} &=& \frac{1}{NM}\sum_{n, m}
    \left[
        \frac{(f_{n,m} - \hat{f}_{n,m})^2}{\sigma_{n,m}^2 + s_m^2}
        + \ln(\sigma_{n,m}^2 + s_m^2)
    \right] \nonumber \\
    && + \frac{\lambda_\ell}{N_c}\sum_{n,l} w_{n,l}(\tilde{\ell}_{n,l} - \tilde{\ell}_{n,l}^{(0)})^2
\end{eqnarray}
In the above, the left term sums over the $n$ stars in our training sample and the $m$ pixels in the spectra, and the right term over the $n$ stars and $l$ labels. Within the loss, $f$ is the observed flux, $\hat{f}$ is our forward modeled flux (eq. \ref{eq:flux_nmf}), $\sigma_{n,m}^2$ is the per-pixel flux variance, $s_m$ is a learned
per-pixel scatter term, $\tilde{\ell}_{n,l}^{(0)}$ are the initial standardized labels,
$w_{n,l}$ are per-label weights, $N_c = \sum_{n,l}\mathbf{1}[w_{n,l}>0]$ is the number of
contributing label terms, and $\lambda_\ell$ controls the strength of label regularization ($\lambda_\ell = 1$ in this implementation).
The log-variance term that comes out of the likelihood in the above prevents
the trivial solution of inflating $s_\lambda$ to absorb all residuals. In the training, $w_{n,l} = 0$ for $[\mathrm{Fe/H}]$ and $[\alpha/\mathrm{M}]$ for the Hot Star group, as these are not defined.

To create a training set for the above, we attempt to uniformly sample the parameter space. Specifically, we bin the data sources from Section \ref{sec:data} in $T_\mathrm{eff}$ ([2800, 6500] K), $\log g$ ([1, 5.25] dex) and $[\alpha/\mathrm{M}]$ ([-0.1, 0.5] dex)
 for 40 bins in each dimension. We randomly select 7 stars per bin to 
add to the training set. If there are fewer than 7, we include all stars 
in that bin in the training set. For stars in the Hot Star group, we 
randomly select 80\% of the sources to 
add to the training set. This scheme prevents over-sampling particular groups 
across the HR diagram, for example, the low$-\alpha$ thin disk relative to the high$-\alpha$ thick disk. This means the basis vectors from the NMF will be more representative of a wider variety of sources.

\subsection{Inference}\label{sec:model_inference}

Next we must infer the parameters for stars with unknown labels. In this step, only the stellar labels are left as free parameters (the unknown quantity), and the loss becomes:
\begin{eqnarray}
    \mathcal{L} &=& \frac{1}{2}\sum_{n,m}
    \left[
        \frac{(f_{n,m} - \hat{f}_{n,m})^2}{\sigma_{n,m}^2 + s_m^2}
    \right]
\end{eqnarray}
The polynomial mapping terms ($\boldsymbol{\Theta}$), NMF basis spectra ($\mathbf{H}$), and per-pixel scatter ($s_\lambda$) are all fixed based on the optimization from the training step (Section \ref{sec:model_train}). The inference of the labels is then done in multiple steps.

First, we train a Multi-layer Perceptron (MLP) regressor \citep{MLP, scikit-learn} with the training set to estimate labels from the basis weights of the spectra (done by solving for $\mathbf{W}$ in eq. \ref{eq:flux_nmf}).  The MLP has hidden layers of sizes $(512, 256, 128, 64)$, uses the tanh activation for each layer, sets the strength of the L2 regularization term to $10^{-5}$, and trains for a maximum of 1000 iterations. This provides a good initial guess for the optimizer.

Then, adam is used with a cosine decay schedule with an initial learning rate of 0.01 for 500 iterations, with a starting value based on the prediction from the MLP regressor. Finally, we run an additional 100 iterations with L-BFGS-B \citep{byrd1995limited, zhu1997algorithm} to settle on the optimal labels for the spectrum.

\begin{figure*}
    \centering
    \includegraphics[width=\textwidth]{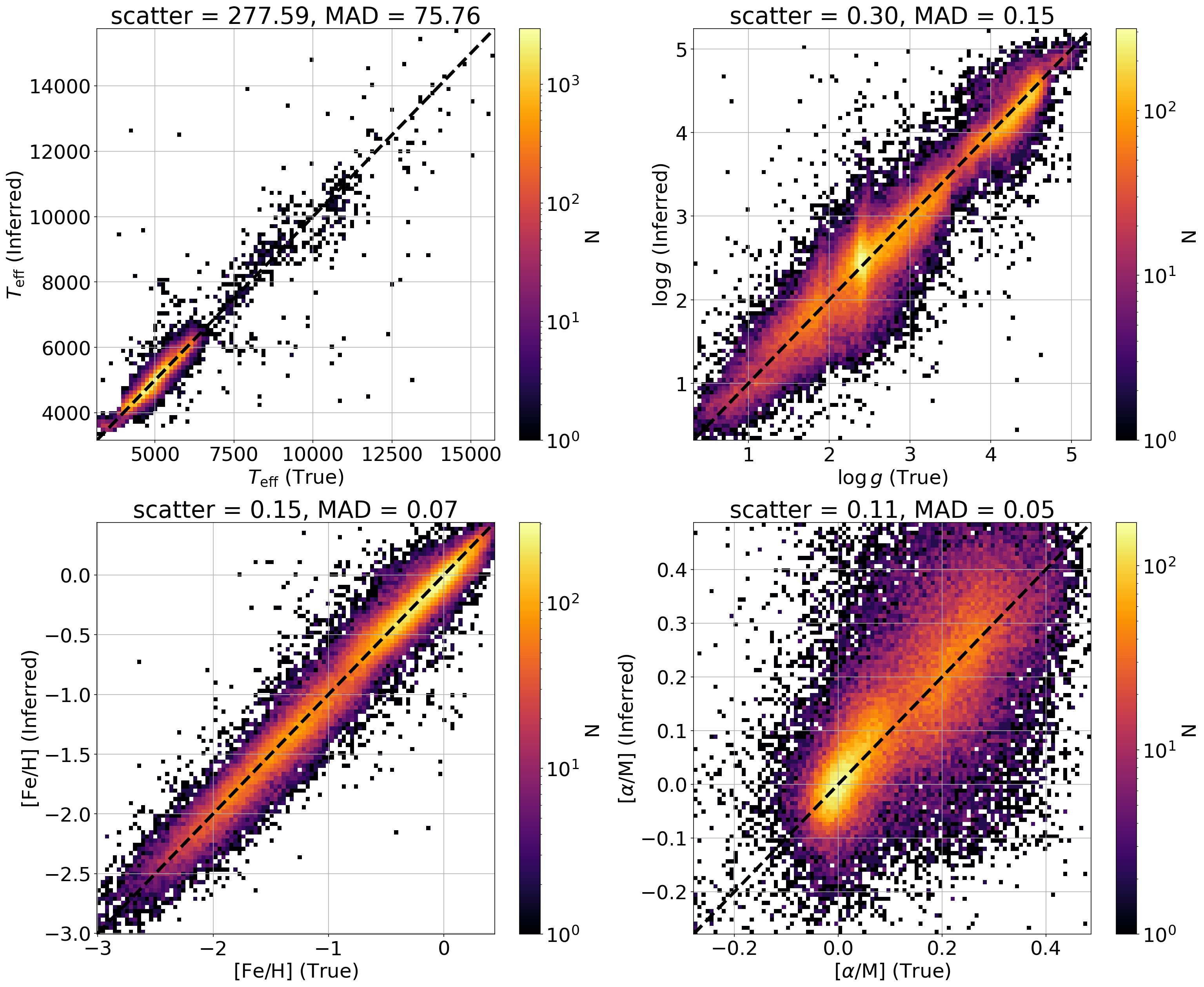}
    \caption{Inference results for the testing set from the BOSS-CLAM pipeline. On all plots, the x-axis shows the true parameters and the y-axis is the BOSS-CLAM parameters. The black dashed line is the one-to-one relation. The title for each panel gives the scatter and median absolute deviation between the true and inferred parameters. In all cases, a good agreement is observed between the true and inferred parameters.}
    \label{fig:clam_results}
\end{figure*}

We run this inference step on all of the data from our four groups (see Section \ref{sec:data}). Figure \ref{fig:clam_results} shows the comparison between the inferred and true parameters for the data. In all cases, a good agreement is observed between the parameters across the full range. Additionally, above each panel, we list the scatter and median absolute deviation between the true and inferred parameters.

Overall, we find the method performs well across a wide range of parameter space, with some exceptions. Most notably, the M dwarfs perform poorly, which is especially apparent in $T_\mathrm{eff}$. This is because these stars are quite underrepresented in the training set and cover a small dynamic range. Future work is needed to include more labels for this class of objects.

Additionally, the scatter seems larger for the $\alpha-$rich stars. As we will show further in this paper, there is also some degeneracy with $[\mathrm{Fe/H}]$ in this regime. The implications of this will be discussed later in this work.

\begin{sidewaysfigure*}
    \centering
    \textbf{Nominal: ASPCAP --- Dwarf}\\
    \includegraphics[width=0.89\textwidth]{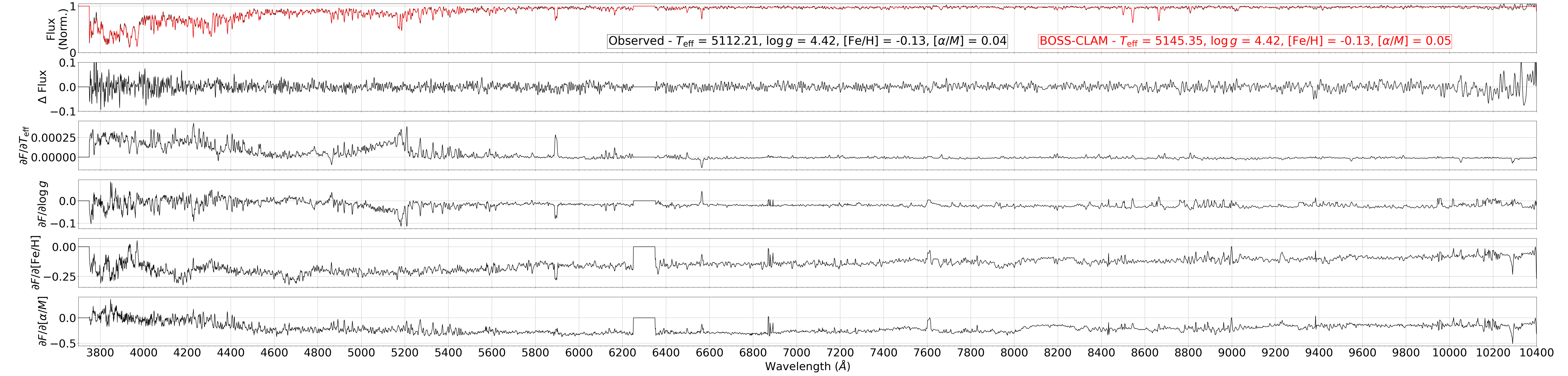}
    \includegraphics[width=0.103\textwidth]{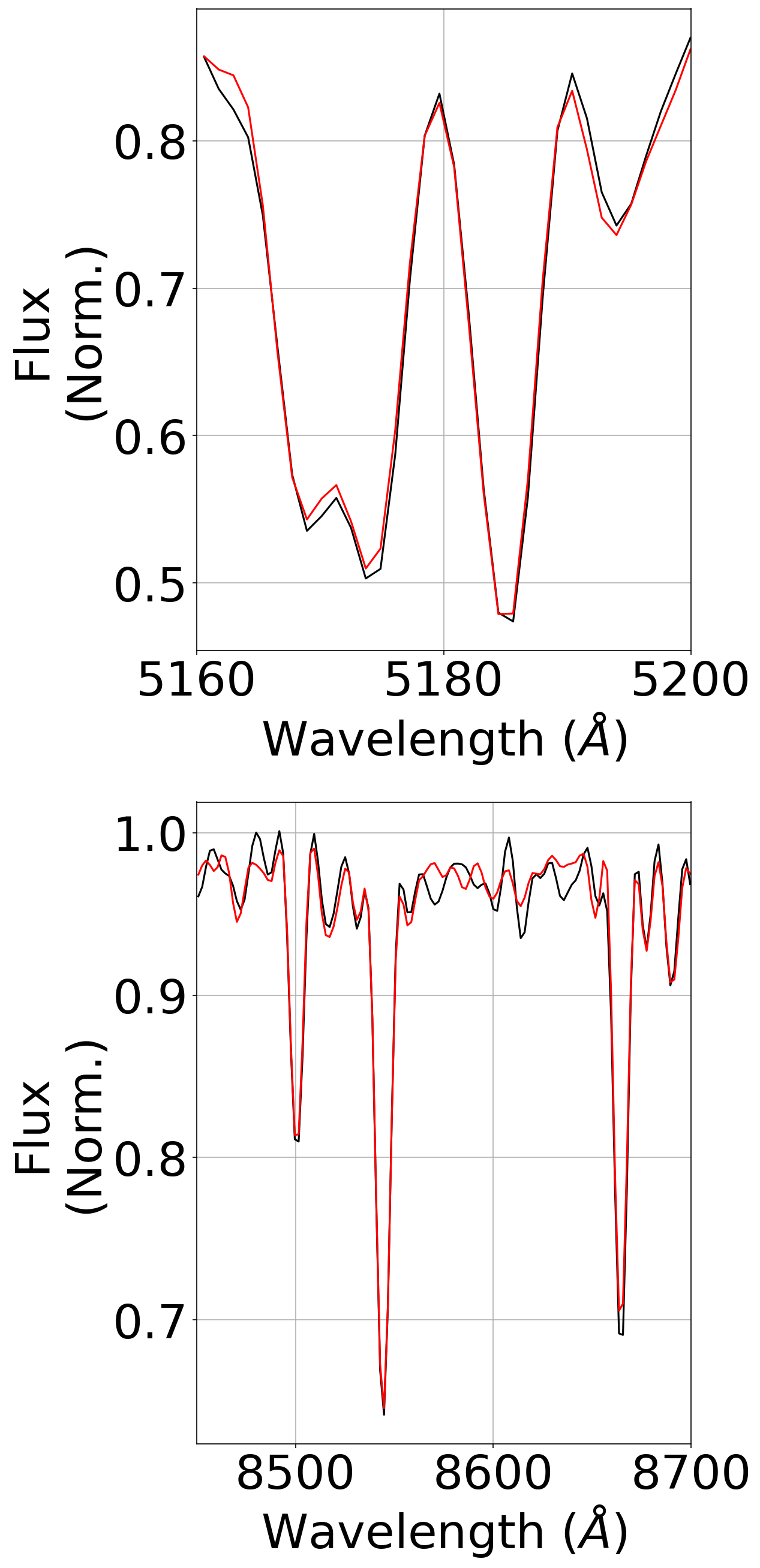}
    \textbf{Nominal: ASPCAP --- Giant}\\
    \includegraphics[width=0.89\textwidth]{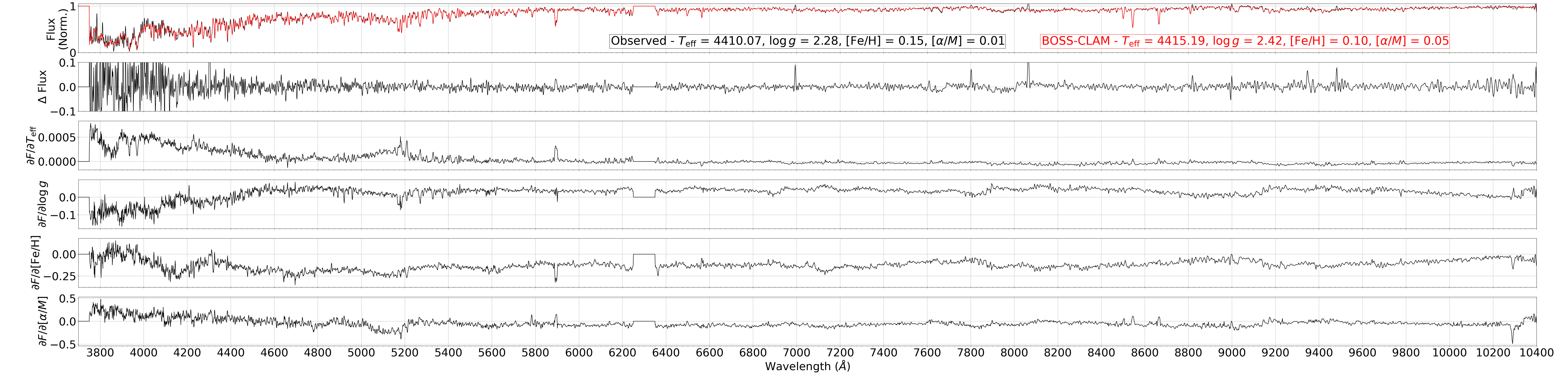}
    \includegraphics[width=0.103\textwidth]{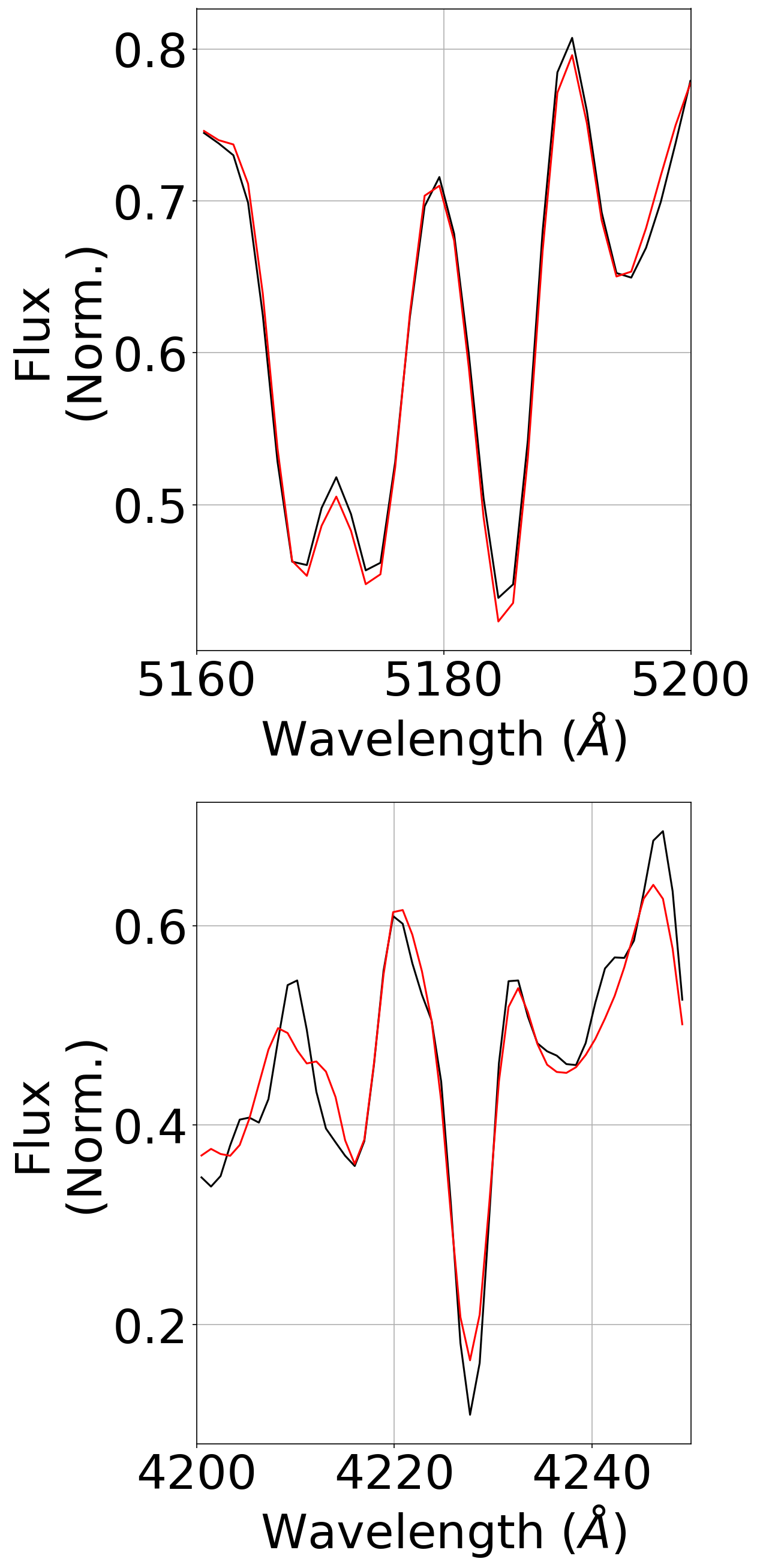}
    \textbf{\texttt{BOSS-MINESweeper}--- Metal-Poor Giant}\\
    \includegraphics[width=0.89\textwidth]{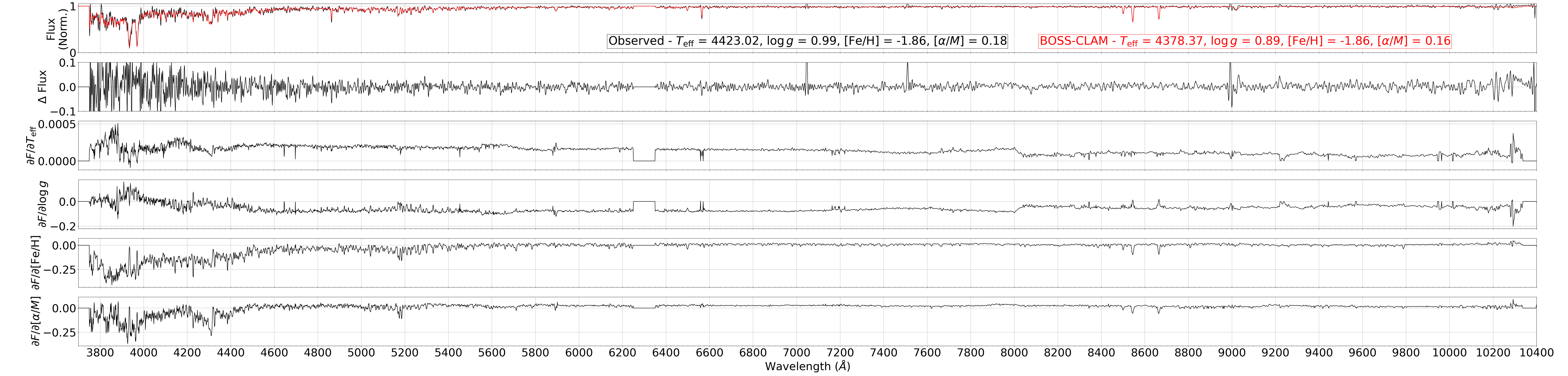}
    \includegraphics[width=0.103\textwidth]{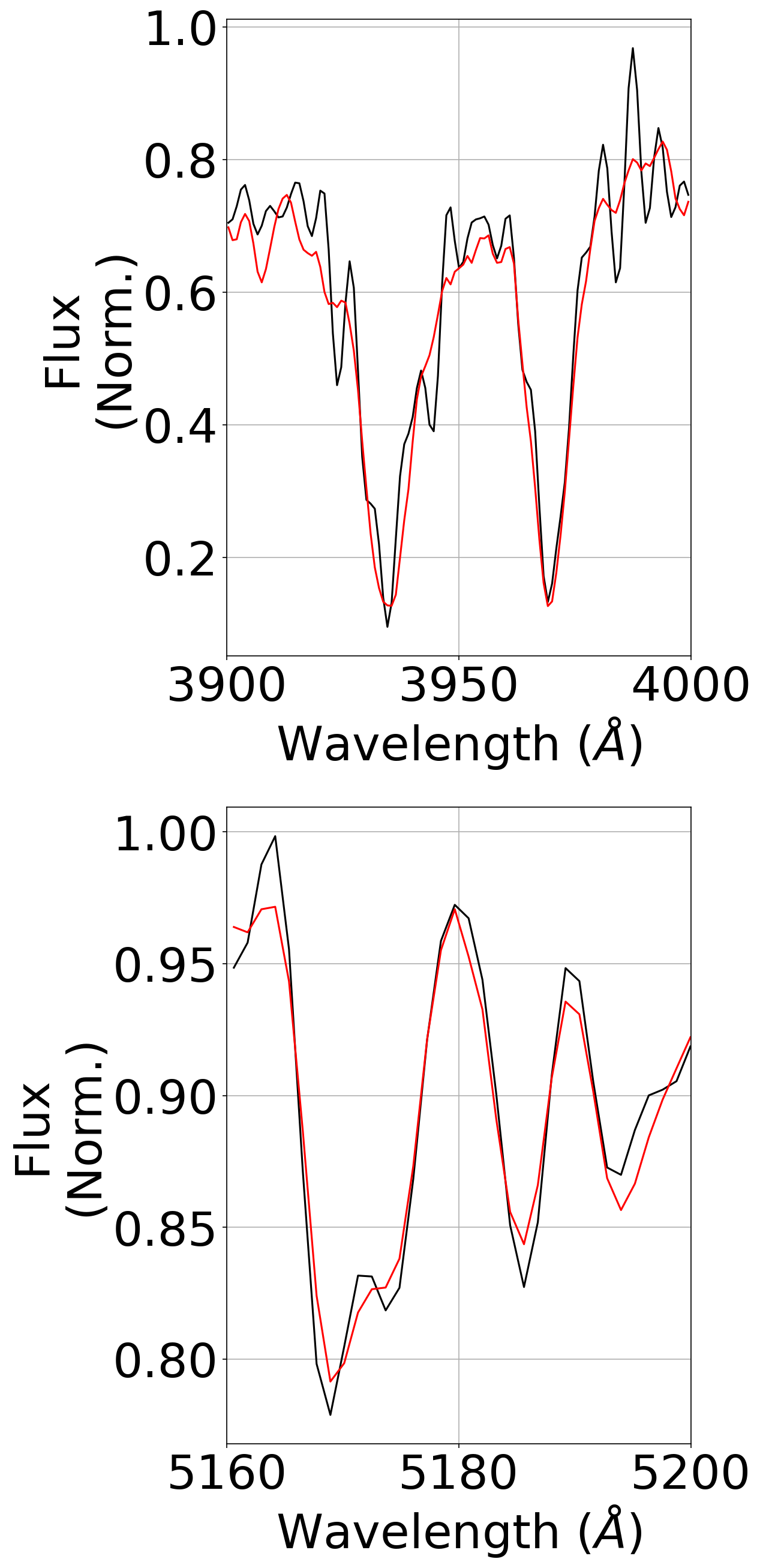}
    \caption{Example spectrum fits from the inference step for different classes of stars in the training sample. In each section, the top panel shows the observed (black) and best fit forward modeled (red) spectra, the second panel is the residual, and the bottom four panels are the forward model spectrum derivatives with respect to each parameter. The right of each section shows a zoom-in on interesting spectral fits for each class of star.}
    \label{fig:clam_spec_results}
\end{sidewaysfigure*}

\begin{sidewaysfigure*}
    \centering
    \textbf{Hot Star}\\
    \includegraphics[width=0.89\textwidth]{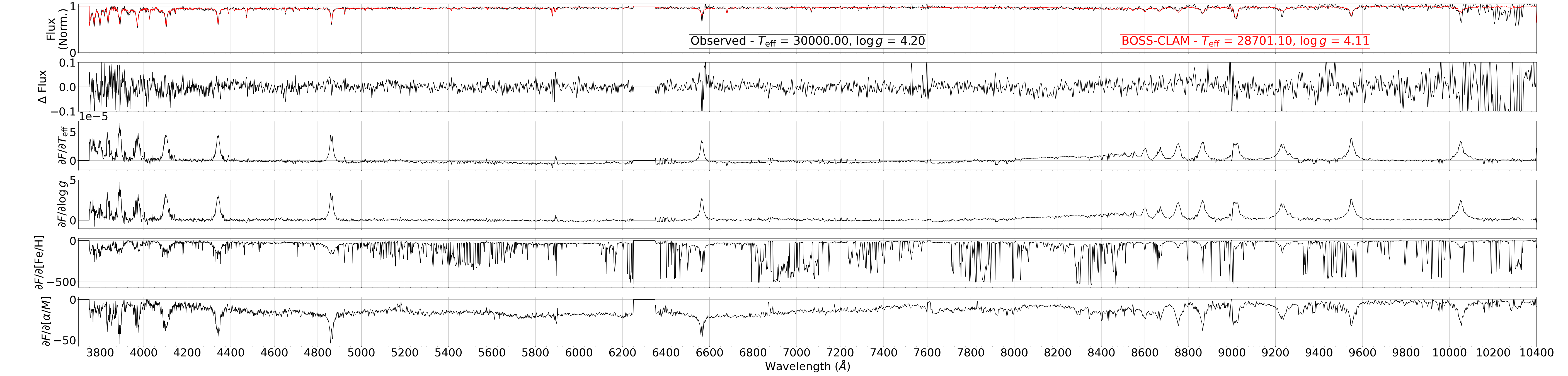}
    \includegraphics[width=0.103\textwidth]{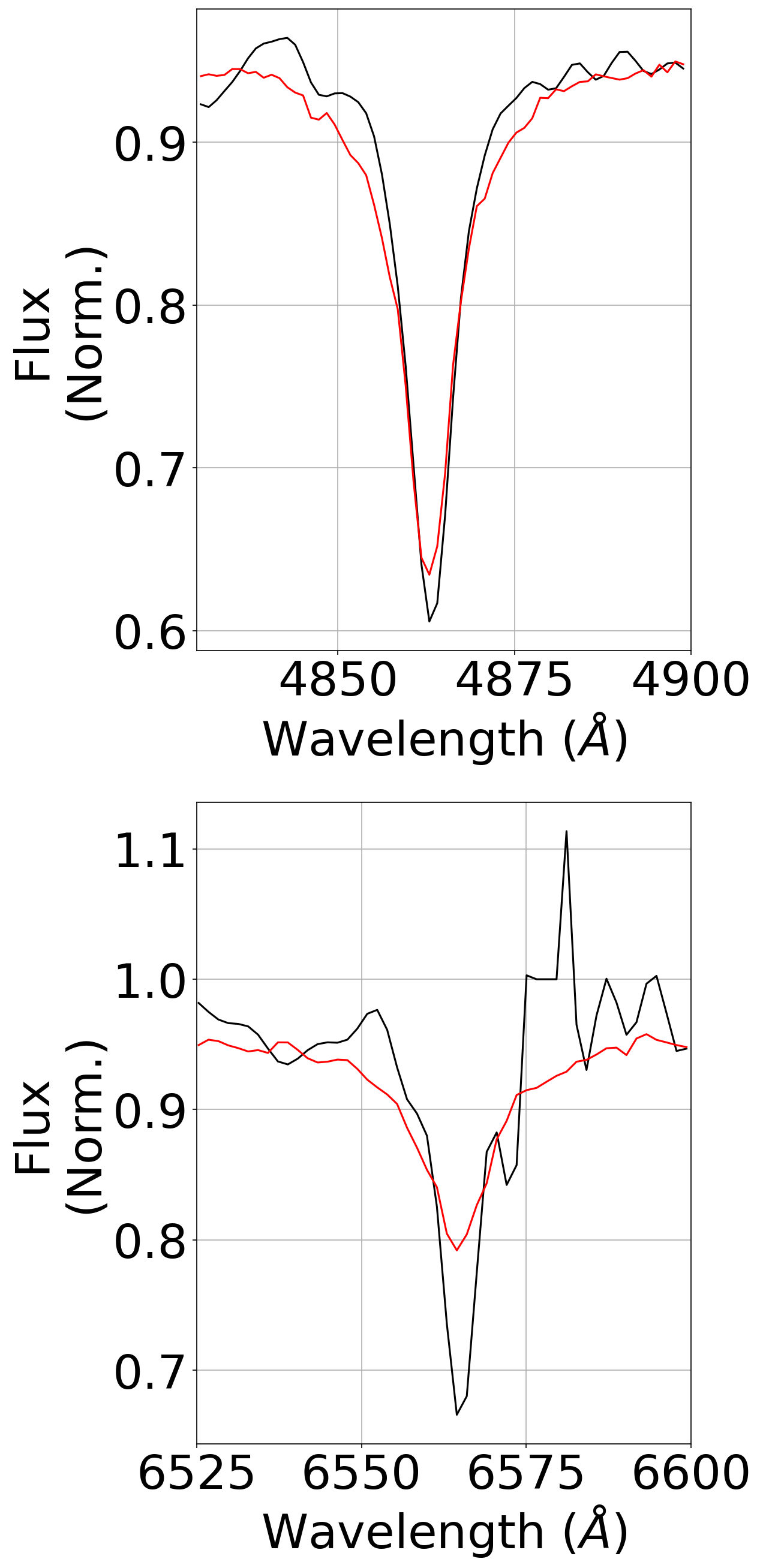}
    \textbf{Wide Binaries --- M Dwarf}\\
    \includegraphics[width=0.89\textwidth]{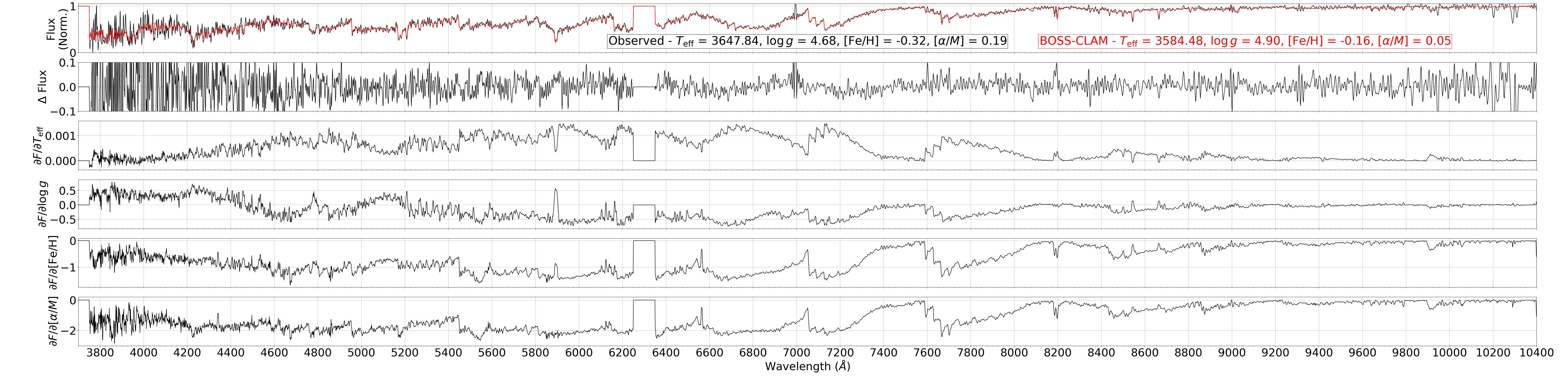}
    \includegraphics[width=0.103\textwidth]{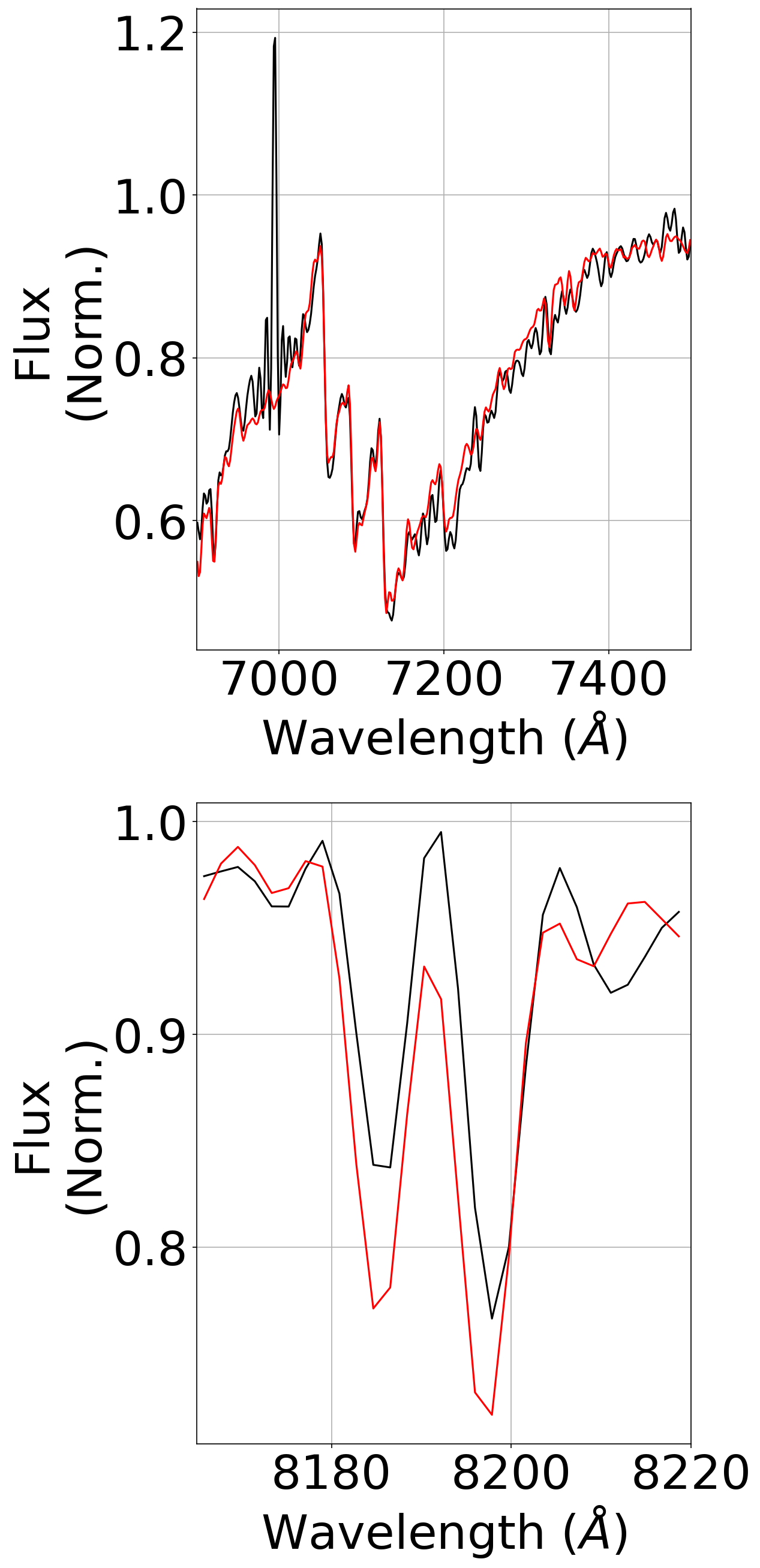}
    \caption{Same layout as Figure \ref{fig:clam_spec_results}.}
    \label{fig:clam_spec_results2}
\end{sidewaysfigure*}

Finally, as BOSS-CLAM is a generative model, it is important to inspect the forward-model spectra produced by the fits. \added{Figures \ref{fig:clam_spec_results} and \ref{fig:clam_spec_results2}} shows the spectrum fits (first row) and residuals (second row) for different classes of stars that cover the wide range of our training sample.

The first two examples are from the nominal, ASPCAP group, where one is a dwarf and the other a subgiant. In both cases, the parameters are well fit, and the residuals are minimal, except at the blue end of the spectrum where line blanketing is densest. Importantly, in the zoomed-in regions to the right of these spectra, we see, for example, the Mg triplet. The line depths clearly vary between dwarfs and giants, and this is well captured by the BOSS-CLAM spectra. Additionally, we can examine the derivatives of the BOSS-CLAM spectra at the best-fit parameters (bottom four rows in each panel). We see that the temperature change is largest in the blue, which is expected for this class of star, and that there is a prominent Mg feature in the $\log g$ derivative. The $[\mathrm{Fe/H}]$ and $[\alpha/\mathrm{M}]$ derivatives also differ from one another in both examples. The metallicity changes broadly, as we expect it to affect the opacity. The $[\alpha/\mathrm{M}]$  derivative, on the other hand, shows specific line features, which we will discuss in more detail in the following section.

To contrast these ``nominal" examples, we also include a metal-poor giant from \texttt{BOSS-MINESweeper}. Here, the fit is also quite good, as demonstrated by the residuals, but the derivatives are quite different. Here we see that the metallicity derivative is dominant at the blue end, where CaHK and a forest of metal lines dominate. We also find that there are strong similarities between the $[\mathrm{Fe/H}]$ and $[\alpha/\mathrm{M}]$ derivatives for this metal-poor source, demonstrating a likely degeneracy. The main difference seems to be the CH band at $\sim4300 \ \AA$. While $[\alpha/\mathrm{M}]$ is correlated with carbon, this may indicate underlying correlations in the training sample; e.g. much of our low-alpha sample may come from relatively low $\log g$ stars where carbon is depleted due to mixing on the RGB. This demonstrates a possible limitation of the method. In e.g., \texttt{BOSS-MINESweeper}, strong isochrone priors are needed to break some of these degeneracies and correlations. So, while BOSS-CLAM can correctly predict the spectra and metallicity of metal-poor giants like these, there may be larger, systematic errors in the results that should be considered.

The last two stars are a stark contrast to the above. The hot star is a worse fit than the above, where some hydrogen line features can be seen in the residual. This is confirmed by the right panel, where the model appears to struggle in fitting the H$\beta$ profile correctly. Despite this, the derivative spectra for $T_\mathrm{eff}$ and $\log g$ correctly reflect the physics of hot stars. The temperature derivative increases line depths, and the gravity derivative increases the line widths via pressure broadening for the Balmer series. As there were no metallicity or alpha labels in the training set, these derivatives then just increase the noise of the spectrum.

Also, the M dwarf example is quite distinct compared to the above. Here, we again see the fit is suboptimal, with some molecular features in the residuals. We do see that the derivatives trace the dominant molecular features of the spectrum though. Specifically for $T_\mathrm{eff}$, at these temperatures, these features greatly change the spectrum, where the opacity from molecules in the atmosphere increases with decreasing temperature. This is well reflected in the derivative. For $[\mathrm{Fe/H}]$ and $[\alpha/\mathrm{M}]$, however, we see that the derivative spectra are nearly identical, demonstrating a strong degeneracy in this temperature regime.

Overall, we find that the model performs quite well and achieves good precision across the majority of the stellar regime considered. There are some exceptions, as shown above, which we will explore more for the remainder of this paper. With this in mind, we will use this trained model to infer the stellar parameters of the broader SDSS-V MWM sample.

\section{BOSS-CLAM Stellar Parameters from DR20 Spectra}\label{sec:result}

Using the method described in Section \ref{sec:model}, we can now infer the stellar parameters for all BOSS spectra in SDSS-V. Specifically, we infer parameters using the \texttt{mwmStar} data products from DR20 for all spectra with \texttt{snr > 10}. For these spectra, we use the same inference routine as described in Section \ref{sec:model_inference}. This results in stellar parameters for 1,708,214 spectra. These parameters are released as part of an SDSS-V DR20 VAC and can be accessed on the SAS\footnote{\url{https://data.sdss.org/sas/dr20/vac/mwm/stellar-params/boss_clam}} after the DR20 release date. The methods outlined here specifically relate to version 1.0.0 of the VAC, and the underlying code and model used to infer these parameters can be found on GitHub\footnote{\url{https://github.com/andycasey/clam-boss}}.

These data cover a much wider range than what is in the training set, in both types of stars and the quality of the data. For example, objects like white dwarfs or young stellar objects are not included in our training set. Additionally, as will be shown below, certain ranges in e.g., temperature that were under-represented affect a large number of the standards used in the survey. This can lead to clear failure modes in the optimization. Such instances can be captured in either the targeting information or the inference results themselves.

We found that a key indicator for failures are the parameter error correlations from the covariance matrices, which are determined from the inverse of the Hessian found during optimization. For example, physically we would expect that the error correlation between $T_\mathrm{eff}$ and $\log g$ should be generally positive. The trend should be quite positive for giants, where temperature and gravity decrease together. For dwarfs, the correlation is very shallow, but not extremely negative. Failure modes where the inference is capturing non-physical correlations would be if this was instead negative. This is likely occurring if either 1) there are issues with the underlying data or 2) the true stellar parameters of the underlying spectra or in a region of parameter space not covered by the training set. In either case, such instances need to be identified.

To address both of the above points, we developed a flagging system for these results:
\begin{itemize}
    \item \textbf{\texttt{clam\_flags}:} Flags set on the parameters, with bits:
    \begin{itemize}
        \item \textbf{\texttt{IN\_WD} (0):} Star targeted as part of the \texttt{mwm\_wd} program, so a likely white dwarf that is not captured by the training set.
        \item \textbf{\texttt{IN\_YSO} (1):} Star targeted as part of the \texttt{mwm\_yso} program, so a likely young stellar object that is not captured by the training set.
        \item \textbf{\texttt{IN\_OB} (2):} Star targeted as part of the \texttt{mwm\_ob} program, so a likely hot star that is in a region of parameter space where no training data had reliable $[\mathrm{Fe/H}]$ or  $[\alpha/\mathrm{M}]$ values.
        \item \textbf{\texttt{FE\_H\_ALPHA\_SUSPECT} (3):} $T_\mathrm{eff} > 6500$ K, indicating a region of parameter space where no training data had reliable $[\mathrm{Fe/H}]$ or  $[\alpha/\mathrm{M}]$ values.
        \item \textbf{\texttt{BAD\_SPEC\_FIT} (4):} $\chi^2_\nu > 1$ between the observed and forward modeled spectrum, indicating a poor fit.
        \item \textbf{\texttt{LOW\_MASS\_SUSPECT} (5):} $T_\mathrm{eff} < 4000$ K and $[\mathrm{Fe/H}] < -1$, so in the low-mass regime where there were very few metal-poor stars in the training set
        \item \textbf{\texttt{FE\_H\_ALPHA\_CORR} (6):} The BOSS-CLAM results typically have a strong negative correlation between $[\mathrm{Fe/H}]$ and $[\alpha/\mathrm{M}]$. A subset of bad $[\alpha/\mathrm{M}]$ estimates do not have such a correlation, but still a typical correlation between $T_\mathrm{eff}$ and $[\mathrm{Fe/H}]$. This flag is set if $\mathrm{corr}([\mathrm{Fe/H}], [\alpha/\mathrm{M}]) > -0.4$ and $\mathrm{corr}([\mathrm{Fe/H}], T_\mathrm{eff}) > -0.6$.
        \item \textbf{\texttt{TEFF\_LOGG\_CORR} (7):} $T_\mathrm{eff}$ and $\log g$ should be positively correlated. This flags any parameters with $\mathrm{corr}(T_\mathrm{eff}, \log g) < 0$.
        \item \textbf{\texttt{LOW\_MASS\_TEFF\_LOGG\_CORR} (8):} Similar to the above, but we apply a more aggressive flag for the low-mass stars, as a large subset seem to decrease in $\log g$ with decreasing $T_\mathrm{eff}$. This flag is then set for $\mathrm{corr}(T_\mathrm{eff}, \log g) < 0.4$ and $T_\mathrm{eff} < 4200$ K.
        \item \textbf{\texttt{EXTREME\_CORR} (9):} There are a subset of stars where the $T_\mathrm{eff}$ and $[\alpha/\mathrm{M}]$, or $\log g$ and $[\alpha/\mathrm{M}]$ are too strongly correlated. These are flagged as $|\mathrm{corr}(T_\mathrm{eff}, [\alpha/\mathrm{M}])| > 0.95$ or $|\mathrm{corr}(\log g, [\alpha/\mathrm{M}])| > 0.95$.
        \item \textbf{\texttt{ALTER\_CORR} (10):} Finally, we found a subset of outliers showed alternating correlations compared to the bulk of the data in the following dimensions: $\mathrm{corr}(T_\mathrm{eff}, [\mathrm{Fe/H}]) < -0.4$ and $\mathrm{corr}(T_\mathrm{eff}, [\alpha/\mathrm{M}]) > 0.2$ and $\mathrm{corr}(\log g, [\alpha/\mathrm{M}]) > 0.2$.
    \end{itemize}
    \item \textbf{\texttt{flag\_warn}:} If \texttt{IN\_OB} or \texttt{FE\_H\_ALPHA\_SUSPECT} set, so should not trust resulting $[\mathrm{Fe/H}]$ or $[\alpha/\mathrm{M}]$ estimates.
    \item \textbf{\texttt{flag\_bad}:} If \texttt{IN\_WD}, \texttt{IN\_YSO}, \texttt{BAD\_SPEC\_FIT}, \texttt{LOW\_MASS\_SUSPECT}, \texttt{FE\_H\_ALPHA\_CORR}, \texttt{TEFF\_LOGG\_CORR}, \texttt{LOW\_MASS\_TEFF\_LOGG\_CORR}, \texttt{EXTREME\_CORR} or \texttt{ALTER\_CORR} are set, so should not trust any of resulting parameters.
\end{itemize}
For the following results plots, if we are examining only $T_\mathrm{eff}$ and $\log g$, we will only include sources with \texttt{flag\_bad = False}, which includes a total of 
977,202 sources. If we also include $[\mathrm{Fe/H}]$ and/or $[\alpha/\mathrm{M}]$, we will only include sources with \texttt{clam\_flags = 0}, which includes a total of 915,514 sources. These are the most conservative cuts that can be made on this catalog. The full catalog includes all flags, covariance matrices, and forward model spectra, so users can make less conservative cuts if desired.

\begin{figure*}
    \centering
    \includegraphics[width=\textwidth]{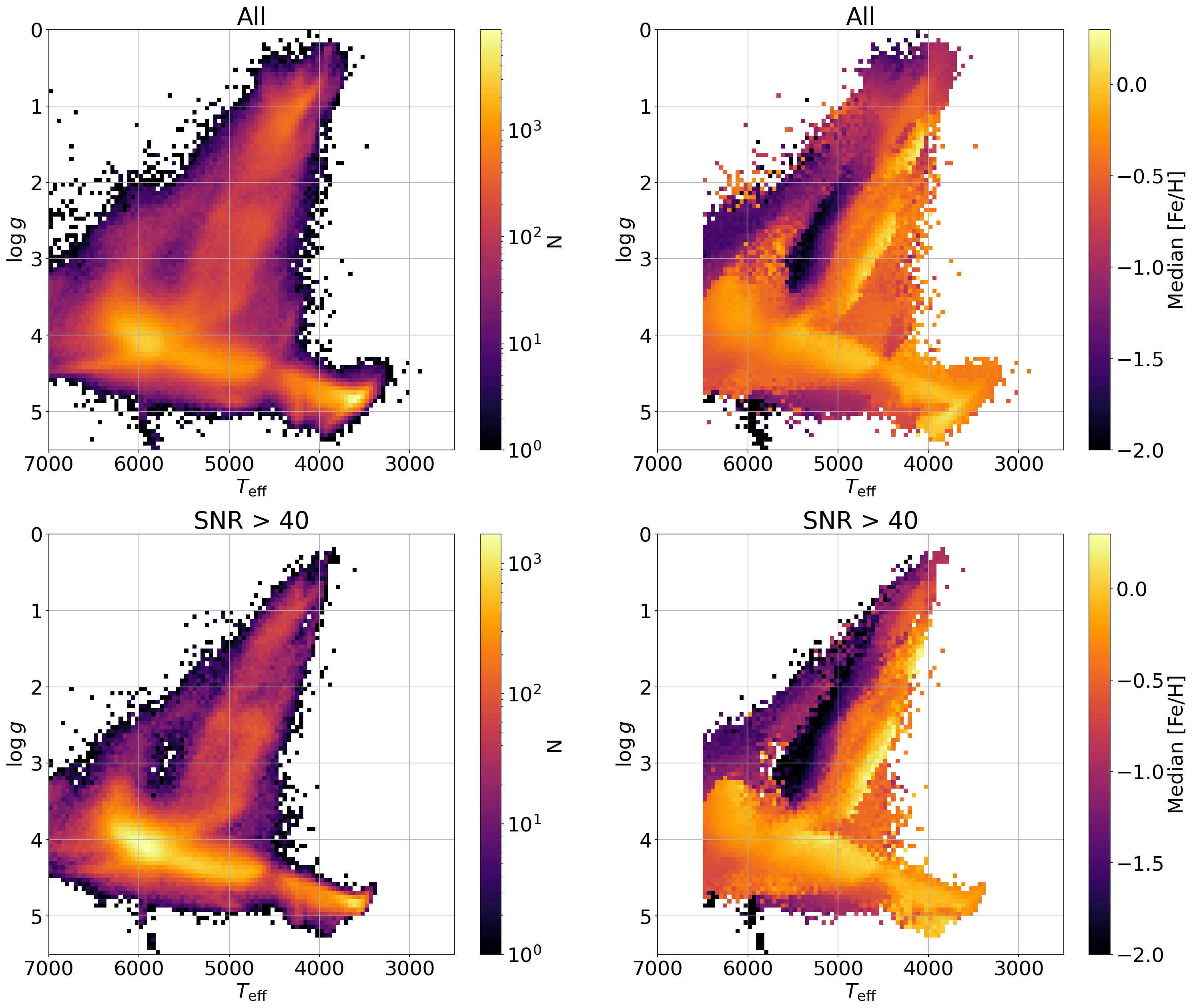}
    \caption{Kiel diagram showing the inferred parameters from the BOSS-CLAM. The left column shows the number counts and the right the median $[\mathrm{Fe/H}]$. The top row shows the Kiel diagram for all parameters that meet our quality cuts (\texttt{flag\_bad = False} for number counts and \texttt{clam\_flags = 0} for the median $[\mathrm{Fe/H}]$) and the bottom row for spectra that also have SNR $ > 40$.}
    \label{fig:kiel_results}
\end{figure*}

Figure \ref{fig:kiel_results} shows the Kiel diagram from all of the spectra that meet our quality cuts (top row) and for spectra that also have SNR $ > 40$ (bottom row). In both plots, we see that the broad trends match the expectations from the higher resolution data in the training set. Additionally, we see a clear trend in metallicity (right panels), for e.g., the giant branch, and that we reach fairly low metallicity values in the full dataset. Finally, when we further cut on signal-to-noise, it is clear that the scatter is greatly reduced. This demonstrates that the parameters inferred are not overfit and indeed the precision likely tracks with the noise in the spectrum.

\begin{figure*}
    \centering
    \includegraphics[width=\textwidth]{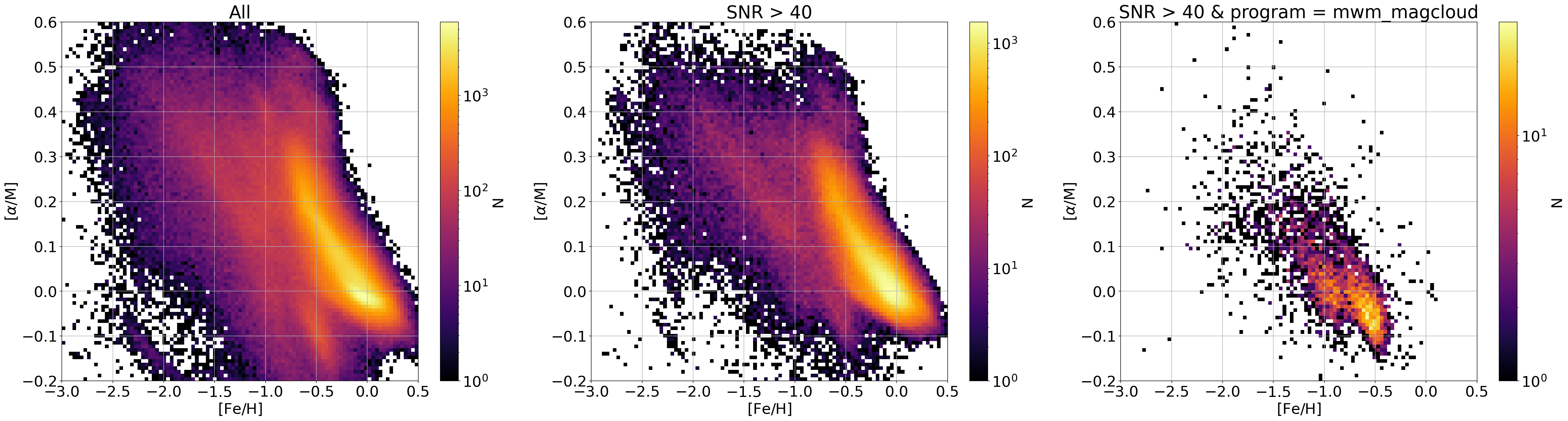}
    \caption{Distribution of $[\alpha/\mathrm{M}]$ vs.~$[\mathrm{Fe/H}]$ using the inferred parameters from the BOSS-CLAM. The left panel shows the distribution for all parameters that meet our quality cuts (\texttt{clam\_flags = 0}), the middle panel for spectra that also have SNR $ > 40$, and the right panel for spectra that also have SNR $ > 40$ and are in the program \texttt{mwm\_magcloud}.}
    \label{fig:alpha_fe_results}
\end{figure*}

To further investigate the results, Figure \ref{fig:alpha_fe_results} shows the distribution of $[\alpha/\mathrm{M}]$ vs.~$[\mathrm{Fe/H}]$ for the spectra that meet our quality cuts (left) and for spectra that also have SNR $ > 40$ (middle). Similar to the above, we find that the sequences are more defined in the high signal-to-noise subset and that broadly the distribution matches the training set. We note that the thin/thick disk separation is not as well-defined as in the high-resolution, training data. Indeed, when examining the covariances in the inferred parameters, we find a degeneracy with $[\alpha/\mathrm{M}]$ vs.~$[\mathrm{Fe/H}]$. This leads to the streak between the high- and low-$\alpha$ sequences observed in the results. This degeneracy should be considered when using the BOSS-CLAM parameters.

\begin{figure*}
    \centering
    \includegraphics[width=\textwidth]{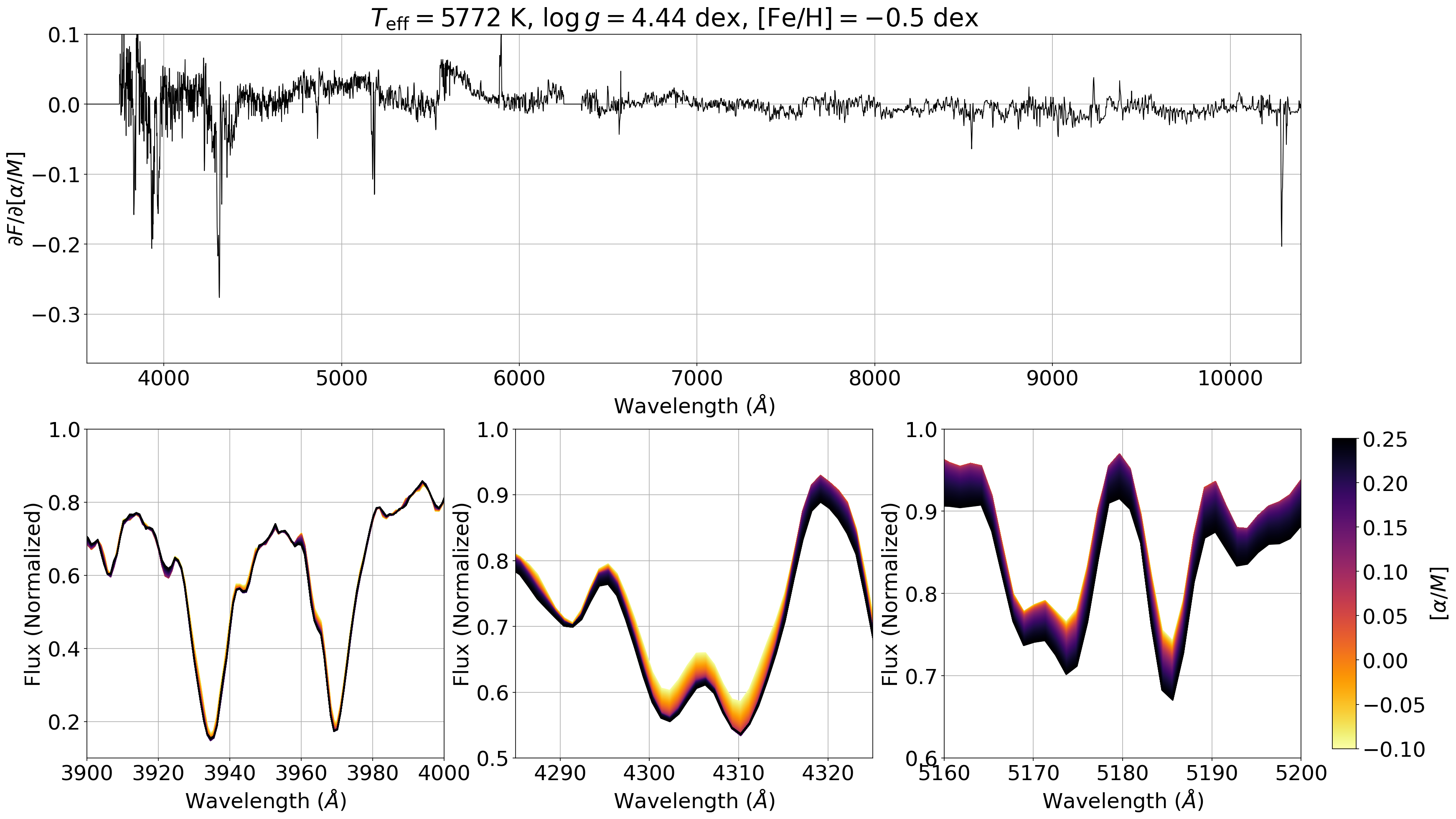}
    \caption{Derivative of forward modeled BOSS-CLAM spectra with respective to $[\alpha/\mathrm{M}]$ for a Solar-like star (top row). The derivative is relative to $[\alpha/\mathrm{M}] = 0$ dex. The bottom rows show regions of the normalized spectra where we see the largest changes in the derivative, as shown in the top row. Each color corresponds to the normalized spectrum with varying $[\alpha/\mathrm{M}]$.}
    \label{fig:alpha_deriv}
\end{figure*}

We find the model is learning real $\alpha-$element features in the spectra, though, so while the degeneracy exists, the model is not simply learning correlations with $[\mathrm{Fe/H}]$. This is best demonstrated by looking at the derivative of forward modeled BOSS-CLAM spectra with respect to $[\alpha/\mathrm{M}]$ for a Solar-like star (top row, Figure \ref{fig:alpha_deriv}). Here we find that there are clear Mgb and MgH lines at $\sim5100 \ \AA$, Ca H+K and Ca Triplet, and some absorption from $4300-4500 \ \AA$ from Ti and CH. The bottom row of Figure \ref{fig:alpha_deriv} shows these regions of the spectra and how the normalized flux varies with $[\alpha/\mathrm{M}]$ for our forward model. Clear changes with $[\alpha/\mathrm{M}]$ are observed. Additionally, we do not see line blanketing in the derivative that would be associated with changes in metallicity, except at the highest $[\alpha/\mathrm{M}]$ values. So, while there is some degeneracy in the parameters that should be considered, especially for $\alpha-$rich sources, the features it is learning are real for a wide range of the parameter space.

This is further confirmed by looking at the $[\alpha/\mathrm{M}]$ vs.~$[\mathrm{Fe/H}]$ for stars in the Magellanic Clouds (right panel; Figure \ref{fig:alpha_fe_results}). Here we see a distribution that is quite different compared to the Milky Way and matches recent studies \citep{Nidever2020, Hasselquist2021}. The difference in distribution for this unique population of stars further demonstrates that the BOSS-CLAM is not simply learning correlations with metallicity that are observed in the Milky Way.

\added{Finally, the BOSS-CLAM results do include estimates of the statistical errors on the parameters through the full covariance matrices. These are estimated from the inverse of the Hessian matrix of the negative log-likelihood evaluated at the best-fit solution, equivalent to a local Gaussian (Laplace) approximation to the posterior about its mode. A more accurate representation of the posterior can be estimated from MCMC, but this is computationally expensive for the full catalog. As a test, we did run full MCMC for a subset of 9,954 stars (Appendix \ref{app:mcmc}). There are two findings to report from this test.}

\begin{figure*}
    \centering
    \includegraphics[width=\textwidth]{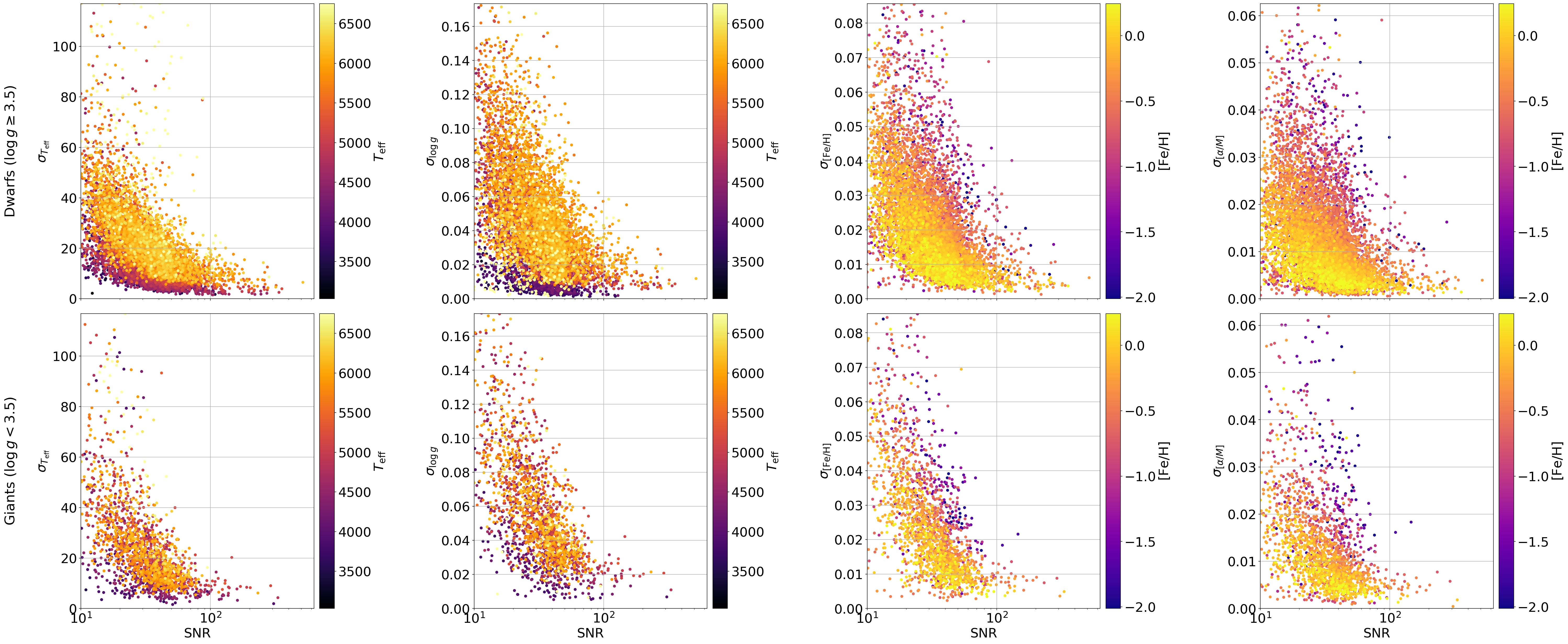}
    \caption{The standard deviation of the posterior on the stellar parameters versus SNR for a subset of 9,954 stars in our MCMC test (Appendix \ref{app:mcmc}). The top row is for dwarfs ($\log g \geq 3.5$) and the bottom row for giants ($\log g < 3.5$). In the first two columns, the standard deviations are colored by the BOSS-CLAM $T_\mathrm{eff}$, while the last two columns are by $[\mathrm{Fe/H}]$, as these seemed to track best with the errors.}
    \label{fig:mcmc_avg_errs}
\end{figure*}

\added{First, as the test sample covered a representative set of stars in SNR, $T_\mathrm{eff}$, $\log g$ and $[\mathrm{Fe/H}]$, we can demonstrate the expected, average errors as a function of the stellar parameters. Figure \ref{fig:mcmc_avg_errs} shows the standard deviation of the posterior on the stellar parameters versus SNR for this subset. For both the dwarfs (top row) and giants (bottom row) we find that these errors are well behaved with SNR and $T_\mathrm{eff}$ or $[\mathrm{Fe/H}]$. This provides users of this data with a quick reference for the magnitude of the statistical errors on the BOSS-CLAM parameters.}

\added{More crucial is the second finding; the covariance matrices estimated from the inverse of the Hessian matrix well match the MCMC posteriors (Table \ref{tab:hessian_mcmc_comparison}). This includes not only the variances along the diagonal, but the covariances as well. This demonstrates that the approximate covariance matrices produced in this work are perfectly adequate and can be used to determine the statistical uncertainties on the parameters. We note that included in the code released with this work is a function to perform MCMC sampling on a result. This can be used if more detailed posteriors are needed. Regardless of these results, we do note that systematic errors are likely larger than these statistical ones and will dominate the results. This will be shown for some areas of parameter space in later sections.}

\subsection{Comparing BOSS-CLAM Parameters to Other Surveys}

\begin{figure*}
    \centering
    \includegraphics[width=\linewidth]{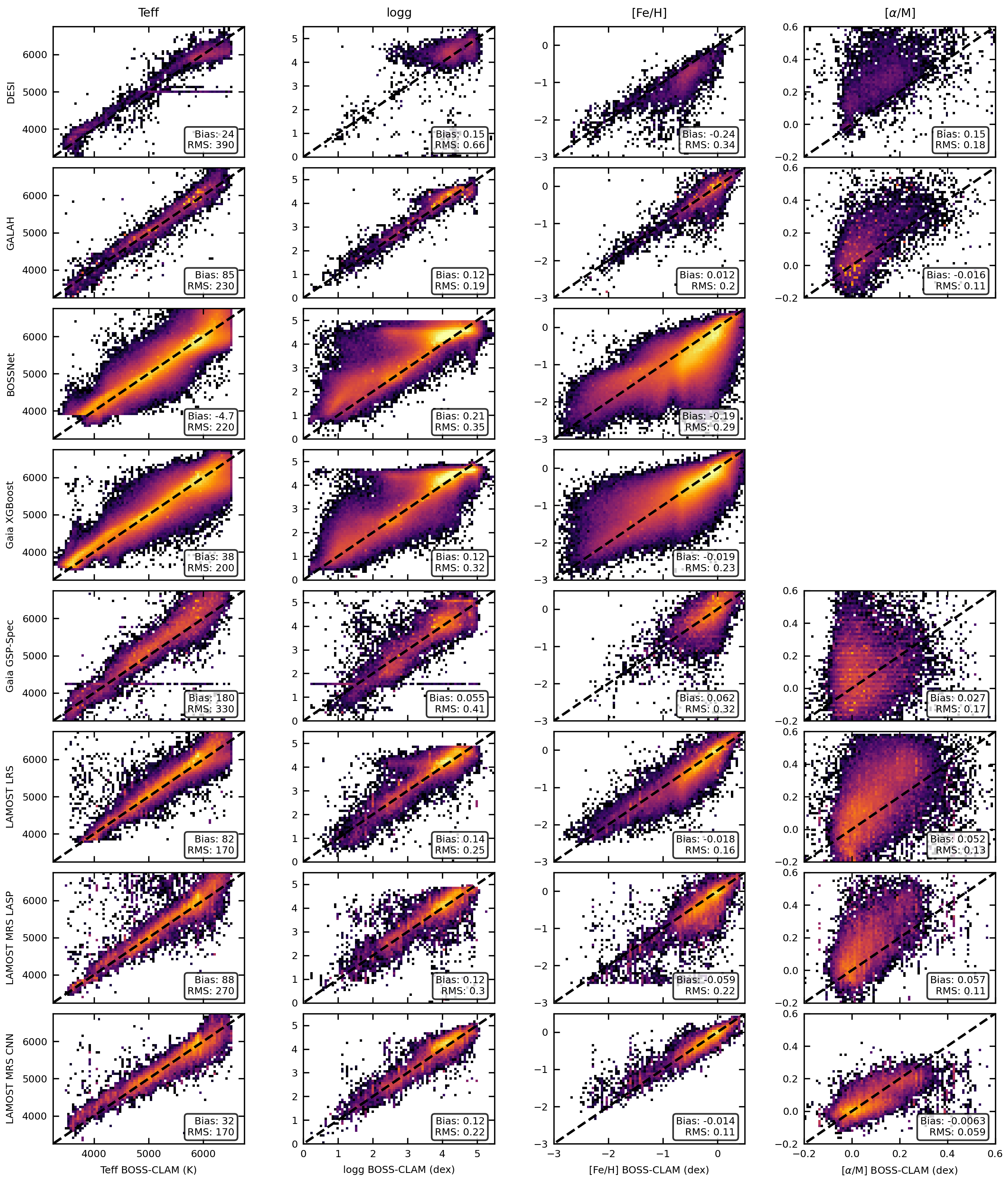}
    \caption{A comparison of the BOSS-CLAM stellar parameters to those measured from 8 other methods. Columns are grouped into four parameters: effective temperature, surface gravity, metallicity, and $\alpha$-abundance. Each row corresponds to a different catalog which is labeled on the far left. In cases where a catalog does not provide $\alpha$-abundance, we leave the plot empty. A one-to-one dashed line is plotted and the comparison's bias and root mean square (RMS) are shown in the bottom right of each panel.}
    \label{fig:xsurvey_compare}
\end{figure*}

Now that we have derived parameters, it is beneficial to examine how our results compare with those of other optical or near-infrared surveys. The surveys we include are: DESI's Milky Way Survey DR1 \citep[$R\sim5,000$;][]{DESI2023_mws},
GALAH DR4 \citep[$R\sim28,000$;][]{GALAH2025},
SDSS-V's BOSSNet \citep[$R\sim2,600$;][]{bossnet},
\textit{Gaia} DR3's XGBoost catalog from the XP spectra \citep[$R\sim50$;][]{Andrae2023_xgboost},
\textit{Gaia} DR3's GSP-Spec from the RVS spectra \citep[$R\sim11,500$;][]{Gaia2023GSPSpec}, and 
LAMOST DR11's Low-Resolution (LRS; $R\sim1,800$) and Medium-Resolution (MRS; $R\sim7,500$) Stellar Parameter Catalogues \citep{LAMOST2026DR11}\footnote{https://www.lamost.org/dr11/v2.0/catalogue}. In LAMOST's MRS catalog, we compare the results from the LAMOST stellar parameter (LASP) and convolutional neural network (CNN) pipelines.  We match to each catalog based on the reported \textit{Gaia} DR3 \texttt{source\_id} and filter out stars in BOSS-CLAM with \texttt{flag\_bad = False} or \texttt{clam\_flags = 0}. The resulting comparison is shown in Figure \ref{fig:xsurvey_compare}.

Our parameters agree in large part with other surveys. For effective temperature ($T_\mathrm{eff}$), surface gravity ($\log g$), and metallicity ($[\mathrm{Fe/H}]$), there is scatter which is consistent with our own precision discussed below in Section \ref{sec:valid}. When comparing to other surveys that measure $[\alpha/\mathrm{M}]$, we see larger scatter and bias, which we attribute to our relatively lower-resolution spectra; this parameter is just more difficult to measure. 

\added{Some surveys report multiple versions, or calibrations for their parameters and are worth clearly differentiating. For DESI, we use parameters from their \textsc{SP} pipeline. The iron abundances from this catalog require a calibration found on the survey's GitHub repository\footnote{\url{https://github.com/desimilkyway/dr1_vac_code/}}, which is what we compare to in Figure \ref{fig:xsurvey_compare}. In GALAH there are two surface gravities reported, one which is calibrated based on parallax (\textsc{logg}) and one which directly comes from the spectroscopic fit (\textsc{logg\_spec}). We choose to compare to the former because it is what is recommended by the GALAH DR4 paper.}

In particular, we observe strong agreement with the higher-resolution GALAH DR4, \textit{Gaia} GSP-Spec, and the LAMOST MRS measurements. This supports that the model is likely predicting the correct parameters because these external surveys can actually resolve some of the lines which are blended in BOSS. We agree with the other low-resolution surveys, \textit{Gaia} XGBoost and LAMOST's LRS parameters, with some expected scatter due to fewer features being resolved. We do, however, see some disagreement with other low-resolution catalogs, such as BOSSNet and DESI. \added{While we discuss our comparison to BOSSNet in depth in Section~\ref{sec:intro}, the discrepancy with DESI may be explained by a few different reasons. It may be that DESI uses a different method to measure stellar parameters. Similar to ASPCAP in SDSS-V, they use the code FERRE \citep{AllendePrieto2006FERRE} to fit the spectra with models generated from PHOENIX \citep{Husser2013_PHOENIX} model atmospheres. Their method first performs a grid search over stellar parameters before fitting individual abundances within windows. Since our parameters are learned from the APOGEE infrared abundances, this may signal a difference in abundance measurements in the optical and infrared or that the models employed by each survey significantly disagree. This discrepancy may also arise from the different underlying selection of stars. Our match between the BOSS and DESI data is largely composed of main sequence stars, as can be seen in the surface gravity comparison. In their first data release, DESI shows strong systematic trends in this region (see Figure 5 of \citet{DESI2023_mws}).}

\subsection{Understanding Failure Modes of the BOSS-CLAM Model}

The above examines the results for the cleaned BOSS-CLAM dataset. Stars removed by our conservative flagging make up a large proportion of the DR20 dataset, though. So, these objects warrant further discussion.

\begin{figure*}
    \includegraphics[width=\textwidth]{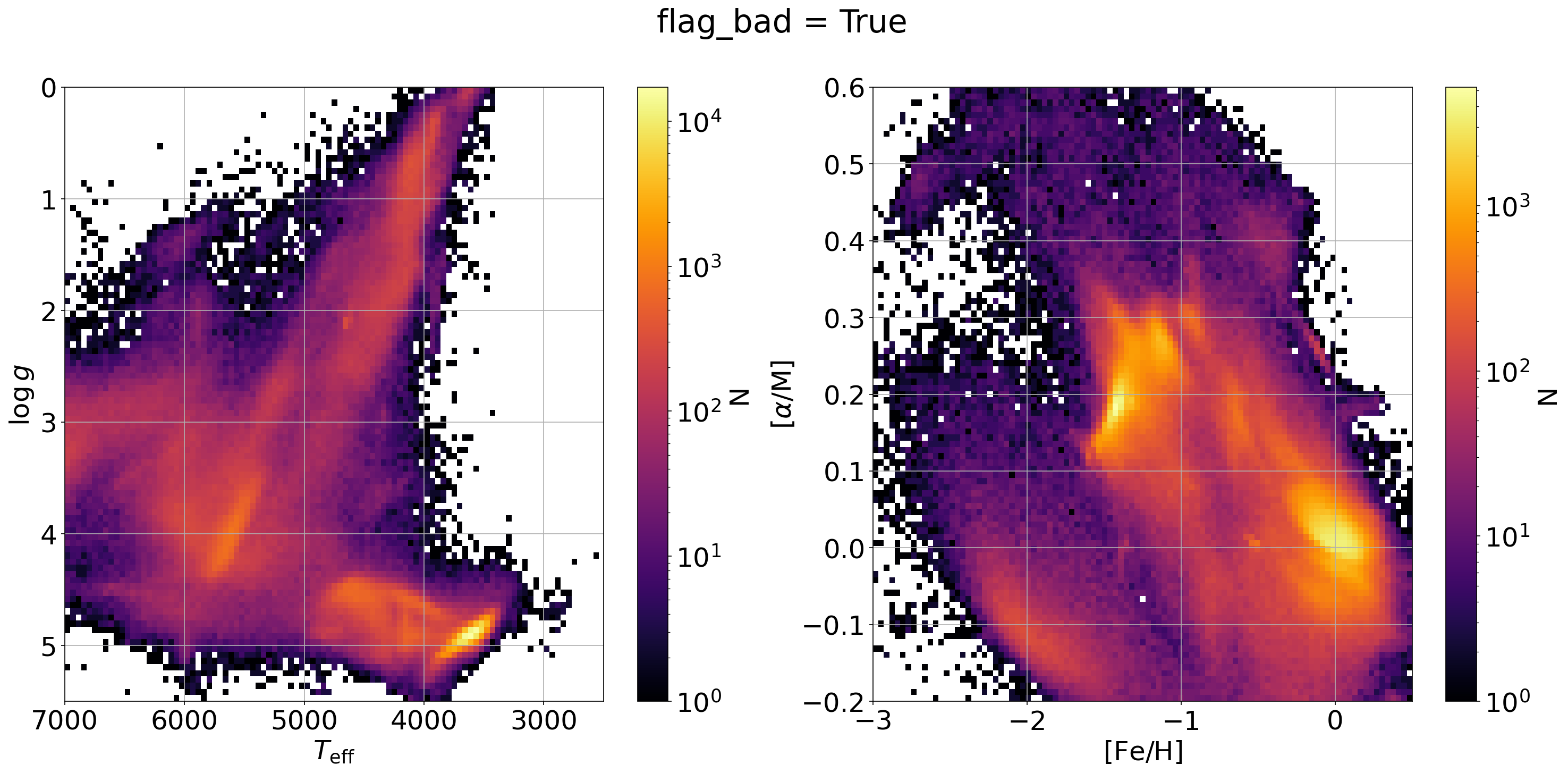}
    \includegraphics[width=0.9\textwidth]{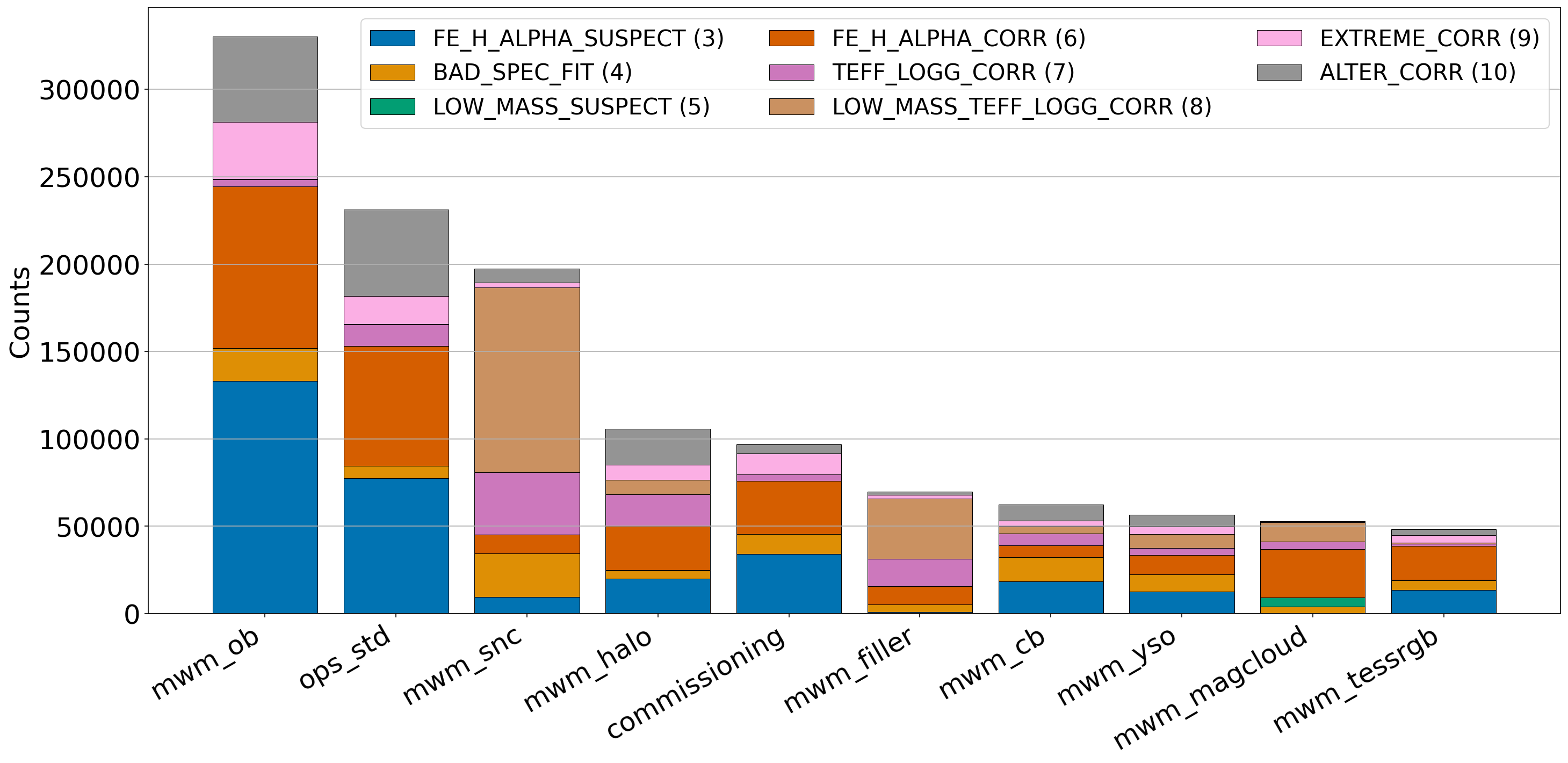}
    \caption{\textbf{Top Row:} Kiel diagram (left) and distribution of $[\alpha/\mathrm{M}]$ vs.~$[\mathrm{Fe/H}]$ (right) for inferred parameters from the BOSS-CLAM with \texttt{flag\_bad = True}. \textbf{Bottom Row:} Bar graph showing the counts of stars within various SDSS-V programs that have the various \texttt{clam\_flags} bits set. This only shows the ten most numerous programs. Also, we should note that a star can be within multiple programs and have multiple flags set.}
    \label{fig:bad_results}
\end{figure*}

The top row of Figure \ref{fig:bad_results} shows both the Kiel and $[\alpha/\mathrm{M}]$ vs.~$[\mathrm{Fe/H}]$ diagrams for stars with \texttt{flag\_bad = True}. This shows where the failure modes for our model are. Largely, things that are flagged are stars with peculiar chemistry or low-mass stars that have the incorrect $T_\mathrm{eff}$ and $\log g$ correlation. The former are failures of our model for various reasons (discussed below). The latter is due to low-mass stars being extremely underrepresented in the training set.

Of course, this only shows the inferred BOSS-CLAM parameters, which are likely wrong and do not reflect the true parameters for the sources. To better understand which classes of stars the model fails with, the bottom row of Figure \ref{fig:bad_results} shows a bar graph of the counts of stars within the top ten SDSS-V programs that have the various \texttt{clam\_flags} bits set. We should note that Open Fiber (\texttt{open\_fiber}) is only included in these statistics if a star is only requested by an Open Fiber carton. This is because the vast majority of stars are primarily requested by an MWM or operations program. This caveat causes \texttt{open\_fiber} to not appear in this plot. Finally, a star can be included in multiple programs, so in such cases this star will contribute to multiple bars in the graph. 

From the bar graph, it is clear that, outside of the white dwarfs and young stellar objects that we already flag, there are only a few types of stars that dominate our bad fits. The main class is OB stars (\texttt{mwm\_ob}), which are normally associated with flags indicating a dubious $[\mathrm{Fe/H}]$ or $[\alpha/\mathrm{M}]$. This makes sense given the lack of these parameters in the training set for hot stars. Next are the standard stars used for MWM (\texttt{ops\_std}), which are primarily F-type stars. The more luminous and hotter of these stars lie within the temperature range where we have very little training data ($6500 < T_\mathrm{eff}< 7200$). Similarly, the Solar Neighborhood Census (\texttt{mwm\_snc}) is made up primarily of low-mass stars, which also make up a considerable portion of all MWM BOSS data, and are again very underrepresented in our training data. Future work is needed for these types of stars in the form of additional training labels to improve the model.

For some of the other programs, while they are generally represented in the training set, they likely contain a subset of chemically peculiar stars that would cause the inference to fail. For example, the halo sample (\texttt{mwm\_halo}) likely has sources with unusually enhanced carbon features. As this is not included in the fit, the $[\alpha/\mathrm{M}]$ may be suspect. Additionally, the method is failing with candidate compact binaries (\texttt{mwm\_cb}), which likely have polluted atmospheres. Similarly, some hot stars likely have polluted atmospheres.

Finally, while this is not captured in the above, we should note that poor continuum normalization can result in bad stellar parameters. These will not necessarily be flagged as such in these results (see Section \ref{sec:apo_lco}), but changes to the continuum can appear as increased opacity, thus changing the metallicity derived from BOSS-CLAM. These changes can be subtle and difficult to flag, so they should be considered when using these data.

Overall, this targeting information demonstrates that 1) the flags are correctly capturing objects that we know should fail with our method, and 2) that these ``bad" parameters are largely isolated to these targeting classes. If, in the future, better labels for these types of stars are collected, the BOSS-CLAM could be retrained to improve these results.

\section{BOSS-CLAM Parameter Validation}\label{sec:valid}

The above results were analyzed only in the context of the BOSS-CLAM parameters, optimization results, and target programs. Here, we perform a series of validation tests on the stellar parameters using information from additional sources. These will demonstrate the accuracy and precision of the results, and 
how they compare to current SDSS-V pipelines. Similar to Section \ref{sec:result}, when examining the parameters, if we are examining only $T_\mathrm{eff}$ and $\log g$, we will only include sources with \texttt{flag\_bad = False}. If we also include $[\mathrm{Fe/H}]$ and/or $[\alpha/\mathrm{M}]$, we will only include sources with \texttt{clam\_flags = 0}. 

\subsection{Star Clusters}\label{sec:valid_clusters}


Star clusters serve as an important validation for stellar parameter pipelines, as you expect the abundances to be homogeneous across the Kiel diagram. To test this, we use the 
Open Cluster Chemical Abundance and Mapping (OCCAM) survey \citep{otto_26} and the catalog of \citet{VasilievBaumgardt2021} to select confident members of both OCs and GCs for our validation tests. 

The validation catalog of open cluster stars was made by making use of the existing OCCAM pipeline, first developed by \citet{donor2018} and last updated by \citet{otto_26}. 
We start by selecting stars in the \citet{Hunt2024ImprovingCatalogue} catalog with BOSS parameters.
A Gaussian kernel smoothing routine is used to determine (1) the bulk cluster radial velocity (RV) and (2) the associated RV membership probabilities for each star in the cluster.
For this work, we only kept clusters with 20 member stars that had high RV membership probability (i.e., RV Prob $> 0.05$).  Of those clusters that passed this criterion, five were clusters with known [Fe/H] abundances as determined by \citet{otto_26} using APOGEE spectra, with an additional six clusters without previous APOGEE-based [Fe/H] abundances.


From the globular cluster catalog of \citet{VasilievBaumgardt2021}, we perform an additional filtering of stars based on their radial velocities. We start by selecting all members associated with a cluster and then follow a similar approach to \citet{SchiavonDR17VAC}, building a two-component Gaussian Mixture Model (GMM) to select members in radial velocity space\footnote{One change we make compared to \citet{SchiavonDR17VAC} is the use of Extreme Deconvolution \citep{BovyXDGMM} instead of a plain GMM, which allows us to naturally account for the uncertainties in the RV measurements.}. After constructing the RV membership model, we select all stars with a membership probability greater than $0.9$ in both the radial velocity membership model and the astrometric membership probability from \citet{VasilievBaumgardt2021}. Finally, we drop clusters with fewer than 10 remaining confident members. We use both of these catalogs for the following validation tests.

\begin{figure*}
    \centering
    \includegraphics[width=\textwidth]{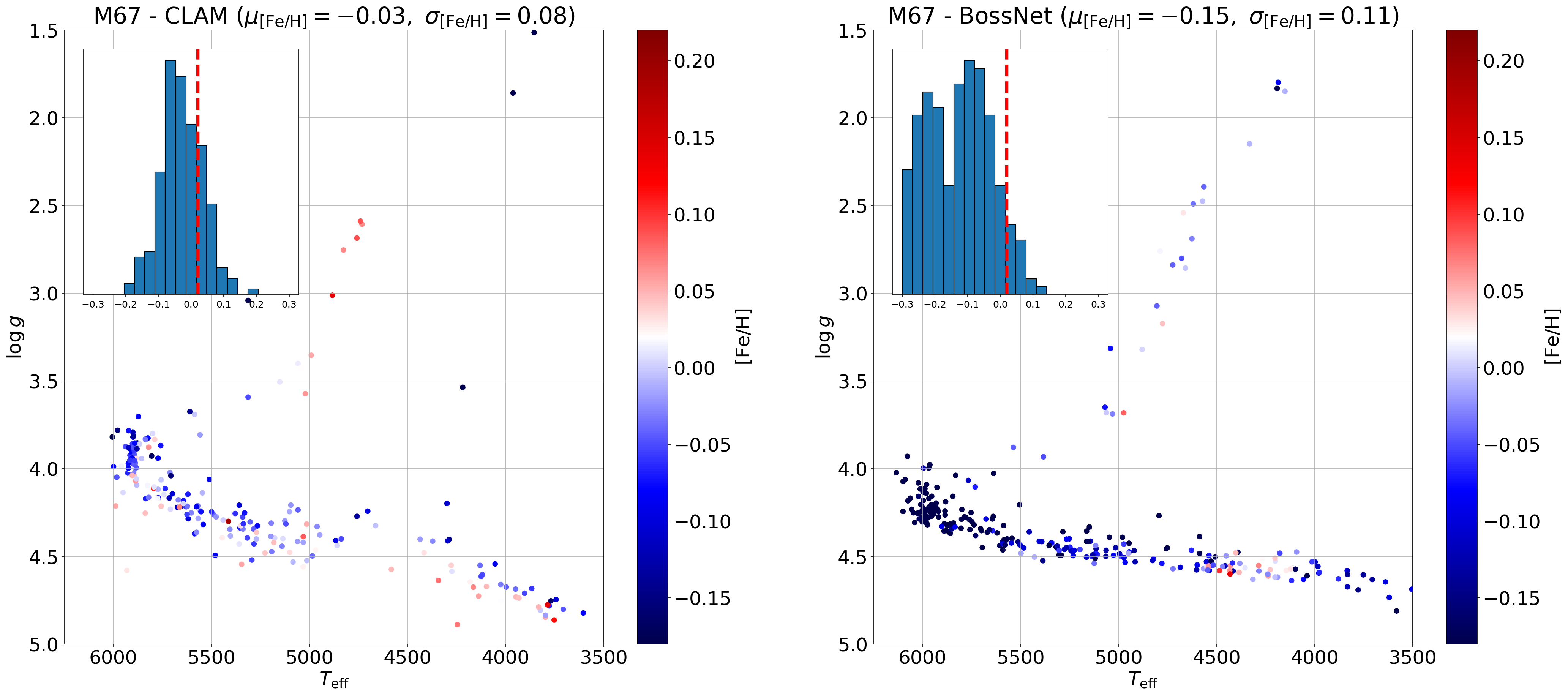}
    \includegraphics[width=\textwidth]{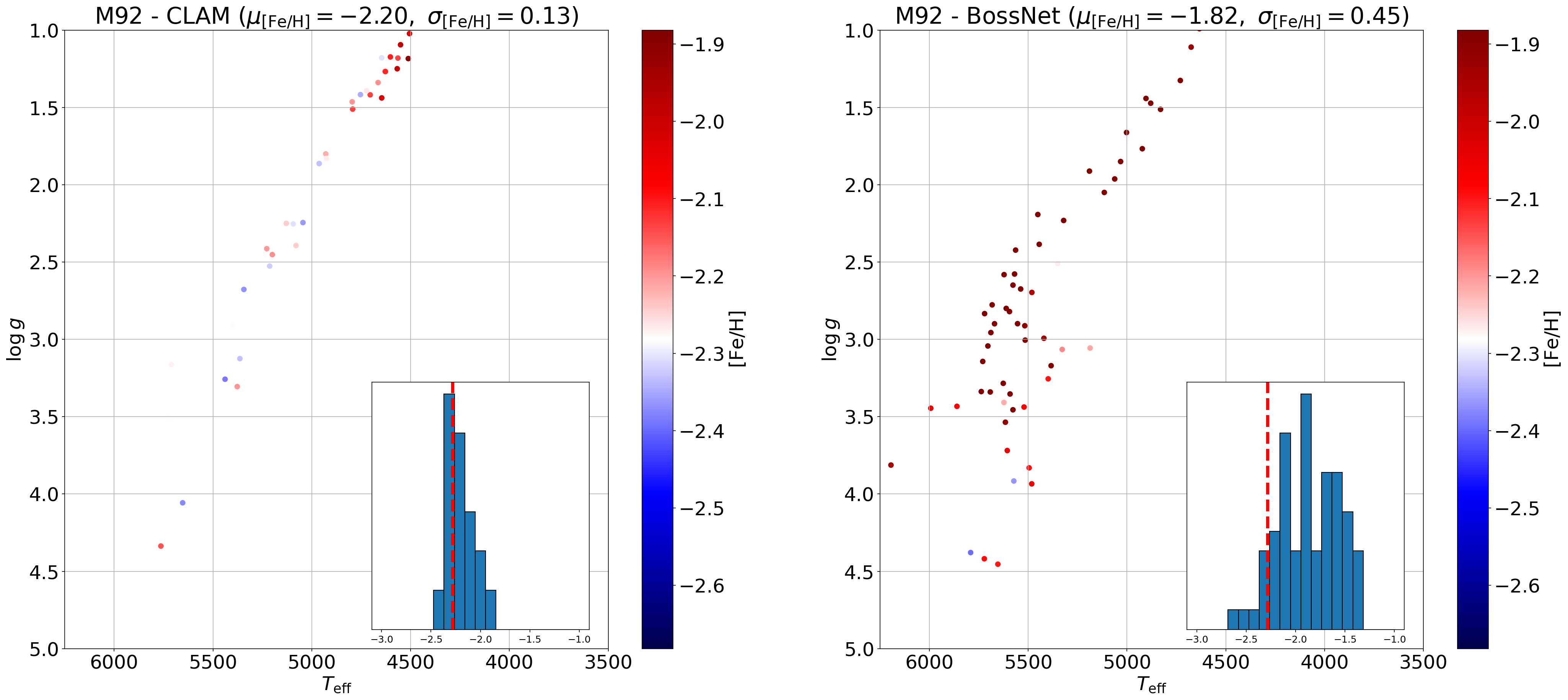}
    \caption{Kiel diagram for M67 (top row) and M92 (bottom row) using the BOSS-CLAM parameters (left column) and BOSSNet (right panel). The scatter points are colored by $[\mathrm{Fe/H}]$ with the colorbar centered on $[\mathrm{Fe/H}]=0.02$ for M67 \citep{Heiter2014} and $[\mathrm{Fe/H}]=-2.282$ for M92 \citep{lee2023}. The inlay shows the distribution of the $[\mathrm{Fe/H}]$ values from each pipeline, and the red dashed line is again the approximate metallicity of the cluster from the literature.}
    \label{fig:clusters}
\end{figure*}

The top row of Figure \ref{fig:clusters} shows the Kiel diagram for M67. Here we compare the results from the BOSS-CLAM to BOSSNet \citep{bossnet}. M67 serves as an important reference as a wide range of $T_\mathrm{eff}$ and $\log g$ is covered in SDSS-V, and it is around Solar metallicity. For the current BOSSNet results, there are clear systematics with $[\mathrm{Fe/H}]$ as a function of $T_\mathrm{eff}$ and $\log g$. Additionally, the mean metallicity, according to BOSSNet, is significantly below Solar. The BOSS-CLAM, on the other hand, correctly predicts the metallicity of the cluster, within $1\sigma$. Additionally, we find that across the full Kiel diagram, stars are roughly normally distributed around this value, with only some slight differences around the sub-giants. This difference near the turnoff in M67 is expected and identified in other studies \citep[e.g.,][]{M67_Diff_2014,Souto2018,Souto2019}. The APOGEE training set, including M67, recovers the effects from diffusion processes near the turn-off, resulting in a shift of $\leq -0.1$ dex \citep{Souto2019}, similar to what is seen here.

Focusing on a more metal-poor globular cluster, M92, the bottom row of Figure \ref{fig:clusters} shows how BOSSNet is not able to predict values metal-poor enough. This is a trend that has been observed in previous studies \citep{minesweeper}. The BOSS-CLAM on the other hand can predict giants this metal-poor, thanks to the inclusion of \texttt{BOSS-MINESweeper} in the training set. Similar to M67, we also find that across the giant branch, we can consistently predict the cluster's metallicity with relatively low scatter.

\begin{figure*}
    \centering
    \includegraphics[width=\textwidth]{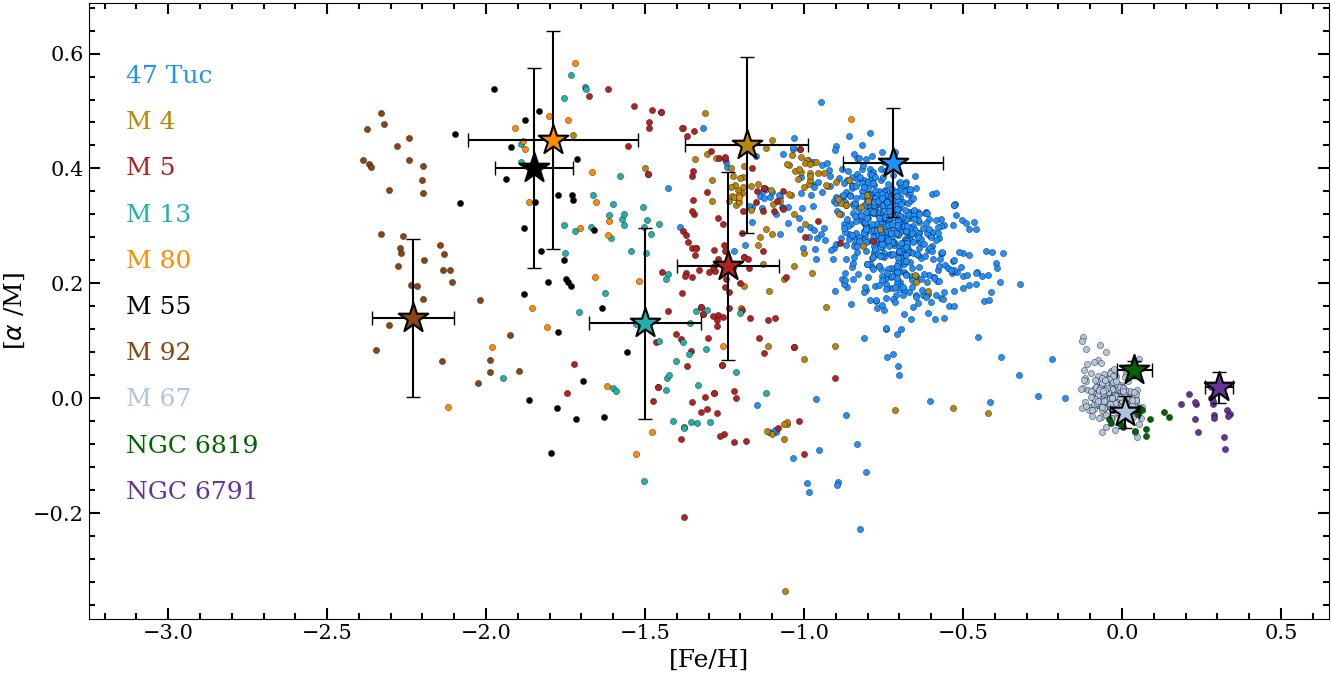}
    \caption{ Cluster member stars plotted in the [Fe/H] versus [$\alpha$/M] plane for several globular and open clusters. 
    Colored stars show the mean [Fe/H] and [$\alpha$/M] value from the literature, with the error bars showing the 1$\sigma$ error. 
    For the globular cluster literature abundances, we use the list from \citet{minesweeper}, which was collated from a variety of sources, including: \citet{roediger2014} (47 Tuc), \citet{ivans1999} (M 4), \citet{meszaros2015} (M 5, M 13), \citet{carretta2015} (M 80), and \citet{vandenberg2018} (M 55). 
    Literature abundance values for the open clusters (M 67, NGC 6819, and NGC 6791) were taken from \citet{otto_26}.}
    \label{fig:all_clusters}
\end{figure*}

Next, we can examine a wide range of clusters in both [Fe/H] and [$\alpha$/M] to assess the accuracy of these parameters. Figure \ref{fig:all_clusters} shows the various cluster members from our globular and open clusters catalogs plotted in [Fe/H] versus [$\alpha$/M]. The scatter points show the BOSS-CLAM values, where we do see an increasing spread with decreasing metallicity. The colored stars show the mean [Fe/H] and [$\alpha$/M] values from the literature, with error bars showing the 1$\sigma$ errors. We find that across all clusters, the BOSS-CLAM metallicity aligns with the literature value, though the scatter tends to be larger. For the [$\alpha$/M] values, we find that BOSS-CLAM can under- or over-estimate the value relative to the literature for all clusters. Such systematics in [$\alpha$/M] should be considered when working with these data.

\begin{figure*}
    \centering
    \includegraphics[width=\textwidth]{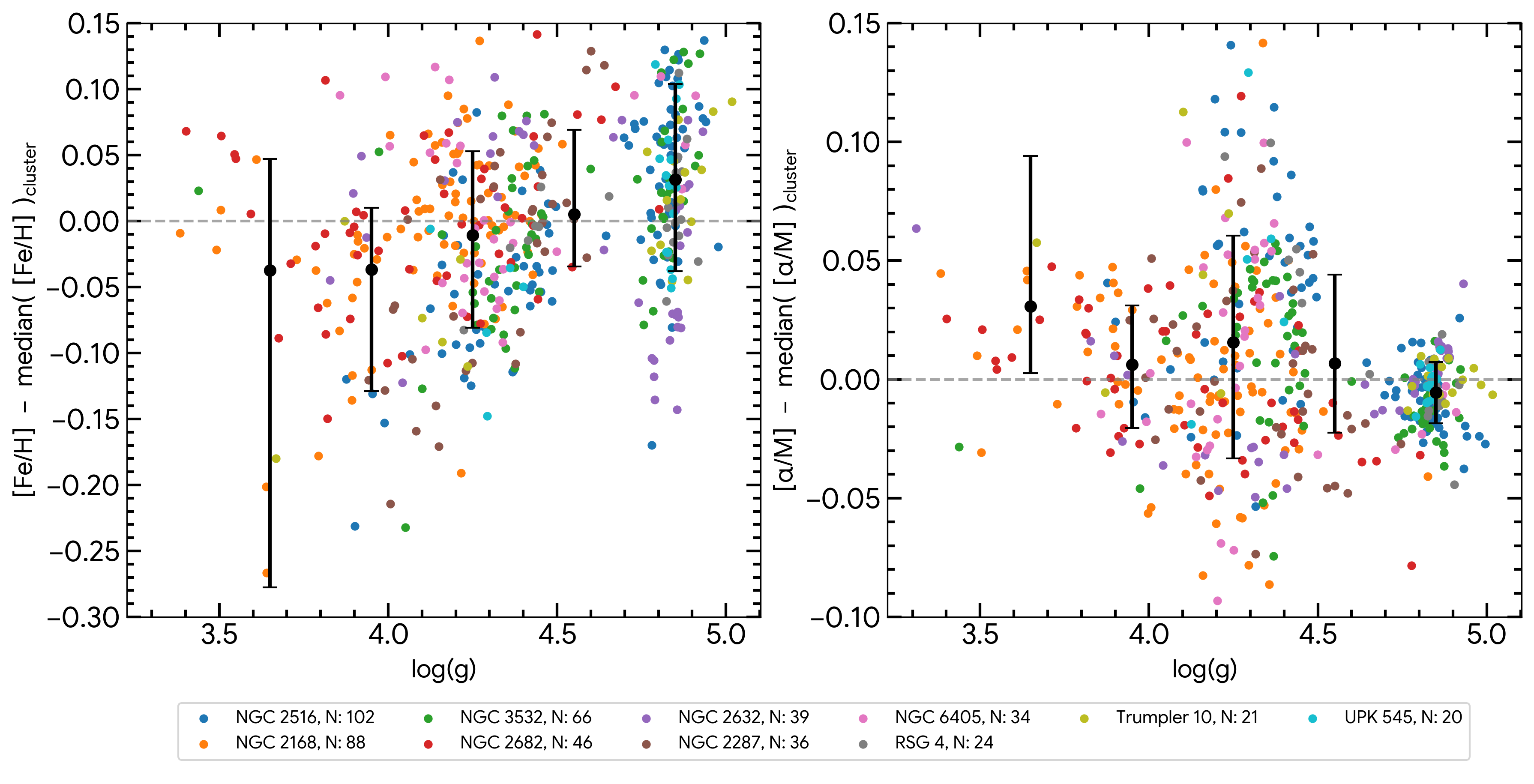}
    \caption{Residual BOSS-CLAM $[\mathrm{Fe/H}]$ (left) and $[\alpha/\mathrm{M}]$ (right) relative to the respective cluster median values are shown against $\log g$ for 10 open clusters from the \citet{Hunt2024ImprovingCatalogue} catalog. Black error bars denote the 16$^{\mathrm{th}}$, median, and 84$^{\mathrm{th}}$ percentiles in $\log g$ bins of width 0.3 from 3.5 to 5. The legend indicates the number of members with $\log g > 3.3$ and membership probability $\geq0.8$ with BOSS-CLAM parameters shown for each cluster. }
    \label{fig:hunt_validation}
\end{figure*}

As a final test of intra-cluster BOSS-CLAM parameter homogeneity across the Kiel diagram, specifically for dwarfs and subgiants, we additionally utilize the cluster catalog from \citet{Hunt2024ImprovingCatalogue}. Figure \ref{fig:hunt_validation} shows the residual [Fe/H] and $[\alpha/\mathrm{M}]$ relative to the median values for 10 prominent open clusters, chosen such that they have $\geq20$ members with BOSS-CLAM parameters with membership probability $\geq0.8$ and $\log g\geq3.3$. The latter is imposed due to the sparser sampling of BOSS-CLAM results for lower-gravity cluster members. Within these clusters, we find the $[\mathrm{Fe/H}]$ spread is of order $\sim0.14$ dex across the dwarfs and subgiants. The $[\alpha/\mathrm{M}]$ spread, however, markedly decreases from $\sim0.08$ in the subgiant regime to $\sim0.03$ in the dwarfs.

Overall, these tests demonstrate that the BOSS-CLAM abundances are consistent over a wide range of $T_\mathrm{eff}$ and $\log g$. Additionally, BOSS-CLAM can accurately report parameters even at low metallicities. Some discrepancies are observed in [$\alpha$/M] from these tests, which should be considered. We note that the addition of C and N labels in the model could break such degeneracies, and help to better understand the [$\alpha$/M] spreads in the globular clusters and whether they are caused by the multiple stellar population phenomenon. Such changes will be considered in future work (see Section \ref{sec:improve}). But overall, this demonstrates the clear utility of these parameters and how they improve on results from other pipelines.

\subsection{Wide Binaries}\label{sec:valid_wbs}


Wide binaries provide another useful resource for testing the accuracy and precision of the stellar parameters from our model. In particular, we can look at the bias in abundances for stars of different temperatures and surface gravities, and look at the scatter in the parameters for stars of similar temperatures and surface gravities. To probe this, we use the wide binaries from \citet{elbadry2021} that meet the same quality cuts as in Section \ref{sec:data_wb}.

\begin{figure*}
    \centering
    \includegraphics[width=\textwidth]{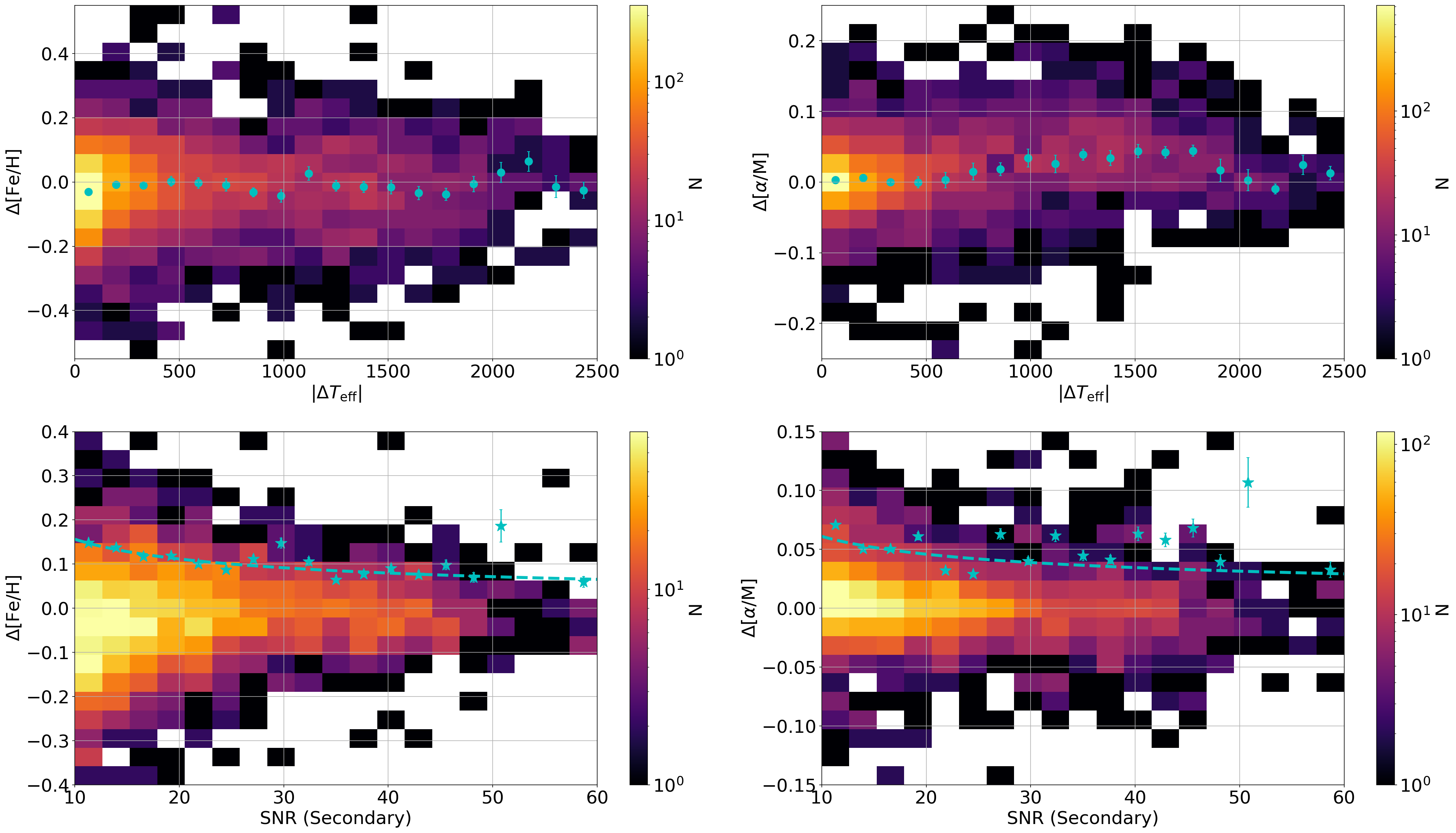}
    \caption{Comparison of abundances from the BOSS-CLAM for wide binary pairs that pass the quality cuts outlined in Section \ref{sec:data_wb}. The top row shows the difference in $[\mathrm{Fe/H}]$ and $[\alpha/\mathrm{M}]$ as a function of the difference in temperature between the primary and secondary. The bottom row shows the difference in $[\mathrm{Fe/H}]$ and/or $[\alpha/\mathrm{M}]$ as a function of SNR of the secondary. Only binaries with $|\Delta T_\mathrm{eff}| < 500$ K are included in the bottom row. In the top row, the cyan points show the mean difference in abundances, tracing the bias in the parameters as a function of the temperature difference. In the bottom row, the cyan stars show the standard deviation of the difference as a function of SNR to trace the average error on the parameters. We also fit these data as $\sigma = A \times SNR^{-B}$, where we find for $[\mathrm{Fe/H}]$ $A=0.479$ and $B=0.486$, and for $[\alpha/\mathrm{M}]$ $A=0.157$ and $B=0.411$, and is shown as the dashed cyan line.}
    \label{fig:widebinary_compare}
\end{figure*}

Figure \ref{fig:widebinary_compare} shows the comparison of abundances from the BOSS-CLAM for these binaries. Specifically, the top row shows the difference in $[\mathrm{Fe/H}]$ and $[\alpha/\mathrm{M}]$ as a function of the difference in temperature between the primary and secondary. We include the average difference as a function of temperature as the cyan data points. Here we see that the bias as you go to pairs of differing temperatures is minimal. Some small trends are observed, but the differences are small compared to the scatter.

The bottom row of Figure \ref{fig:widebinary_compare} shows the difference in $[\mathrm{Fe/H}]$ and/or $[\alpha/\mathrm{M}]$ as a function SNR of the secondary for pairs with $|\Delta T_\mathrm{eff}| < 500$ K. This importantly probes the average scatter as a function of SNR for stars with similar spectra. The cyan stars show the scatter as a function of SNR, where we additionally fit the points with $\sigma = A \times SNR^{-B}$. here we find that for $[\mathrm{Fe/H}]$ $A=0.479$ and $B=0.486$, and for $[\alpha/\mathrm{M}]$ $A=0.157$ and $B=0.411$, implying that $\sigma_{[\mathrm{Fe/H}]} \approx 0.15$ dex and $\sigma_{[\alpha/\mathrm{M}]} \approx 0.06$ at SNR = 10. This statistical uncertainty can be used to approximate the errors on the abundances from the BOSS-CLAM as a function of SNR. We note that this is simply an average, and the errors will likely vary with $T_\mathrm{eff}$ and $\log g$, which requires a more detailed error analysis.

\subsection{APO vs.~LCO Spectra}\label{sec:apo_lco}

SDSS-V is being conducted at two observatories to facilitate all-sky coverage. This allows for some overlap in the fields from the two observatories, meaning that some stars are observed by both. This provides a compelling validation, as we can see how consistently the model can recover parameters from spectra of the same star at the two observatories.

\begin{figure*}
    \centering
    \includegraphics[width=\textwidth]{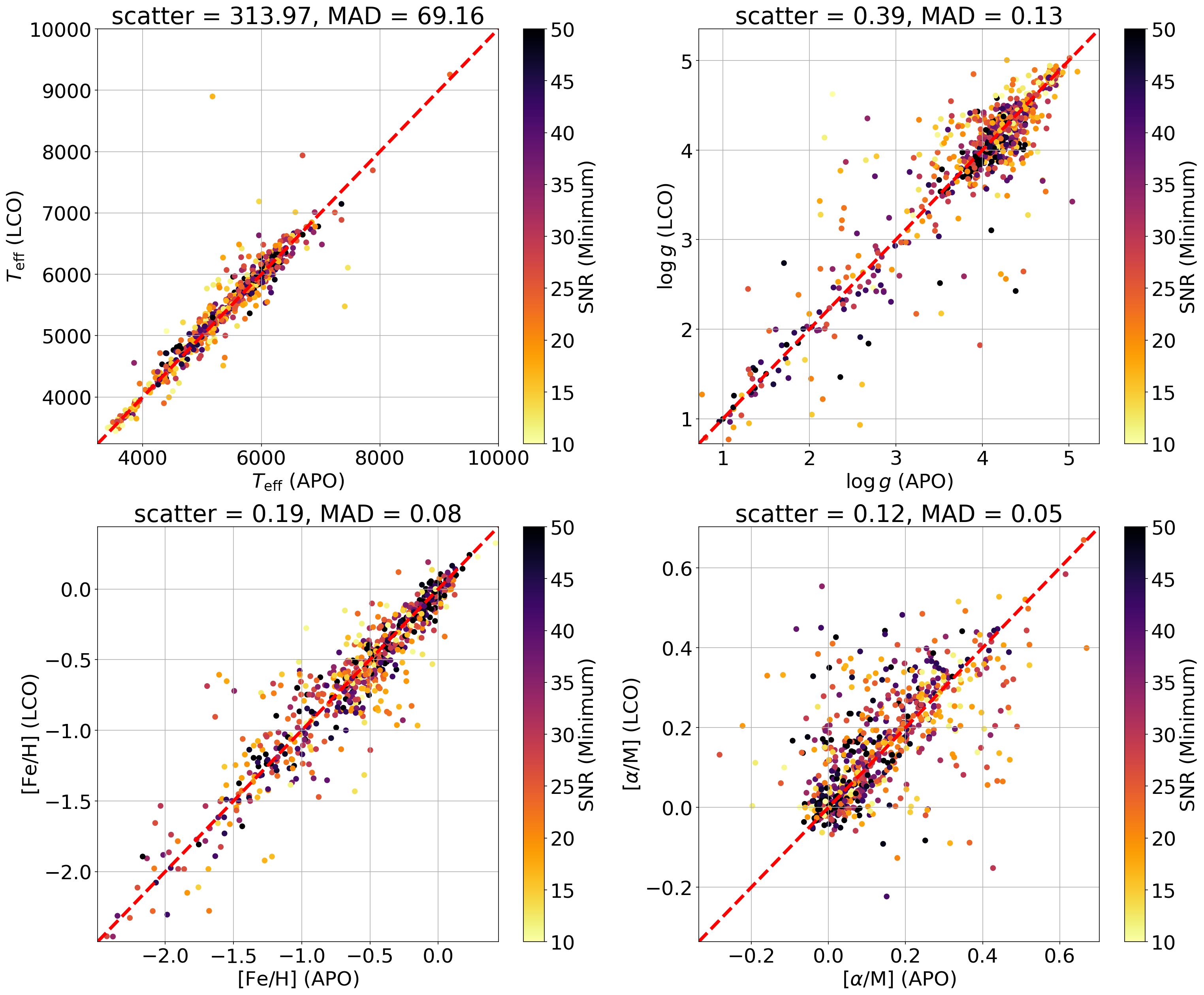}
    \caption{Comparison of parameters from the BOSS-CLAM for 719 stars with spectra taken at both APO and LCO. The points are color-coded by the minimum SNR from the two spectra (cut at SNR to better show the dynamic range). The title above each plot shows the scatter and median absolute deviation between the parameters derived from each spectrum.}
    \label{fig:apolco_compare}
\end{figure*}

Figure \ref{fig:apolco_compare} shows the 719 stars that have spectra from both observatories and pass our BOSS-CLAM quality cuts. The points are color-coded by the minimum SNR from the two spectra (cut at SNR$=50$ to better show the dynamic range), and the title above each panel shows overall scatter and median absolute deviation (MAD) between the parameters derived from each spectrum. As expected, we find that lower-SNR spectra have higher scatter than higher-SNR spectra. Additionally, the scatter and MAD match what we found in the results from the testing data (Figure \ref{fig:clam_results}).

Clearly, there are some stars, even at high SNR, that have greatly differing parameters, though. This is likely due to flux calibration issues. In some cases, the BOSS spectra can have large flux calibration issues that greatly change the resulting SED of the source \citep{medan2025}. The CLAM, which calculates the continuum for the SDSS-V spectra, is meant to account for such issues by jointly fitting absorption, continuum, and flux calibration errors. When such errors are large and the SNR is high, the joint model can get slightly different solutions for the underlying thermal continuum for the same source. This will result in different continuum-normalized spectra and thus different stellar parameters.

This shows that there are likely some systematic errors in the BOSS-CLAM parameters due to these flux calibration issues. This test demonstrates that it occurs in low levels, though, and that the majority of sources are not affected. Such issues should be considered when interpreting these results, especially at the single-object level.

\section{Improvements to Method}\label{sec:improve}

While we have shown above that the parameters resulting from the BOSS-CLAM produce results consistent with the training set (Section \ref{sec:result}) and pass a series of external validation tests (Section \ref{sec:valid}), there are still some improvements that could be made to the method. These improvements are planned to be included in subsequent iterations of the BOSS-CLAM and released with future data releases of SDSS-V. 

First, there could be improvements in the $T_\mathrm{eff}$ and $\log g$ sequences resulting from the BOSS-CLAM. Primarily, we found in Figure \ref{fig:clusters} that the isochrone sequence, compared to BOSSNet, has a much wider spread, especially around the main-sequence turnoff. This can be improved by having an isochrone prior on the parameters \citep[see e.g.,][]{minesweeper} or by adding photometry into the analysis to reduce degeneracies between these parameters \citep[see e.g.,][]{Zhang2023}. Either path would likely greatly improve the parameters, though this requires more implementation and testing of the method and its results.

Similarly, there are strong degeneracies with $[\mathrm{Fe/H}]$ and $[\alpha/\mathrm{M}]$, especially for $\alpha-$rich stars (Figure \ref{fig:alpha_fe_results}). This is likely the result of fitting the entire spectrum. At the moderate resolution of BOSS, the added continuum opacity from more metals in the star's atmosphere dominates over the individual lines in the spectrum, especially when equal weight is applied to all spectral pixels. This means $[\mathrm{Fe/H}]$ and $[\alpha/\mathrm{M}]$ can be traded in equal amounts and result in similar spectrum shapes. This will dominate the $\chi^2$ compared to the individual lines, which may not be fit as well as a result. In the future, we could give greater weight to spectral windows known to contain lines sensitive to the star's metallicity. This would force the model to fit these features rather than the overall shape of the spectrum, thereby removing this degeneracy. This is complicated by the fact that such windows may change slightly with $T_\mathrm{eff}$ and $\log g$. Careful work is needed to define such windows over a wide range of the HR diagram.

Next, as mentioned multiple times in this paper, there are a few classes of objects for which the trained model, in its current form, fails: some OB stars, hot F dwarfs, cool M dwarfs, and some likely carbon- or nitrogen-enhanced stars. This is especially problematic for the M dwarfs, which make up about 1/3 of all BOSS MWM spectra \citep{medan2025}. Additionally, it is unfortunate that interesting stars in e.g. the halo sample with peculiar chemistry are removed by our quality cuts. This likely requires two additions to the method. First, more training data is needed in this regime to produce better parameters for these classes of objects. Additionally, the model would benefit from more dimensions to the label set, such as including carbon and nitrogen abundances, which again requires good labels from various sources. Such additions would greatly expand the number of BOSS spectra for which the model can accurately predict parameters.

Finally, a more robust error analysis would greatly enhance the results. In the current work, we have provided an estimate of the statistical uncertainty through the inverse of the Hessian (see Section \ref{sec:result}) and an average of the systematic uncertainties (Figure \ref{fig:widebinary_compare}). In future iterations, this could be improved. Firstly, with a forward model, the full posterior of the parameters can be determined via MCMC. With such a large dataset, this is computationally expensive, which is why it was not done in the current iteration. \added{Additionally, we do find this is not needed is the majority of cases (Appendix \ref{app:mcmc}).} We could \added{also} better formalize the systematic uncertainties. This can be done through e.g.~an ensemble of trained models, where each model is trained on a different subset of that data. This ensemble could then be used to infer the parameters, and the variance in results would provide a much better estimate of the systematic uncertainties. Again, this would be computationally expensive at the inference stage for such a large dataset.

All of the above improvements are under consideration for this method and will continue to be developed in future iterations of BOSS-CLAM. If these improvements can be implemented, future releases of the BOSS-CLAM parameters with SDSS-V will include them, resulting in enhanced results for the community. At this stage, the results in their current form still provide accurate parameters for a very wide range of the HR diagram and can be used for a variety of scientific analyses.

\section{Example Astrophysical Applications of BOSS-CLAM Parameters}\label{sec:applications}


The BOSS-CLAM framework provides precise, scalable stellar parameters across a large fraction of the SDSS-V BOSS sample, enabling a wide range of astrophysical applications. In this section, we outline several representative use cases that highlight the scientific utility of the BOSS-CLAM catalog, particularly in the context of Galactic archaeology, stellar population studies, and data-driven discovery.

\subsection{Galactic Chemical Evolution and Disk Structure}

Large spectroscopic surveys have revealed that the joint distribution of [$\alpha$/M] and [Fe/H] encodes the star formation and enrichment history of the Milky Way disk. In particular, APOGEE studies have demonstrated a bimodal $\alpha$-sequence corresponding to chemically distinct thin and thick disk populations \citep{Nidever2014,Hayden2015}. The APOGEE data is more limited in scope than the BOSS data in SDSS-V. BOSS allows for the observation of much fainter populations, especially at higher Galactic scale heights.

With BOSS-CLAM, such analyses can be extended to the optical BOSS sample at a significantly larger scale. The recovery of distinct chemical sequences, even at modest signal-to-noise ratios, enables the mapping of disk structure across a wide volume. This allows one to (i) trace radial migration through changes in sequence morphology, (ii) identify chemically distinct populations, and (iii) constrain chemical evolution models \citep{Chiappini1997,Minchev2013}.

\begin{figure*}
    \centering
    \includegraphics[width=\textwidth]{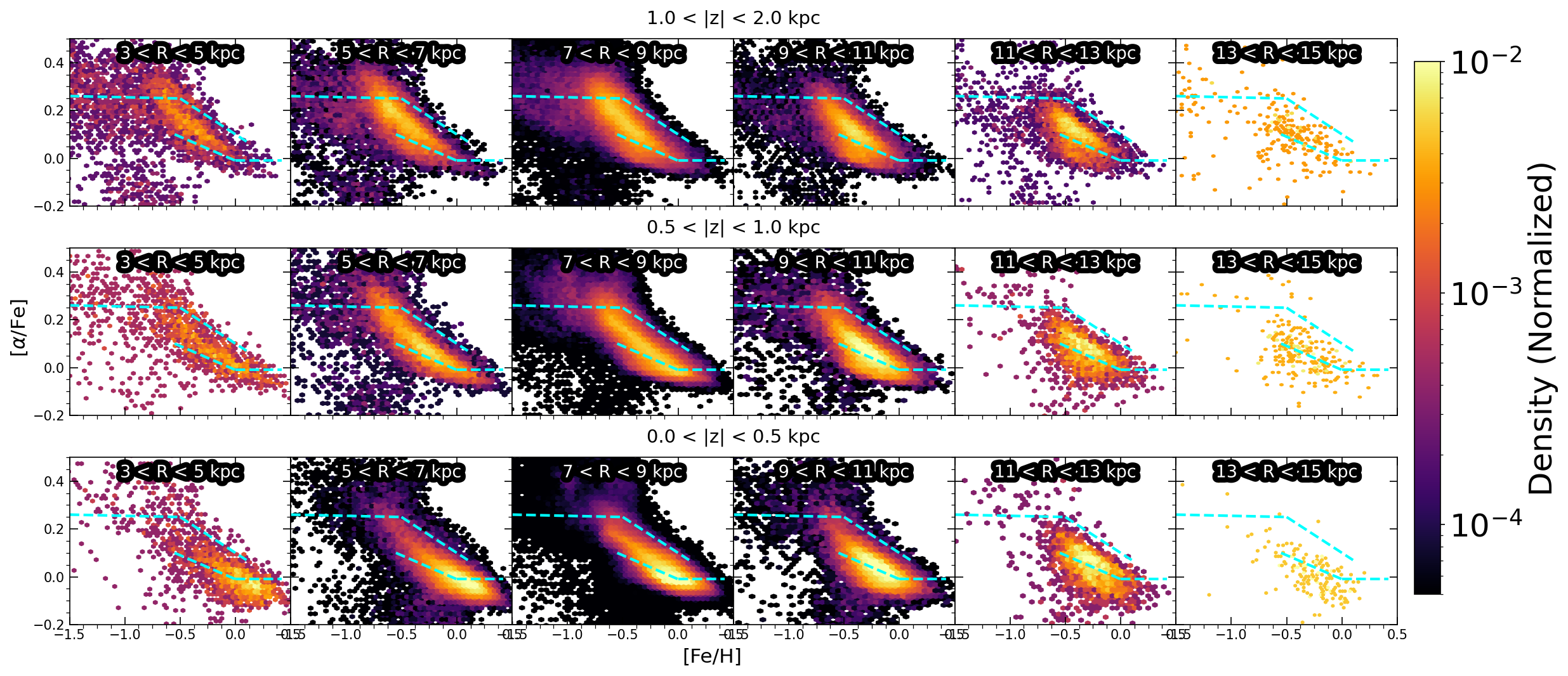}
    \caption{Distribution of [$\alpha$/M] and [Fe/H] using the BOSS-CLAM parameters for different bins of Galactic radius (R) and height (z), similar to the figure from \citet{Hayden2015}. These Galactic coordinates are calculated using the photogeometric distances from \citet{bj2021}. The cyan lines show the approximate location of the high- and low-$\alpha$ sequences from the APOGEE DR19 data.}
    \label{fig:alpha_fe_R_z}
\end{figure*}

These applications closely parallel those enabled by other data-driven spectral models such as \textit{The Cannon} \citep{Ness2015} and \textit{The Payne} \citep{Ting2019}, but benefit from the expanded parameter coverage and flexibility of BOSS-CLAM. As an example of this utility, Figure \ref{fig:alpha_fe_R_z} shows the distribution of [$\alpha$/M] and [Fe/H] for different bins of Galactic radius (R) and height (z), similar to the figure from \citet{Hayden2015}. These Galactic coordinates are calculated from the photogeometric distances from \citet{bj2021} and have the same quality cuts as the rest of this work. The cyan lines show the approximate locations of the high- and low-$\alpha$ sequences in the APOGEE DR19 data and are added to guide the reader through the observed trends.

In this plot, we observe the expected trends for the chemical structure of the Milky Way. At low Galactic heights (bottom row), we observe a clear metallicity gradient moving from, on average, high to low metallicity as we increase in Galactic radius. Additionally, at the Solar circle ($7 < R < 9$ kpc), as we increase in Galactic height, we transition from the thin disk to thick disk and correspondingly see a change from low- to high-$\alpha$ sources. Similarly, at the high Galactic heights (top row), we see that from the inner to outer Galaxy we move from the high- to low-$\alpha$. All of these trends have been observed in previous studies \citep[e.g.,][]{Nidever2014,Hayden2015}, and our method correctly identifies them as well. This indicates these data will be useful for similar studies of the evolution and structure of the Milky Way.

\subsection{Chemical Tagging and Stellar Association Recovery}

Chemical tagging aims to identify stars that formed in the same birth environment by exploiting structure in chemical abundance space \citep{Freeman2002}. When combined with kinematics, even low-dimensional abundance spaces can recover coeval populations.

The BOSS-CLAM catalog enables large-scale chemical tagging analyses by providing homogeneous abundance estimates for millions of stars. Combined with Gaia kinematics, this facilitates the identification of disrupted clusters, stellar streams, and accreted halo structures such as Gaia-Enceladus \citep{helmi2018}.

Recent studies have shown that machine learning methods applied to spectroscopic data can recover chemo-dynamical substructure in the Milky Way halo \citep{Berni2024}, and that data-driven abundance pipelines preserve subtle chemical signatures necessary for such analyses \citep{Ting2019}. Additionally, these data can be used to better refine candidates in star clusters, which are crucial benchmarks in stellar astrophysics.
For example, J. M. Otto et al. ({\em in prep}) use the BOSS-CLAM catalog to enhance {\it Gaia}-based 5D astrometry cluster membership probabilities. 
The addition of RV and [Fe/H] membership probabilities allows for purer cluster membership catalogs, limiting the contamination from moving groups and interlopers.
Stellar cluster catalogs with spectroscopic information are well suited for calibrating large spectroscopic surveys as well as recovering large-scale Galactic properties \citep[e.g.][]{aspcapdr19,lamostOCs2025,otto_26}.

\subsection{Stellar Population Modeling and Forward Modeling of Surveys}

Forward modeling of stellar populations requires accurate stellar parameters across a wide range of evolutionary phases. The BOSS-CLAM model can be used as a generative tool to map stellar labels into spectral space, enabling the construction of synthetic surveys.

This enables: (i) generation of mock spectra for arbitrary stellar populations, (ii) testing of survey selection functions, and (iii) validation of population synthesis models. This approach is conceptually similar to generative spectral models such as \textit{The Payne} \citep{Ting2019}, but BOSS-CLAM’s non-negative matrix factorization basis provides an interpretable representation of spectral variation \citep{Lee2000}.

\begin{figure}
    \centering
    \includegraphics[width=\columnwidth]{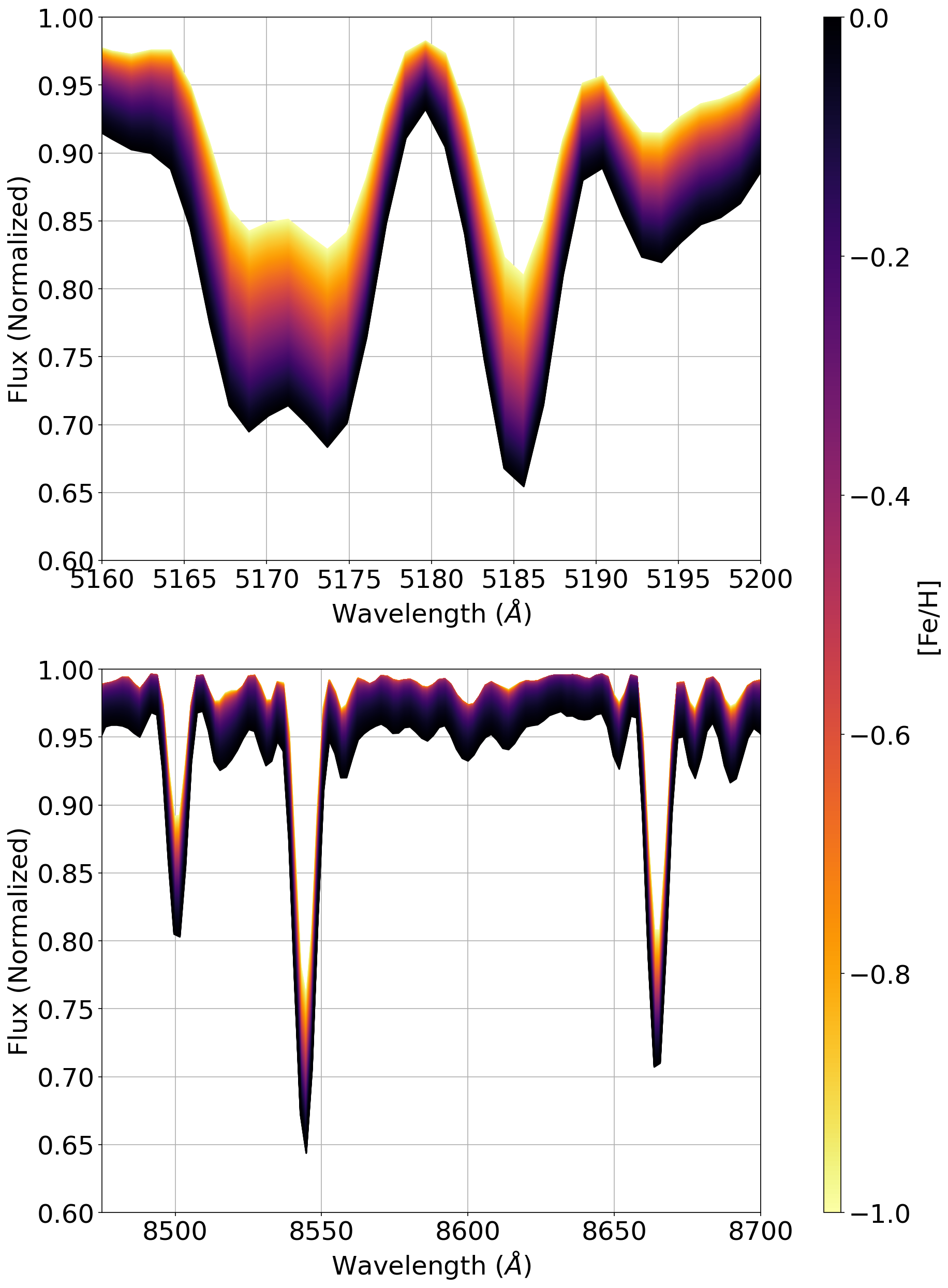}
    \caption{BOSS-CLAM forward modeled spectra for $T_\mathrm{eff} = 5772$ K, $\log g = 4.44$ dex and [$\alpha$/M]$=0$, for various values of [Fe/H]. The top panel is zoomed in on the Mg triplet region, and the bottom panel on the Ca triplet; two metallicity indicators in the BOSS wavelength range that are prominent in Solar-type stars.}
    \label{fig:forward_mod_spec}
\end{figure}

Not only are we releasing the parameters resulting from the BOSS-CLAM in the form of an SDSS-V VAC, but we are also including the underlying code and the model used to infer these parameters\footnote{\url{https://github.com/andycasey/clam-boss}}. This means the model can be used to simply forward model BOSS-like spectra for arbitrary stellar parameters. For example, Figure \ref{fig:forward_mod_spec} shows the BOSS-CLAM forward model spectra for a star at Solar parameters, except we vary [Fe/H]. This figure shows the Mg triplet (top panel) and the Ca triplet (bottom panel) zoomed in. Clear changes in line depths with changes in metallicity are observed. This shows the additional features that the BOSS-CLAM model itself can provide for population modeling using the SDSS-V dataset.

\subsection{Low Signal-to-Noise Regime and Survey Optimization}

A key advantage of data-driven models is their ability to extract information from low signal-to-noise (SNR) spectra. BOSS-CLAM maintains useful abundance precision even at SNR $\sim 10$ (Figure \ref{fig:widebinary_compare}), enabling the inclusion of fainter stellar populations in Galactic studies.

This capability supports survey optimization, including depth-versus-precision trade-offs and target prioritization. Similar advantages have been demonstrated in previous data-driven pipelines \citep{Ness2015}, but BOSS-CLAM extends these capabilities across a broader parameter space. Empirical relations between abundance precision and SNR can be directly incorporated into survey design and downstream analyses with these results.

\section{Summary and Future Work}\label{sec:summary}

In this work, we develop a generative model to infer $T_\mathrm{eff}$, $\log g$, $[\mathrm{Fe/H}]$, $[\alpha/\mathrm{M}]$ from SDSS-V BOSS spectra in DR20. The aim was to develop a model that covered a wide range of the HR diagram, improved on other pipelines used in SDSS-V, and was well-suited for the lower-resolution BOSS dataset. To accomplish this, we developed BOSS-CLAM. BOSS-CLAM forward models spectra by mapper stellar labels to NMF basis vector weights, which are combined with NMF basis vectors to represent BOSS spectra. This mapping and these basis vectors are optimized using a training set consisting of labels from ASPCAP, \texttt{MINESweeper}, wide binaries, and a hot star validation sample. The trained model is able to infer parameters from a wide range of parameter spaces with a high level of accuracy and precision.

With this trained model, we predict the stellar parameters for 1,708,214 BOSS spectra in SDSS-V DR20. These data are released as a VAC in DR20\footnote{\url{https://data.sdss5.org/sas/dr20/vac/mwm/stellar-params/boss_clam/}} and include a comprehensive flagging system to clean the dataset. The recommended clean dataset consists of 915,514 sources with \texttt{clam\_flags = 0}. We demonstrate that the vast majority of the ``bad" sources are restricted to a few classes of objects: hot O/B stars, hot F stars used as standards, and cool M dwarfs. In future iterations, the training labels in these regimes will be added to improve the results.

We demonstrate through a series of validation tests that the parameters resulting from BOSS-CLAM are indeed accurate and to a decent level of precision. We examine open and globular clusters from \citet{VasilievBaumgardt2021, Hunt2024ImprovingCatalogue, otto_26} and find that BOSS-CLAM produces consistent results across the Kiel Diagram for clusters over a wide range of metallicity. We also show that the parameters provide a substantial improvement over BOSSNet in these clusters. Additionally, we utilize wide binaries from \cite{elbadry2021} to probe the bias and scatter in the results. We find both a minimal and an average scatter of $\sigma_{[\mathrm{Fe/H}]} \approx 0.15$ dex and $\sigma_{[\alpha/\mathrm{M}]} \approx 0.06$ at SNR = 10. Finally, we compare spectra of the same star at APO and LCO. The average scatter is comparable to the test data, though some significant outliers remain. We discuss how this is due to flux calibration issues in the BOSS spectra and how such systematic errors should be considered when using these results.

We also discuss multiple improvements to the method that are outside the scope of the present work. Primarily, we have discuss the addition of better priors, weighting of specific windows of the spectra, data to the training set, and a more robust error analysis, which would all greatly increase the results from the BOSS-CLAM for SDSS-V. These added features are planned for inclusion in future data releases of SDSS-V.

Finally, we conclude the paper with some possible applications of these results. We discuss how these results will greatly enhance work in Galactic structure and evolution, chemical tagging and stellar association recovery, stellar population modeling in SDSS-V, and survey optimization. Notably, all of this can be done across a wide range of the HR diagram and in the low signal-to-noise regime, thereby greatly increasing the sample size for these analyses. These BOSS-CLAM parameters will then be a significant contribution to many areas of stellar astronomy that rely on data from large, all-sky surveys like SDSS-V.

\begin{acknowledgments}

JMO, NRM, and PMF acknowledge support for this research from the National Science Foundation Astronomy and Astrophysics grant AST-2206541.
APJ acknowledges support from NSF grant AST-2206264, AST-2307599 and the Alfred P. Sloan Foundation.

Funding for the Sloan Digital Sky Survey V has been provided by the Alfred P. Sloan Foundation, the Heising-Simons Foundation, the National Science Foundation, and the Participating Institutions. SDSS acknowledges support and resources from the Center for High-Performance Computing at the University of Utah. SDSS telescopes are located at Apache Point Observatory, funded by the Astrophysical Research Consortium and operated by New Mexico State University, and at Las Campanas Observatory, operated by the Carnegie Institution for Science. The SDSS website is \url{www.sdss.org}.

SDSS is managed by the Astrophysical Research Consortium for the Participating Institutions of the SDSS Collaboration, including Caltech, The Carnegie Institution for Science, Chilean National Time Allocation Committee (CNTAC) ratified researchers, The Flatiron Institute, the Gotham Participation Group, Harvard University, Heidelberg University, The Johns Hopkins University, L’Ecole polytechnique f\'{e}d\'{e}rale de Lausanne (EPFL), Leibniz-Institut f\"{u}r Astrophysik Potsdam (AIP), Max-Planck-Institut f\"{u}r Astronomie (MPIA Heidelberg), Max-Planck-Institut f\"{u}r Extraterrestrische Physik (MPE), Nanjing University, National Astronomical Observatories of China (NAOC), New Mexico State University, The Ohio State University, Pennsylvania State University, Smithsonian Astrophysical Observatory, Space Telescope Science Institute (STScI), the Stellar Astrophysics Participation Group, Universidad Nacional Aut\'{o}noma de M\'{e}xico, University of Arizona, University of Colorado Boulder, University of Illinois at Urbana-Champaign, University of Toronto, University of Utah, University of Virginia, Yale University, and Yunnan University.

The research behind these results has (partially) received funding from the BELgian federal Science Policy Office (BELSPO) through PRODEX grant PLATO (ZKE8588) and from the Flemish Government under the long-term structural Methusalem funding program by means of the project SOUL: Stellar evolution in full glory, grant METH/24/012 at KU Leuven.

\end{acknowledgments}

\software{Astropy \citep{2013A&A...558A..33A,2018AJ....156..123A, astropy:2022}, Numpy \citep{numpy}, Matplotlib \citep{matplotlib}, JAX \citep{jax2018github}, JAXopt \citep{jaxopt_implicit_diff}, Optax \citep{deepmind2020jax}, Scikit-learn \citep{scikit-learn}}


\bibliography{main}{}

@ARTICLE{sdssVdr19,
       author = {{SDSS Collaboration} and {Adamane Pallathadka}, Gautham and {Aghakhanloo}, Mojgan and {Aird}, James and {Almeida}, Andr{\'e}s and {Amrita}, Singh and {Anders}, Friedrich and {Anderson}, Scott F. and {Arseneau}, Stefan and {Gonz{\'a}lez Avila}, Consuelo and {Aviram}, Shir and {Aydar}, Catarina and {Badenes}, Carles and {Barrera-Ballesteros}, Jorge K. and {Bauer}, Franz E. and {Behmard}, Aida and {Berg}, Michelle and {Besser}, F. and {Moni Bidin}, Christian and {Bizyaev}, Dmitry and {Blanc}, Guillermo and {Blanton}, Michael R. and {Bovy}, Jo and {Brandt}, William Nielsen and {Brownstein}, Joel R. and {Buchner}, Johannes and {Bulbul}, Esra and {Burchett}, Joseph N. and {Carigi}, Leticia and {Carlberg}, Joleen K. and {Casey}, Andrew R. and {Chakraborty}, Priyanka and {Chanam{\'e}}, Julio and {Chandra}, Vedant and {Chiappini}, Cristina and {Chilingarian}, Igor and {Comparat}, Johan and {Covey}, Kevin and {Crumpler}, Nicole and {Cunha}, Katia and {D'Onghia}, Elena and {Dai}, Xinyu and {Darling}, Jeremy and {Davis}, Megan and {De Lee}, Nathan and {Deacon}, Niall and {M{\'e}ndez Delgado}, Jos{\'e} Eduardo and {Demasi}, Sebastian and {Demianenko}, Mariia and {Demke}, Delvin and {Donor}, John and {Drory}, Niv and {Villa Durango}, Monica Alejandra and {Dwelly}, Tom and {Egorov}, Oleg and {Egorova}, Evgeniya and {El-Badry}, Kareem and {Eracleous}, Mike and {Fan}, Xiaohui and {Farr}, Emily and {Finkbeiner}, Douglas P. and {Fries}, Logan and {Frinchaboy}, Peter and {Gentile Fusillo}, Nicola Pietro and {Serrano F{\'e}lix}, Luis Daniel and {Gaensicke}, Boris and {Galligan}, Emma and {Garc{\'\i}a}, Pablo and {Gelfand}, Joseph and {Grabowski}, Katie and {Grebel}, Eva and {Green}, Paul J and {Greve}, Hannah and {Grier}, Catherine and {Griffith}, Emily and {Guetzoyan}, Paloma and {Gupta}, Pramod and {Hackshaw}, Zoe and {Hall}, Patrick B. and {Hawkins}, Keith and {Heged{\H{u}}s}, Viola and {Hekker}, Saskia and {Herbst}, T.~M. and {Hermes}, J.~J. and {Hern{\'a}ndez-Garc{\'\i}a}, Lorena and {Hiremath}, Pranavi and {Hogg}, David W and {Holtzman}, Jon and {Horne}, Keith and {Horta}, Danny and {Huang}, Yang and {Hutchinson}, Brian and {H{\"a}berle}, Maximilian and {Ibarra-Medel}, Hector Javier and {Ji}, Alexander P. and {Jofre}, Paula and {Johnson}, James W. and {Johnson}, Jennifer and {Johnston}, Evelyn J. and {Kaldor}, Mary and {Katkov}, Ivan and {Khalatyan}, Arman and {Khoperskov}, Sergey and {Klessen}, Ralf and {Kluge}, Matthias and {Koekemoer}, Anton M. and {Kollmeier}, Juna A. and {Kounkel}, Marina and {Kreckel}, Kathryn and {Krishnarao}, Dhanesh and {Krumpe}, Mirko and {Lacerna}, Ivan and {Laporte}, Chervin and {Lepine}, Sebastien and {Li}, Jing and {Liang}, Fu-Heng and {Limberg}, Guilherme and {Liu}, Xin and {Loebman}, Sarah and {Long}, Knox and {Lu}, Yuxi and {Lucey}, Madeline and {Lugo-Aranda}, Alejandra Z. and {Mart{\'\i}nez Martinez-Aldama}, Mary Loli and {McKinnon}, Kevin and {Medan}, Ilija and {Merloni}, Andrea and {Morrison}, Sean and {Myers}, Natalie and {M{\'e}sz{\'a}ros}, Szabolcs and {M{\"u}ller-Horn}, Johanna and {Nepal}, Samir and {Ness}, Melissa and {Nidever}, David and {Nitschelm}, Christian and {Oravetz}, Audrey and {Otto}, Jonah and {Pan}, Kaike and {P{\'e}rez Paolino}, Facundo and {Negrete Pe{\~n}aloza}, Castalia Alenka and {Pinsonneault}, Marc and {Taghizadeh Popp}, Manuchehr and {Price-Whelan}, Adrian and {Pulatova}, Nadiia and {Queiroz}, Anna Barbara and {Raddick}, Jordan and {Rankine}, Amy and {Rix}, Hans-Walter and {Rom{\'a}n-Z{\'u}{\~n}iga}, Carlos and {Fern{\'a}ndez Rosso}, Daniela and {Runnoe}, Jessie and {Mahmud Saad}, Serat and {Salvato}, Mara and {Sanchez}, Sebastian F. and {Sattler}, Natascha and {Saydjari}, Andrew and {Sayres}, Conor and {Schlaufman}, Kevin and {Schneider}, Donald P. and {Schwope}, Axel and {Seaton}, Lucas M. and {Seeburger}, Rhys and {Serna}, Javier and {Sharma}, Sanjib and {Shen}, Yue and {Sinha}, Amaya and {Sizemore}, Brian and {Sniegowska}, Marzena and {Song}, Yingyi and {Souto}, Diogo and {Stassun}, Keivan and {Steinmetz}, Matthias and {Stone}, Zachary and {Stone-Martinez}, Alexander and {Stringfellow}, Guy S. and {Mata S{\'a}nchez}, Aurora and {S{\'a}nchez-Gallego}, Jos{\'e} and {Tan}, Jonathan and {Tayar}, Jamie and {Thai}, Riley and {Thakar}, Ani and {Thibodeaux}, Pierre and {Ting}, Yuan-Sen and {Tkachenko}, Andrew and {Trakhtenbrot}, Benny and {Fernandez Trincado}, Jose G. and {Troup}, Nicholas and {Trump}, Jonathan R. and {Ulloa}, Natalie and {Van der Marel}, Roeland P. and {Vera}, Pablo and {Villanova}, Sandro and {Villase{\~n}or}, Jaime and {Wang}, Ji and {Way}, Zachary and {Weijmans}, Anne-Marie and {Wheeler}, Adam and {Wilson}, John C. and {Wofford}, Aida and {Wong}, Tony},
        title = "{The Nineteenth Data Release of the Sloan Digital Sky Survey}",
      journal = {arXiv e-prints},
         year = 2025,
        month = jul,
          eid = {arXiv:2507.07093},
        pages = {arXiv:2507.07093},
          doi = {10.48550/arXiv.2507.07093},
archivePrefix = {arXiv},
       eprint = {2507.07093},
 primaryClass = {astro-ph.GA},
       adsurl = {https://ui.adsabs.harvard.edu/abs/2025arXiv250707093S}
}

@ARTICLE{aspcapdr19,
       author = {{M{\'e}sz{\'a}ros}, Szabolcs and {Jofr{\'e}}, Paula and {Johnson}, Jennifer A. and {Bird}, Jonathan C. and {Bovy}, Jo and {Casey}, Andrew R. and {Chanam{\'e}}, Julio and {Cunha}, Katia and {De Lee}, Nathan and {Frinchaboy}, Peter and {Guiglion}, Guillaume and {Heged{\H{u}}s}, Viola and {Ji}, Alex P. and {Kollmeier}, Juna A. and {Ness}, Melissa K. and {Otto}, Jonah and {Pinsonneault}, Marc H. and {Roman-Lopes}, Alexandre and {Saydjari}, Andrew and {Sinha}, Amaya and {Song}, Ying-Yi and {Stringfellow}, Guy S. and {Stassun}, Keivan G. and {Tayar}, Jamie and {Tkachenko}, Andrew and {Valentini}, Marica and {Way}, Zachary and {Weingrill}, J{\"o}rg},
        title = "{SDSS-V Milky Way Mapper (MWM): ASPCAP Stellar Parameters and Abundances in SDSS-V Data Release 19}",
      journal = {\aj},
         year = 2025,
        month = aug,
       volume = {170},
       number = {2},
          eid = {96},
        pages = {96},
          doi = {10.3847/1538-3881/ade4b9},
archivePrefix = {arXiv},
       eprint = {2506.07845},
 primaryClass = {astro-ph.SR},
       adsurl = {https://ui.adsabs.harvard.edu/abs/2025AJ....170...96M}
}

@ARTICLE{minesweeper,
doi = {10.3847/1538-4357/ae448a},
url = {https://doi.org/10.3847/1538-4357/ae448a},
year = {2026},
month = {mar},
publisher = {The American Astronomical Society},
volume = {1000},
number = {2},
pages = {283},
author = {Chandra, Vedant and Cargile, Phillip A. and Ji, Alexander P. and Conroy, Charlie and Rix, Hans-Walter and Cunningham, Emily and Dias, Bruno and Laporte, Chervin and Cerny, William and Limberg, Guilherme and Bandyopadhyay, Avrajit and Bonaca, Ana and Casey, Andrew R. and Donor, John and Fernández-Trincado, José G. and Frinchaboy, Peter M. and Gupta, Pramod and Hawkins, Keith and Johnson, Jennifer A. and Kollmeier, Juna A. and Lucey, Madeline and Medan, Ilija and Mészáros, Szabolcs and Morrison, Sean and Sánchez-Gallego, José and Saydjari, Andrew K. and Sayres, Conor and Schlaufman, Kevin C. and Stassun, Keivan G. and Tayar, Jamie and Way, Zachary},
title = {Mapping the Distant and Metal-poor Milky Way with SDSS-V},
journal = {The Astrophysical Journal}
}

@ARTICLE{Cargile2020,
       author = {{Cargile}, Phillip A. and {Conroy}, Charlie and {Johnson}, Benjamin D. and {Ting}, Yuan-Sen and {Bonaca}, Ana and {Dotter}, Aaron and {Speagle}, Joshua S.},
        title = "{MINESweeper: Spectrophotometric Modeling of Stars in the Gaia Era}",
      journal = {ApJ},
         year = 2020,
        month = sep,
       volume = {900},
       number = {1},
          eid = {28},
        pages = {28},
          doi = {10.3847/1538-4357/aba43b},
archivePrefix = {arXiv},
       eprint = {1907.07690},
 primaryClass = {astro-ph.SR},
       adsurl = {https://ui.adsabs.harvard.edu/abs/2020ApJ...900...28C}
}

@ARTICLE{elbadry2021,
       author = {{El-Badry}, Kareem and {Rix}, Hans-Walter and {Heintz}, Tyler M.},
        title = "{A million binaries from Gaia eDR3: sample selection and validation of Gaia parallax uncertainties}",
      journal = {\mnras},
         year = 2021,
        month = sep,
       volume = {506},
       number = {2},
        pages = {2269-2295},
          doi = {10.1093/mnras/stab323},
archivePrefix = {arXiv},
       eprint = {2101.05282},
 primaryClass = {astro-ph.SR},
       adsurl = {https://ui.adsabs.harvard.edu/abs/2021MNRAS.506.2269E}
}

@ARTICLE{mann2015,
       author = {{Mann}, Andrew W. and {Feiden}, Gregory A. and {Gaidos}, Eric and {Boyajian}, Tabetha and {von Braun}, Kaspar},
        title = "{How to Constrain Your M Dwarf: Measuring Effective Temperature, Bolometric Luminosity, Mass, and Radius}",
      journal = {\apj},
         year = 2015,
        month = may,
       volume = {804},
       number = {1},
          eid = {64},
        pages = {64},
          doi = {10.1088/0004-637X/804/1/64},
archivePrefix = {arXiv},
       eprint = {1501.01635},
 primaryClass = {astro-ph.SR},
       adsurl = {https://ui.adsabs.harvard.edu/abs/2015ApJ...804...64M}
}

@ARTICLE{mann2019,
       author = {{Mann}, Andrew W. and {Dupuy}, Trent and {Kraus}, Adam L. and {Gaidos}, Eric and {Ansdell}, Megan and {Ireland}, Michael and {Rizzuto}, Aaron C. and {Hung}, Chao-Ling and {Dittmann}, Jason and {Factor}, Samuel and {Feiden}, Gregory and {Martinez}, Raquel A. and {Ru{\'\i}z-Rodr{\'\i}guez}, Dary and {Thao}, Pa Chia},
        title = "{How to Constrain Your M Dwarf. II. The Mass-Luminosity-Metallicity Relation from 0.075 to 0.70 Solar Masses}",
      journal = {\apj},
         year = 2019,
        month = jan,
       volume = {871},
       number = {1},
          eid = {63},
        pages = {63},
          doi = {10.3847/1538-4357/aaf3bc},
archivePrefix = {arXiv},
       eprint = {1811.06938},
 primaryClass = {astro-ph.SR},
       adsurl = {https://ui.adsabs.harvard.edu/abs/2019ApJ...871...63M}
}

@unpublished{ Tkachenko2026,
  author = {{Tkachenko}, A. and others},
  year = {in prep}
}

@ARTICLE{casey2026,
       author = {{Casey}, Andrew R. and {Wheeler}, Adam and {Bedell}, Megan and {Hogg}, David W. and {Saydjari}, Andrew and {Zhao}, Lily},
        title = "{A Constrained Linear Model for Continuum Normalization of Stellar Spectra}",
      journal = {\apj},
         year = 2026,
        month = feb,
       volume = {998},
       number = {2},
          eid = {192},
        pages = {192},
          doi = {10.3847/1538-4357/ae3bd6},
archivePrefix = {arXiv},
       eprint = {2601.19973},
 primaryClass = {astro-ph.SR},
       adsurl = {https://ui.adsabs.harvard.edu/abs/2026ApJ...998..192C}
}

@article{kingma2014adam,
  title={Adam: A method for stochastic optimization},
  author={Kingma, Diederik P and Ba, Jimmy},
  journal={arXiv preprint arXiv:1412.6980},
  year={2014}
}

@article{byrd1995limited,
  author  = {Byrd, Richard H. and Lu, Peihuang and Nocedal, Jorge and Zhu, Ciyou},
  title   = {A Limited Memory Algorithm for Bound Constrained Optimization},
  journal = {SIAM Journal on Scientific Computing},
  volume  = {16},
  number  = {5},
  pages   = {1190--1208},
  year    = {1995},
  doi     = {10.1137/0916069}
}

@article{zhu1997algorithm,
  author  = {Zhu, Ciyou and Byrd, Richard H. and Lu, Peihuang and Nocedal, Jorge},
  title   = {Algorithm 778: {L-BFGS-B}: {Fortran} Subroutines for Large-Scale
             Bound-Constrained Optimization},
  journal = {ACM Transactions on Mathematical Software},
  volume  = {23},
  number  = {4},
  pages   = {550--560},
  year    = {1997},
  doi     = {10.1145/279232.279236}
}

@inproceedings{NMF,
 author = {Lee, Daniel and Seung, H. Sebastian},
 booktitle = {Advances in Neural Information Processing Systems},
 editor = {T. Leen and T. Dietterich and V. Tresp},
 pages = {},
 publisher = {MIT Press},
 title = {Algorithms for Non-negative Matrix Factorization},
 url = {https://proceedings.neurips.cc/paper_files/paper/2000/file/f9d1152547c0bde01830b7e8bd60024c-Paper.pdf},
 volume = {13},
 year = {2000}
}

@ARTICLE{bossnet,
       author = {{Sizemore}, Logan and {Llanes}, Diego and {Kounkel}, Marina and {Hutchinson}, Brian and {Stassun}, Keivan G. and {Chandra}, Vedant},
        title = "{A Self-consistent Data-driven Model for Determining Stellar Parameters from Optical and Near-infrared Spectra}",
      journal = {\aj},
         year = 2024,
        month = apr,
       volume = {167},
       number = {4},
          eid = {173},
        pages = {173},
          doi = {10.3847/1538-3881/ad291d},
archivePrefix = {arXiv},
       eprint = {2402.05184},
 primaryClass = {astro-ph.SR},
       adsurl = {https://ui.adsabs.harvard.edu/abs/2024AJ....167..173S}
}

@ARTICLE{Nidever2020,
       author = {{Nidever}, David L. and {Hasselquist}, Sten and {Hayes}, Christian R. and {Hawkins}, Keith and {Povick}, Joshua and {Majewski}, Steven R. and {Smith}, Verne V. and {Anguiano}, Borja and {Stringfellow}, Guy S. and {Sobeck}, Jennifer S. and {Cunha}, Katia and {Beers}, Timothy C. and {Bestenlehner}, Joachim M. and {Cohen}, Roger E. and {Garcia-Hernandez}, D.~A. and {J{\"o}nsson}, Henrik and {Nitschelm}, Christian and {Shetrone}, Matthew and {Lacerna}, Ivan and {Allende Prieto}, Carlos and {Beaton}, Rachael L. and {Dell'Agli}, Flavia and {Fern{\'a}ndez-Trincado}, Jos{\'e} G. and {Feuillet}, Diane and {Gallart}, Carme and {Hearty}, Fred R. and {Holtzman}, Jon and {Manchado}, Arturo and {Mu{\~n}oz}, Ricardo R. and {O'Connell}, Robert and {Rosado}, Margarita},
        title = "{The Lazy Giants: APOGEE Abundances Reveal Low Star Formation Efficiencies in the Magellanic Clouds}",
      journal = {\apj},
         year = 2020,
        month = jun,
       volume = {895},
       number = {2},
          eid = {88},
        pages = {88},
          doi = {10.3847/1538-4357/ab7305},
archivePrefix = {arXiv},
       eprint = {1901.03448},
 primaryClass = {astro-ph.GA},
       adsurl = {https://ui.adsabs.harvard.edu/abs/2020ApJ...895...88N}
}

@ARTICLE{Hasselquist2021,
       author = {{Hasselquist}, Sten and {Hayes}, Christian R. and {Lian}, Jianhui and {Weinberg}, David H. and {Zasowski}, Gail and {Horta}, Danny and {Beaton}, Rachael and {Feuillet}, Diane K. and {Garro}, Elisa R. and {Gallart}, Carme and {Smith}, Verne V. and {Holtzman}, Jon A. and {Minniti}, Dante and {Lacerna}, Ivan and {Shetrone}, Matthew and {J{\"o}nsson}, Henrik and {Cioni}, Maria-Rosa L. and {Fillingham}, Sean P. and {Cunha}, Katia and {O'Connell}, Robert and {Fern{\'a}ndez-Trincado}, Jos{\'e} G. and {Mu{\~n}oz}, Ricardo R. and {Schiavon}, Ricardo and {Almeida}, Andres and {Anguiano}, Borja and {Beers}, Timothy C. and {Bizyaev}, Dmitry and {Brownstein}, Joel R. and {Cohen}, Roger E. and {Frinchaboy}, Peter and {Garc{\'\i}a-Hern{\'a}ndez}, D.~A. and {Geisler}, Doug and {Lane}, Richard R. and {Majewski}, Steven R. and {Nidever}, David L. and {Nitschelm}, Christian and {Povick}, Joshua and {Price-Whelan}, Adrian and {Roman-Lopes}, Alexandre and {Rosado}, Margarita and {Sobeck}, Jennifer and {Stringfellow}, Guy and {Valenzuela}, Octavio and {Villanova}, Sandro and {Vincenzo}, Fiorenzo},
        title = "{APOGEE Chemical Abundance Patterns of the Massive Milky Way Satellites}",
      journal = {\apj},
         year = 2021,
        month = dec,
       volume = {923},
       number = {2},
          eid = {172},
        pages = {172},
          doi = {10.3847/1538-4357/ac25f9},
archivePrefix = {arXiv},
       eprint = {2109.05130},
 primaryClass = {astro-ph.GA},
       adsurl = {https://ui.adsabs.harvard.edu/abs/2021ApJ...923..172H}
}

@ARTICLE{Zhang2023,
       author = {{Zhang}, Xiangyu and {Green}, Gregory M. and {Rix}, Hans-Walter},
        title = "{Parameters of 220 million stars from Gaia BP/RP spectra}",
      journal = {\mnras},
         year = 2023,
        month = sep,
       volume = {524},
       number = {2},
        pages = {1855-1884},
          doi = {10.1093/mnras/stad1941},
archivePrefix = {arXiv},
       eprint = {2303.03420},
 primaryClass = {astro-ph.SR},
       adsurl = {https://ui.adsabs.harvard.edu/abs/2023MNRAS.524.1855Z}
}

@ARTICLE{Souto2018,
       author = {{Souto}, Diogo and {Cunha}, Katia and {Smith}, Verne V. and {Allende Prieto}, C. and {Garc{\'\i}a-Hern{\'a}ndez}, D.~A. and {Pinsonneault}, Marc and {Holzer}, Parker and {Frinchaboy}, Peter and {Holtzman}, Jon and {Johnson}, J.~A. and {J{\"o}nsson}, Henrik and {Majewski}, Steven R. and {Shetrone}, Matthew and {Sobeck}, Jennifer and {Stringfellow}, Guy and {Teske}, Johanna and {Zamora}, Olga and {Zasowski}, Gail and {Carrera}, Ricardo and {Stassun}, Keivan and {Fernandez-Trincado}, J.~G. and {Villanova}, Sandro and {Minniti}, Dante and {Santana}, Felipe},
        title = "{Chemical Abundances of Main-sequence, Turnoff, Subgiant, and Red Giant Stars from APOGEE Spectra. I. Signatures of Diffusion in the Open Cluster M67}",
      journal = {\apj},
         year = 2018,
        month = apr,
       volume = {857},
       number = {1},
          eid = {14},
        pages = {14},
          doi = {10.3847/1538-4357/aab612},
archivePrefix = {arXiv},
       eprint = {1803.04461},
 primaryClass = {astro-ph.SR},
       adsurl = {https://ui.adsabs.harvard.edu/abs/2018ApJ...857...14S}
}

@ARTICLE{Souto2019,
       author = {{Souto}, Diogo and {Allende Prieto}, C. and {Cunha}, Katia and {Pinsonneault}, Marc and {Smith}, Verne V. and {Garcia-Dias}, R. and {Bovy}, Jo and {Garc{\'\i}a-Hern{\'a}ndez}, D.~A. and {Holtzman}, Jon and {Johnson}, J.~A. and {J{\"o}nsson}, Henrik and {Majewski}, Steve R. and {Shetrone}, Matthew and {Sobeck}, Jennifer and {Zamora}, Olga and {Pan}, Kaike and {Nitschelm}, Christian},
        title = "{Chemical Abundances of Main-sequence, Turnoff, Subgiant, and Red Giant Stars from APOGEE Spectra. II. Atomic Diffusion in M67 Stars}",
      journal = {\apj},
         year = 2019,
        month = mar,
       volume = {874},
       number = {1},
          eid = {97},
        pages = {97},
          doi = {10.3847/1538-4357/ab0b43},
archivePrefix = {arXiv},
       eprint = {1902.10199},
 primaryClass = {astro-ph.SR},
       adsurl = {https://ui.adsabs.harvard.edu/abs/2019ApJ...874...97S}
}

@ARTICLE{M67_Diff_2014,
       author = {{{\"O}nehag}, Anna and {Gustafsson}, Bengt and {Korn}, Andreas},
        title = "{Abundances and possible diffusion of elements in M 67 stars}",
      journal = {\aap},
         year = 2014,
        month = feb,
       volume = {562},
          eid = {A102},
        pages = {A102},
          doi = {10.1051/0004-6361/201322663},
archivePrefix = {arXiv},
       eprint = {1310.6297},
 primaryClass = {astro-ph.SR},
       adsurl = {https://ui.adsabs.harvard.edu/abs/2014A&A...562A.102O}
}

@ARTICLE{otto_26,
       author = {{Otto}, Jonah M. and {Frinchaboy}, Peter M. and {Myers}, Natalie R. and {Johnson}, James W. and {Donor}, John and {Hossain}, Ahabar and {M{\'e}sz{\'a}ros}, Szabolcs and {Wallace}, Hailey and {Cunha}, Katia and {Bhattarai}, Binod and et al.},
        title = "{The Open Cluster Chemical Abundances and Mapping Survey. VIII. Galactic Chemical Gradient and Azimuthal Analysis from SDSS/MWM DR19}",
      journal = {\aj},
         year = 2026,
        month = feb,
       volume = {171},
       number = {2},
          eid = {91},
        pages = {91},
          doi = {10.3847/1538-3881/ae28d8},
archivePrefix = {arXiv},
       eprint = {2507.07264},
 primaryClass = {astro-ph.GA},
       adsurl = {https://ui.adsabs.harvard.edu/abs/2026AJ....171...91O}
}

@article{MLP,
  title={Connectionist Learning Procedures},
  author={Geoffrey E. Hinton},
  journal={Artif. Intell.},
  year={1989},
  volume={40},
  pages={185-234},
  url={https://api.semanticscholar.org/CorpusID:7840452}
}

@article{scikit-learn,
  title={Scikit-learn: Machine Learning in {P}ython},
  author={Pedregosa, F. and Varoquaux, G. and Gramfort, A. and Michel, V.
          and Thirion, B. and Grisel, O. and Blondel, M. and Prettenhofer, P.
          and Weiss, R. and Dubourg, V. and Vanderplas, J. and Passos, A. and
          Cournapeau, D. and Brucher, M. and Perrot, M. and Duchesnay, E.},
  journal={Journal of Machine Learning Research},
  volume={12},
  pages={2825--2830},
  year={2011}
}

@article{Ness2015,
  author = {Ness, M. and Hogg, D. W. and Rix, H.-W. and Ho, A. Y. Q. and Zasowski, G.},
  year = {2015},
  title = {The Cannon: A data-driven approach to stellar label determination},
  journal = {The Astrophysical Journal},
  volume = {808},
  number = {1},
  pages = {16},
  doi = {10.1088/0004-637X/808/1/16}
}

@article{Ting2019,
  author = {Ting, Y.-S. and Conroy, C. and Rix, H.-W. and Cargile, P.},
  year = {2019},
  title = {The Payne: Self-consistent abundance fitting of stellar spectra},
  journal = {The Astrophysical Journal},
  volume = {879},
  number = {2},
  pages = {69},
  doi = {10.3847/1538-4357/ab2331}
}

@article{Hayden2015,
  author = {Hayden, M. R. and Bovy, J. and Holtzman, J. A. and et al.},
  year = {2015},
  title = {Chemical Cartography with APOGEE},
  journal = {The Astrophysical Journal},
  volume = {808},
  number = {2},
  pages = {132},
  doi = {10.1088/0004-637X/808/2/132}
}

@article{Nidever2014,
  author = {Nidever, D. L. and Bovy, J. and Bird, J. C. and et al.},
  year = {2014},
  title = {Tracing chemical evolution with APOGEE},
  journal = {The Astrophysical Journal},
  volume = {796},
  number = {1},
  pages = {38},
  doi = {10.1088/0004-637X/796/1/38}
}

@article{Chiappini1997,
  author = {Chiappini, C. and Matteucci, F. and Gratton, R.},
  year = {1997},
  title = {The chemical evolution of the Galaxy},
  journal = {The Astrophysical Journal},
  volume = {477},
  pages = {765},
  doi = {10.1086/303726}
}

@article{Minchev2013,
  author = {Minchev, I. and Chiappini, C. and Martig, M.},
  year = {2013},
  title = {Chemodynamical evolution of the Milky Way disk},
  journal = {Astronomy \& Astrophysics},
  volume = {558},
  pages = {A9},
  doi = {10.1051/0004-6361/201220386}
}

@article{Freeman2002,
  author = {Freeman, K. and Bland-Hawthorn, J.},
  year = {2002},
  title = {The New Galaxy: Signatures of Its Formation},
  journal = {Annual Review of Astronomy and Astrophysics},
  volume = {40},
  pages = {487},
  doi = {10.1146/annurev.astro.40.060401.093840}
}

@article{Lee2000,
  author = {Lee, D. D. and Seung, H. S.},
  year = {2000},
  title = {Algorithms for non-negative matrix factorization},
  journal = {Advances in Neural Information Processing Systems},
  volume = {13}
}

@article{Berni2024,
  author = {Berni, L.},
  year = {2024},
  title = {Searching for chemo-kinematic structures in the Milky Way halo with deep clustering algorithms},
  journal = {arXiv e-prints},
  eprint = {2409.11429}
}

@ARTICLE{sdssV,
        author = {{Kollmeier}, Juna A. and {Rix}, Hans-Walter and {Aerts}, Conny and {Aird}, James and {Vera Alfaro}, Pablo and {Almeida}, Andr{\'e}s and {Anderson}, Scott F. and {Arseneau}, Stefan M. and {Assef}, Roberto J. and {Aviram}, Shir and {Aydar}, Catarina and {Badenes}, Carles and {Bandyopadhyay}, Avrajit and {Barger}, Kat and {Barkhouser}, Robert H. and {Bauer}, Franz E. and {Behmard}, Aida and {Bender}, Chad and {Besser}, Felipe and {Bhattarai}, Binod and {Bilgi}, Pavaman and {Bird}, Jonathan and {Bizyaev}, Dmitry and {Blanc}, Guillermo A. and {Blanton}, Michael R. and {Bochanski}, John and {Bovy}, Jo and {Brandon}, Christopher and {Brandt}, William Nielsen and {Brownstein}, Joel R. and {Buchner}, Johannes and {Burchett}, Joseph N. and {Carlberg}, Joleen and {Casey}, Andrew R. and {Castaneda-Carlos}, Lesly and {Chakraborty}, Priyanka and {Chanam{\'e}}, Julio and {Chandra}, Vedant and {Cherinka}, Brian and {Chilingarian}, Igor and {Comparat}, Johan and {Cosens}, Maren and {Covey}, Kevin and {Crane}, Jeffrey D. and {Crumpler}, Nicole R. and {Cruz-Gonzalez}, Irene and {Cunha}, Katia and {Cunningham}, Tim and {Dai}, Xinyu and {Darling}, Jeremy and {Davidson}, Jr., James W. and {Davis}, Megan C. and {De Lee}, Nathan and {Deacon}, Niall and {M{\'e}ndez Delgado}, Jos{\'e} Eduardo and {Demasi}, Sebastian and {Demianenko}, Mariia and {Derwent}, Mark and {D'Onghia}, Elena and {Di Mille}, Francesco and {Dias}, Bruno and {Donor}, John and {Dow}, Peter N. and {Drory}, Niv and {Dwelly}, Tom and {Egorov}, Oleg and {Egorova}, Evgeniya and {El-Badry}, Kareem and {Engelman}, Mike and {Eracleous}, Mike and {Fan}, Xiaohui and {Farr}, Emily and {Fries}, Logan and {Frinchaboy}, Peter and {Froning}, Cynthia S. and {G{\"a}nsicke}, Boris T. and {Garc{\'\i}a}, Pablo and {Gelfand}, Joseph and {Gentile Fusillo}, Nicola Pietro and {Glover}, Simon and {Grabowski}, Katie and {Grebel}, Eva K. and {Green}, Paul J. and {Grier}, Catherine and {Gupta}, Pramod and {Gray}, Aidan C. and {H{\"a}berle}, Maximilian and {Hall}, Patrick B. and {Hammond}, Randolph P. and {Hawkins}, Keith and {Harding}, Albert C. and {Heged{\H{u}}s}, Viola and {Herbst}, Tom and {Hermes}, J.~J. and {Rodr{\'\i}guez Hidalgo}, Paola and {Hilder}, Thomas and {Hogg}, David W. and {Holtzman}, Jon A. and {Horta}, Danny and {Huang}, Yang and {Hwang}, Hsiang-Chih and {Ibarra-Medel}, Hector Javier and {Imig}, Julie and {Inight}, Keith and {Jana}, Arghajit and {Ji}, Alexander P. and {Jim{\'e}nez-Arranz}, {\'O}scar and {Jofre}, Paula and {Johns}, Matt and {Johnson}, Jennifer and {Johnson}, James W. and {Johnston}, Evelyn J. and {Jones}, Amy M. and {Katkov}, Ivan and {Knapp}, Gillian R. and {Koekemoer}, Anton M. and {Kounkel}, Marina and {Kreckel}, Kathryn and {Krishnarao}, Dhanesh and {Krumpe}, Mirko and {Kumari}, Nimisha and {Kupfer}, Thomas and {Lacerna}, Ivan and {Laporte}, Chervin and {Lepine}, Sebastien and {Li}, Jing and {Liu}, Xin and {Loebman}, Sarah and {Long}, Knox and {Roman-Lopes}, Alexandre and {Lu}, Yuxi and {Majewski}, Steven Raymond and {Maoz}, Dan and {McKinnon}, Kevin A. and {Medan}, Ilija and {Merloni}, Andrea and {Minniti}, Dante and {Morrison}, Sean and {Myers}, Natalie and {M{\'e}sz{\'a}ros}, Szabolcs and {Nandra}, Kirpal and {Nayak}, Prasanta K. and {Ness}, Melissa K. and {Nidever}, David L. and {O'Brien}, Thomas and {Oeur}, Micah and {Oravetz}, Audrey and {Oravetz}, Daniel and {Otto}, Jonah and {Pallathadka}, Gautham Adamane and {Palunas}, Povilas and {Pan}, Kaike and {Pappalardo}, Daniel and {Pandey}, Rakesh and {Negrete Pe{\~n}aloza}, Castalia Alenka and {Pinsonneault}, Marc H. and {Pogge}, Richard W. and {Taghizadeh Popp}, Manuchehr and {Price-Whelan}, Adrian M. and {Pulatova}, Nadiia and {Qiu}, Dan and {Ramirez}, Solange and {Rankine}, Amy and {Ricci}, Claudio and {Runnoe}, Jessie C. and {Sanchez}, Sebastian and {Salvato}, Mara and {Sarbadhicary}, Sumit K. and {Sattler}, Natascha and {Saydjari}, Andrew K. and {Sayres}, Conor and {Schinnerer}, Eva and {Schlaufman}, Kevin C. and {Schneider}, Donald P. and {Schreiber}, Matthias R. and {Schwope}, Axel and {Serna}, Javier and {Shen}, Yue and {Sif{\'o}n}, Crist{\'o}bal and {Singh}, Amrita and {Sinha}, Amaya and {Smee}, Stephen and {Song}, Ying-Yi and {Souto}, Diogo and {Stassun}, Keivan G. and {Steinmetz}, Matthias and {Stone-Martinez}, Alexander and {Stringfellow}, Guy and {Stutz}, Amelia and {S{\'a}nchez-Gallego}, Jos{\'e} and {Tan}, Jonathan C. and {Tayar}, Jamie and {Thai}, Riley and {Thakar}, Ani and {Ting}, Yuan-Sen and {Tkachenko}, Andrew and {Tovmassian}, Gagik and {Trakhtenbrot}, Benny and {Fern{\'a}ndez-Trincado}, Jos{\'e} G. and {Troup}, Nicholas},
        title = "{Sloan Digital Sky Survey. V. Pioneering Panoptic Spectroscopy}",
      journal = {\aj},
         year = 2026,
        month = jan,
       volume = {171},
       number = {1},
          eid = {52},
        pages = {52},
          doi = {10.3847/1538-3881/ae0576},
archivePrefix = {arXiv},
       eprint = {2507.06989},
 primaryClass = {astro-ph.IM},
       adsurl = {https://ui.adsabs.harvard.edu/abs/2026AJ....171...52K}
}

@ARTICLE{APO,
       author = {{Gunn}, James E. and {Siegmund}, Walter A. and {Mannery}, Edward J. and {Owen}, Russell E. and {Hull}, Charles L. and {Leger}, R. French and {Carey}, Larry N. and {Knapp}, Gillian R. and {York}, Donald G. and {Boroski}, William N. and {Kent}, Stephen M. and {Lupton}, Robert H. and {Rockosi}, Constance M. and {Evans}, Michael L. and {Waddell}, Patrick and {Anderson}, John E. and {Annis}, James and {Barentine}, John C. and {Bartoszek}, Larry M. and {Bastian}, Steven and {Bracker}, Stephen B. and {Brewington}, Howard J. and {Briegel}, Charles I. and {Brinkmann}, Jon and {Brown}, Yorke J. and {Carr}, Michael A. and {Czarapata}, Paul C. and {Drennan}, Craig C. and {Dombeck}, Thomas and {Federwitz}, Glenn R. and {Gillespie}, Bruce A. and {Gonzales}, Carlos and {Hansen}, Sten U. and {Harvanek}, Michael and {Hayes}, Jeffrey and {Jordan}, Wendell and {Kinney}, Ellyne and {Klaene}, Mark and {Kleinman}, S.~J. and {Kron}, Richard G. and {Kresinski}, Jurek and {Lee}, Glenn and {Limmongkol}, Siriluk and {Lindenmeyer}, Carl W. and {Long}, Daniel C. and {Loomis}, Craig L. and {McGehee}, Peregrine M. and {Mantsch}, Paul M. and {Neilsen}, Eric H., Jr. and {Neswold}, Richard M. and {Newman}, Peter R. and {Nitta}, Atsuko and {Peoples}, John, Jr. and {Pier}, Jeffrey R. and {Prieto}, Peter S. and {Prosapio}, Angela and {Rivetta}, Claudio and {Schneider}, Donald P. and {Snedden}, Stephanie and {Wang}, Shu-i.},
        title = "{The 2.5 m Telescope of the Sloan Digital Sky Survey}",
      journal = {\aj},
         year = 2006,
        month = apr,
       volume = {131},
       number = {4},
        pages = {2332-2359},
          doi = {10.1086/500975},
archivePrefix = {arXiv},
       eprint = {astro-ph/0602326},
 primaryClass = {astro-ph},
       adsurl = {https://ui.adsabs.harvard.edu/abs/2006AJ....131.2332G}
}

@ARTICLE{LCO,
       author = {{Bowen}, I.~S. and {Vaughan}, A.~H., Jr.},
        title = "{The optical design of the 40-in. telescope and of the Ir{\'e}n{\'e}e DuPont telescope at Las Campanas Observatory, Chile.}",
      journal = {\ao},
         year = 1973,
        month = jan,
       volume = {12},
        pages = {1430-1434},
          doi = {10.1364/AO.12.001430},
       adsurl = {https://ui.adsabs.harvard.edu/abs/1973ApOpt..12.1430B}
}

@INPROCEEDINGS{FPS,
       author = {{Pogge}, Richard W. and {Derwent}, Mark A. and {O'Brien}, Thomas P. and {Jurgenson}, Colby A. and {Pappalardo}, Daniel and {Engelman}, Michael and {Brandon}, Christopher and {Brady}, Julia and {Clawson}, Nicholas and {Shover}, Jon and {Mason}, Jerry and {Kneib}, Jean-Paul and {Araujo}, Ricardo and {Bouri}, Mohamed and {Kronig}, Luzius and {Grossen}, Lo{\"\i}c. and {Gillet}, Denis and {Macktoobian}, Matin and {Tuttle}, Sarah E. and {Farr}, Emily and {S{\'a}nchez-Gallego}, Jos{\'e} and {Sayres}, Conor},
        title = "{A robotic Focal Plane System (FPS) for the Sloan Digital Sky Survey V}",
    booktitle = {Ground-based and Airborne Instrumentation for Astronomy VIII},
         year = 2020,
       editor = {{Evans}, Christopher J. and {Bryant}, Julia J. and {Motohara}, Kentaro},
       series = {Society of Photo-Optical Instrumentation Engineers (SPIE) Conference Series},
       volume = {11447},
        month = dec,
          eid = {1144781},
        pages = {1144781},
          doi = {10.1117/12.2561113},
       adsurl = {https://ui.adsabs.harvard.edu/abs/2020SPIE11447E..81P}
}

@article{ smee13a,
author = {Smee, Stephen A and Gunn, James E and Uomoto, Alan and Roe, 
Natalie and Schlegel, David and Rockosi, Constance M and Carr, Michael A
 and Leger, French and Dawson, Kyle S and Olmstead, Matthew D and 
Brinkmann, Jon and Owen, Russell and Barkhouser, Robert H and Honscheid,
 Klaus and Harding, Paul and Long, Dan and Lupton, Robert H and Loomis, 
Craig and Anderson, Lauren and Annis, James and Bernardi, Marangela and 
Bhardwaj, Vaishali and Bizyaev, Dmitry and bolton, Adam S and 
Brewington, Howard and Briggs, John W and Burles, Scott and Burns, James
 G and Castander, Francisco Javier and Connolly, Andrew and Davenport, 
James R A and Ebelke, Garrett and Epps, Harland and Feldman, Paul D and 
Friedman, Scott D and Frieman, Joshua and Heckman, Timothy and Hull, 
Charles L and Knapp, Gillian R and Lawrence, David M and Loveday, Jon 
and Mannery, Edward J and Malanushenko, Elena and Malanushenko, Viktor 
and Merrelli, Aronne James and Muna, Demitri and Newman, Peter R and 
Nichol, Robert C and Oravetz, Daniel and Pan, Kaike and Pope, Adrian C 
and Ricketts, Paul G and Shelden, Alaina and Sandford, Dale and 
Siegmund, Walter and Simmons, Audrey and Smith, D Shane and Snedden, 
Stephanie and Schneider, Donald P and SubbaRao, Mark and Tremonti, 
Christy and Waddell, Patrick and York, Donald G},
title = {{The Multi-object, Fiber-fed Spectrographs for the Sloan 
Digital Sky Survey and the Baryon Oscillation Spectroscopic Survey}},
journal = {The Astronomical Journal},
year = {2013},
volume = {146},
number = {2},
pages = {32},
month = aug
}

@ARTICLE{ wilson19a,
       author = {{Wilson}, J.~C. and {Hearty}, F.~R. and {Skrutskie}, M.~F. and
         {Majewski}, S.~R. and {Holtzman}, J.~A. and {Eisenstein}, D. and
         {Gunn}, J. and {Blank}, B. and {Henderson}, C. and {Smee}, S. and
         {Nelson}, M. and {Nidever}, D. and {Arns}, J. and {Barkhouser}, R. and
         {Barr}, J. and {Beland}, S. and {Bershady}, M.~A. and {Blanton}, M.~R. and
         {Brunner}, S. and {Burton}, A. and {Carey}, L. and {Carr}, M. and
         {Colque}, J.~P. and {Crane}, J. and {Damke}, G.~J. and
         {Davidson}, J.~W., Jr. and {Dean}, J. and {Di Mille}, F. and
         {Don}, K.~W. and {Ebelke}, G. and {Evans}, M. and {Fitzgerald}, G. and
         {Gillespie}, B. and {Hall}, M. and {Harding}, A. and {Harding}, P. and
         {Hammond}, R. and {Hancock}, D. and {Harrison}, C. and {Hope}, S. and
         {Horne}, T. and {Karakla}, J. and {Lam}, C. and {Leger}, F. and
         {MacDonald}, N. and {Maseman}, P. and {Matsunari}, J. and {Melton}, S. and
         {Mitcheltree}, T. and {O'Brien}, T. and {O'Connell}, R.~W. and
         {Patten}, A. and {Richardson}, W. and {Rieke}, G. and {Rieke}, M. and
         {Roman-Lopes}, A. and {Schiavon}, R.~P. and {Sobeck}, J.~S. and
         {Stolberg}, T. and {Stoll}, R. and {Tembe}, M. and {Trujillo}, J.~D. and
         {Uomoto}, A. and {Vernieri}, M. and {Walker}, E. and {Weinberg}, D.~H. and
         {Young}, E. and {Anthony-Brumfield}, B. and {Bizyaev}, D. and
         {Breslauer}, B. and {De Lee}, N. and {Downey}, J. and {Halverson}, S. and
         {Huehnerhoff}, J. and {Klaene}, M. and {Leon}, E. and {Long}, D. and
         {Mahadevan}, S. and {Malanushenko}, E. and {Nguyen}, D.~C. and
         {Owen}, R. and {S{\'a}nchez-Gallego}, J.~R. and {Sayres}, C. and
         {Shane}, N. and {Shectman}, S.~A. and {Shetrone}, M. and {Skinner}, D. and
         {Stauffer}, F. and {Zhao}, B.},
        title = "{The Apache Point Observatory Galactic Evolution Experiment (APOGEE) Spectrographs}",
      journal = {\pasp},
         year = 2019,
        month = may,
       volume = {131},
       number = {999},
        pages = {055001},
          doi = {10.1088/1538-3873/ab0075},
archivePrefix = {arXiv},
       eprint = {1902.00928},
 primaryClass = {astro-ph.IM},
       adsurl = {https://ui.adsabs.harvard.edu/abs/2019PASP..131e5001W}
}

@ARTICLE{Heiter2014,
       author = {{Heiter}, U. and {Soubiran}, C. and {Netopil}, M. and {Paunzen}, E.},
        title = "{On the metallicity of open clusters. II. Spectroscopy}",
      journal = {\aap},
         year = 2014,
        month = jan,
       volume = {561},
          eid = {A93},
        pages = {A93},
          doi = {10.1051/0004-6361/201322559},
archivePrefix = {arXiv},
       eprint = {1311.2306},
 primaryClass = {astro-ph.GA},
       adsurl = {https://ui.adsabs.harvard.edu/abs/2014A&A...561A..93H}
}

@ARTICLE{lee2023,
       author = {{Lee}, Jae-Woo},
        title = "{M92 (NGC 6341) Is a Metal-complex Globular Cluster with an Atypical Primordial Population}",
      journal = {\apjl},
         year = 2023,
        month = may,
       volume = {948},
       number = {2},
          eid = {L16},
        pages = {L16},
          doi = {10.3847/2041-8213/acd05a},
archivePrefix = {arXiv},
       eprint = {2305.02983},
 primaryClass = {astro-ph.GA},
       adsurl = {https://ui.adsabs.harvard.edu/abs/2023ApJ...948L..16L}
}

@ARTICLE{medan2025,
       author = {{Medan}, Ilija and {Way}, Zachary and {Rojas-Ayala}, B{\'a}rbara and {Stringfellow}, Guy S. and {Sayres}, Conor and {Stassun}, Keivan G. and {Casey}, Andrew R. and {L{\'e}pine}, S{\'e}bastien and {Galligan}, Emma and {Souto}, Diogo and {Saydjari}, Andrew K.},
        title = "{The Importance of Standardizing Spectra in the Era of Large Spectroscopic Surveys: A Case Study of M Dwarfs in SDSS-V}",
      journal = {\aj},
         year = 2025,
        month = dec,
       volume = {170},
       number = {6},
          eid = {302},
        pages = {302},
          doi = {10.3847/1538-3881/ae0a12},
archivePrefix = {arXiv},
       eprint = {2509.15309},
 primaryClass = {astro-ph.SR},
       adsurl = {https://ui.adsabs.harvard.edu/abs/2025AJ....170..302M}
}

@ARTICLE{bj2021,
       author = {{Bailer-Jones}, C.~A.~L. and {Rybizki}, J. and {Fouesneau}, M. and {Demleitner}, M. and {Andrae}, R.},
        title = "{Estimating Distances from Parallaxes. V. Geometric and Photogeometric Distances to 1.47 Billion Stars in Gaia Early Data Release 3}",
      journal = {\aj},
         year = 2021,
        month = mar,
       volume = {161},
       number = {3},
          eid = {147},
        pages = {147},
          doi = {10.3847/1538-3881/abd806},
archivePrefix = {arXiv},
       eprint = {2012.05220},
 primaryClass = {astro-ph.SR},
       adsurl = {https://ui.adsabs.harvard.edu/abs/2021AJ....161..147B}
}

@article{Hunt2024ImprovingCatalogue,
    title = {{Improving the open cluster census: III. Using cluster masses, radii, and dynamics to create a cleaned open cluster catalogue}},
    year = {2024},
    journal = {Astronomy and Astrophysics},
    author = {Hunt, Emily L. and Reffert, Sabine},
    month = {6},
    volume = {686},
    publisher = {EDP Sciences},
    doi = {10.1051/0004-6361/202348662},
    issn = {14320746},
    arxivId = {2403.05143}
}

@ARTICLE{VasilievBaumgardt2021,
       author = {{Vasiliev}, Eugene and {Baumgardt}, Holger},
        title = "{Gaia EDR3 view on galactic globular clusters}",
      journal = {\mnras},
         year = 2021,
        month = aug,
       volume = {505},
       number = {4},
        pages = {5978-6002},
          doi = {10.1093/mnras/stab1475},
archivePrefix = {arXiv},
       eprint = {2102.09568},
 primaryClass = {astro-ph.GA},
       adsurl = {https://ui.adsabs.harvard.edu/abs/2021MNRAS.505.5978V}
}

@ARTICLE{SchiavonDR17VAC,
       author = {{Schiavon}, Ricardo P. and {Phillips}, Si{\^a}n G. and {Myers}, Natalie and {Horta}, Danny and {Minniti}, Dante and {Allende Prieto}, Carlos and {Anguiano}, Borja and {Beaton}, Rachael L. and {Beers}, Timothy C. and {Brownstein}, Joel R. and {Cohen}, Roger E. and {Fern{\'a}ndez-Trincado}, Jos{\'e} G. and {Frinchaboy}, Peter M. and {J{\"o}nsson}, Henrik and {Kisku}, Shobhit and {Lane}, Richard R. and {Majewski}, Steven R. and {Mason}, Andrew C. and {M{\'e}sz{\'a}ros}, Szabolcs and {Stringfellow}, Guy S.},
        title = "{The APOGEE value-added catalogue of Galactic globular cluster stars}",
      journal = {\mnras},
         year = 2024,
        month = feb,
       volume = {528},
       number = {2},
        pages = {1393-1407},
          doi = {10.1093/mnras/stad3020},
archivePrefix = {arXiv},
       eprint = {2310.07764},
 primaryClass = {astro-ph.GA},
       adsurl = {https://ui.adsabs.harvard.edu/abs/2024MNRAS.528.1393S}
}

@ARTICLE{BovyXDGMM,
       author = {{Bovy}, Jo and {Hogg}, David W. and {Roweis}, Sam T.},
        title = "{Extreme deconvolution: Inferring complete distribution functions from noisy, heterogeneous and incomplete observations}",
      journal = {Annals of Applied Statistics},
         year = 2011,
        month = jun,
       volume = {5},
       number = {2},
        pages = {1657-1677},
          doi = {10.1214/10-AOAS439},
archivePrefix = {arXiv},
       eprint = {0905.2979},
 primaryClass = {stat.ME},
       adsurl = {https://ui.adsabs.harvard.edu/abs/2011AnApS...5.1657B}
}

@software{jax2018github,
  author = {James Bradbury and Roy Frostig and Peter Hawkins and Matthew James Johnson and Yash Katariya and Chris Leary and Dougal Maclaurin and George Necula and Adam Paszke and Jake Vander{P}las and Skye Wanderman-{M}ilne and Qiao Zhang},
  title = {{JAX}: composable transformations of {P}ython+{N}um{P}y programs},
  url = {http://github.com/jax-ml/jax},
  version = {0.3.13},
  year = {2018},
}

@article{jaxopt_implicit_diff,
  title={Efficient and Modular Implicit Differentiation},
  author={Blondel, Mathieu and Berthet, Quentin and Cuturi, Marco and Frostig, Roy 
    and Hoyer, Stephan and Llinares-L{\'o}pez, Felipe and Pedregosa, Fabian 
    and Vert, Jean-Philippe},
  journal={arXiv preprint arXiv:2105.15183},
  year={2021}
}

@software{deepmind2020jax,
  title = {The {D}eep{M}ind {JAX} {E}cosystem},
  author = {DeepMind and Babuschkin, Igor and Baumli, Kate and Bell, Alison and Bhupatiraju, Surya and Bruce, Jake and Buchlovsky, Peter and Budden, David and Cai, Trevor and Clark, Aidan and Danihelka, Ivo and Dedieu, Antoine and Fantacci, Claudio and Godwin, Jonathan and Jones, Chris and Hemsley, Ross and Hennigan, Tom and Hessel, Matteo and Hou, Shaobo and Kapturowski, Steven and Keck, Thomas and Kemaev, Iurii and King, Michael and Kunesch, Markus and Martens, Lena and Merzic, Hamza and Mikulik, Vladimir and Norman, Tamara and Papamakarios, George and Quan, John and Ring, Roman and Ruiz, Francisco and Sanchez, Alvaro and Sartran, Laurent and Schneider, Rosalia and Sezener, Eren and Spencer, Stephen and Srinivasan, Srivatsan and Stanojevi\'{c}, Milo\v{s} and Stokowiec, Wojciech and Wang, Luyu and Zhou, Guangyao and Viola, Fabio},
  url = {http://github.com/google-deepmind},
  year = {2020},
}

@ARTICLE{2018AJ....156..123A,
       author = {{Astropy Collaboration} and {Price-Whelan}, A.~M. and {Sip{\H{o}}cz}, B.~M. and {G{\"u}nther}, H.~M. and {Lim}, P.~L. and {Crawford}, S.~M. and {Conseil}, S. and {Shupe}, D.~L. and {Craig}, M.~W. and {Dencheva}, N. and {Ginsburg}, A. and {VanderPlas}, J.~T. and {Bradley}, L.~D. and {P{\'e}rez-Su{\'a}rez}, D. and {de Val-Borro}, M. and {Aldcroft}, T.~L. and {Cruz}, K.~L. and {Robitaille}, T.~P. and {Tollerud}, E.~J. and {Ardelean}, C. and {Babej}, T. and {Bach}, Y.~P. and {Bachetti}, M. and {Bakanov}, A.~V. and {Bamford}, S.~P. and {Barentsen}, G. and {Barmby}, P. and {Baumbach}, A. and {Berry}, K.~L. and {Biscani}, F. and {Boquien}, M. and {Bostroem}, K.~A. and {Bouma}, L.~G. and {Brammer}, G.~B. and {Bray}, E.~M. and {Breytenbach}, H. and {Buddelmeijer}, H. and {Burke}, D.~J. and {Calderone}, G. and {Cano Rodr{\'\i}guez}, J.~L. and {Cara}, M. and {Cardoso}, J.~V.~M. and {Cheedella}, S. and {Copin}, Y. and {Corrales}, L. and {Crichton}, D. and {D'Avella}, D. and {Deil}, C. and {Depagne}, {\'E}. and {Dietrich}, J.~P. and {Donath}, A. and {Droettboom}, M. and {Earl}, N. and {Erben}, T. and {Fabbro}, S. and {Ferreira}, L.~A. and {Finethy}, T. and {Fox}, R.~T. and {Garrison}, L.~H. and {Gibbons}, S.~L.~J. and {Goldstein}, D.~A. and {Gommers}, R. and {Greco}, J.~P. and {Greenfield}, P. and {Groener}, A.~M. and {Grollier}, F. and {Hagen}, A. and {Hirst}, P. and {Homeier}, D. and {Horton}, A.~J. and {Hosseinzadeh}, G. and {Hu}, L. and {Hunkeler}, J.~S. and {Ivezi{\'c}}, {\v{Z}}. and {Jain}, A. and {Jenness}, T. and {Kanarek}, G. and {Kendrew}, S. and {Kern}, N.~S. and {Kerzendorf}, W.~E. and {Khvalko}, A. and {King}, J. and {Kirkby}, D. and {Kulkarni}, A.~M. and {Kumar}, A. and {Lee}, A. and {Lenz}, D. and {Littlefair}, S.~P. and {Ma}, Z. and {Macleod}, D.~M. and {Mastropietro}, M. and {McCully}, C. and {Montagnac}, S. and {Morris}, B.~M. and {Mueller}, M. and {Mumford}, S.~J. and {Muna}, D. and {Murphy}, N.~A. and {Nelson}, S. and {Nguyen}, G.~H. and {Ninan}, J.~P. and {N{\"o}the}, M. and {Ogaz}, S. and {Oh}, S. and {Parejko}, J.~K. and {Parley}, N. and {Pascual}, S. and {Patil}, R. and {Patil}, A.~A. and {Plunkett}, A.~L. and {Prochaska}, J.~X. and {Rastogi}, T. and {Reddy Janga}, V. and {Sabater}, J. and {Sakurikar}, P. and {Seifert}, M. and {Sherbert}, L.~E. and {Sherwood-Taylor}, H. and {Shih}, A.~Y. and {Sick}, J. and {Silbiger}, M.~T. and {Singanamalla}, S. and {Singer}, L.~P. and {Sladen}, P.~H. and {Sooley}, K.~A. and {Sornarajah}, S. and {Streicher}, O. and {Teuben}, P. and {Thomas}, S.~W. and {Tremblay}, G.~R. and {Turner}, J.~E.~H. and {Terr{\'o}n}, V. and {van Kerkwijk}, M.~H. and {de la Vega}, A. and {Watkins}, L.~L. and {Weaver}, B.~A. and {Whitmore}, J.~B. and {Woillez}, J. and {Zabalza}, V. and {Astropy Contributors}},
        title = "{The Astropy Project: Building an Open-science Project and Status of the v2.0 Core Package}",
      journal = {\aj},
         year = 2018,
        month = sep,
       volume = {156},
       number = {3},
          eid = {123},
        pages = {123},
          doi = {10.3847/1538-3881/aabc4f},
archivePrefix = {arXiv},
       eprint = {1801.02634},
 primaryClass = {astro-ph.IM},
       adsurl = {https://ui.adsabs.harvard.edu/abs/2018AJ....156..123A}
}

@ARTICLE{2013A&A...558A..33A,
       author = {{Astropy Collaboration} and {Robitaille}, Thomas P. and
         {Tollerud}, Erik J. and {Greenfield}, Perry and {Droettboom}, Michael and
         {Bray}, Erik and {Aldcroft}, Tom and {Davis}, Matt and
         {Ginsburg}, Adam and {Price-Whelan}, Adrian M. and
         {Kerzendorf}, Wolfgang E. and {Conley}, Alexander and {Crighton}, Neil and
         {Barbary}, Kyle and {Muna}, Demitri and {Ferguson}, Henry and
         {Grollier}, Fr{\'e}d{\'e}ric and {Parikh}, Madhura M. and
         {Nair}, Prasanth H. and {Unther}, Hans M. and {Deil}, Christoph and
         {Woillez}, Julien and {Conseil}, Simon and {Kramer}, Roban and
         {Turner}, James E.~H. and {Singer}, Leo and {Fox}, Ryan and
         {Weaver}, Benjamin A. and {Zabalza}, Victor and {Edwards}, Zachary I. and
         {Azalee Bostroem}, K. and {Burke}, D.~J. and {Casey}, Andrew R. and
         {Crawford}, Steven M. and {Dencheva}, Nadia and {Ely}, Justin and
         {Jenness}, Tim and {Labrie}, Kathleen and {Lim}, Pey Lian and
         {Pierfederici}, Francesco and {Pontzen}, Andrew and {Ptak}, Andy and
         {Refsdal}, Brian and {Servillat}, Mathieu and {Streicher}, Ole},
        title = "{Astropy: A community Python package for astronomy}",
      journal = {\aap},
         year = "2013",
        month = "Oct",
       volume = {558},
          eid = {A33},
        pages = {A33},
          doi = {10.1051/0004-6361/201322068},
archivePrefix = {arXiv},
       eprint = {1307.6212},
 primaryClass = {astro-ph.IM},
       adsurl = {https://ui.adsabs.harvard.edu/abs/2013A&A...558A..33A}
}

@ARTICLE{astropy:2022,
       author = {{Astropy Collaboration} and {Price-Whelan}, Adrian M. and {Lim}, Pey Lian and {Earl}, Nicholas and {Starkman}, Nathaniel and {Bradley}, Larry and {Shupe}, David L. and {Patil}, Aarya A. and {Corrales}, Lia and {Brasseur}, C.~E. and {N{"o}the}, Maximilian and {Donath}, Axel and {Tollerud}, Erik and {Morris}, Brett M. and {Ginsburg}, Adam and {Vaher}, Eero and {Weaver}, Benjamin A. and {Tocknell}, James and {Jamieson}, William and {van Kerkwijk}, Marten H. and {Robitaille}, Thomas P. and {Merry}, Bruce and {Bachetti}, Matteo and {G{"u}nther}, H. Moritz and {Aldcroft}, Thomas L. and {Alvarado-Montes}, Jaime A. and {Archibald}, Anne M. and {B{'o}di}, Attila and {Bapat}, Shreyas and {Barentsen}, Geert and {Baz{'a}n}, Juanjo and {Biswas}, Manish and {Boquien}, M{'e}d{'e}ric and {Burke}, D.~J. and {Cara}, Daria and {Cara}, Mihai and {Conroy}, Kyle E. and {Conseil}, Simon and {Craig}, Matthew W. and {Cross}, Robert M. and {Cruz}, Kelle L. and {D'Eugenio}, Francesco and {Dencheva}, Nadia and {Devillepoix}, Hadrien A.~R. and {Dietrich}, J{"o}rg P. and {Eigenbrot}, Arthur Davis and {Erben}, Thomas and {Ferreira}, Leonardo and {Foreman-Mackey}, Daniel and {Fox}, Ryan and {Freij}, Nabil and {Garg}, Suyog and {Geda}, Robel and {Glattly}, Lauren and {Gondhalekar}, Yash and {Gordon}, Karl D. and {Grant}, David and {Greenfield}, Perry and {Groener}, Austen M. and {Guest}, Steve and {Gurovich}, Sebastian and {Handberg}, Rasmus and {Hart}, Akeem and {Hatfield-Dodds}, Zac and {Homeier}, Derek and {Hosseinzadeh}, Griffin and {Jenness}, Tim and {Jones}, Craig K. and {Joseph}, Prajwel and {Kalmbach}, J. Bryce and {Karamehmetoglu}, Emir and {Ka{l}uszy{'n}ski}, Miko{l}aj and {Kelley}, Michael S.~P. and {Kern}, Nicholas and {Kerzendorf}, Wolfgang E. and {Koch}, Eric W. and {Kulumani}, Shankar and {Lee}, Antony and {Ly}, Chun and {Ma}, Zhiyuan and {MacBride}, Conor and {Maljaars}, Jakob M. and {Muna}, Demitri and {Murphy}, N.~A. and {Norman}, Henrik and {O'Steen}, Richard and {Oman}, Kyle A. and {Pacifici}, Camilla and {Pascual}, Sergio and {Pascual-Granado}, J. and {Patil}, Rohit R. and {Perren}, Gabriel I. and {Pickering}, Timothy E. and {Rastogi}, Tanuj and {Roulston}, Benjamin R. and {Ryan}, Daniel F. and {Rykoff}, Eli S. and {Sabater}, Jose and {Sakurikar}, Parikshit and {Salgado}, Jes{'u}s and {Sanghi}, Aniket and {Saunders}, Nicholas and {Savchenko}, Volodymyr and {Schwardt}, Ludwig and {Seifert-Eckert}, Michael and {Shih}, Albert Y. and {Jain}, Anany Shrey and {Shukla}, Gyanendra and {Sick}, Jonathan and {Simpson}, Chris and {Singanamalla}, Sudheesh and {Singer}, Leo P. and {Singhal}, Jaladh and {Sinha}, Manodeep and {Sip{H{o}}cz}, Brigitta M. and {Spitler}, Lee R. and {Stansby}, David and {Streicher}, Ole and {{␋{S}}umak}, Jani and {Swinbank}, John D. and {Taranu}, Dan S. and {Tewary}, Nikita and {Tremblay}, Grant R. and {Val-Borro}, Miguel de and {Van Kooten}, Samuel J. and {Vasovi{'c}}, Zlatan and {Verma}, Shresth and {de Miranda Cardoso}, Jos{'e} Vin{'i}cius and {Williams}, Peter K.~G. and {Wilson}, Tom J. and {Winkel}, Benjamin and {Wood-Vasey}, W.~M. and {Xue}, Rui and {Yoachim}, Peter and {Zhang}, Chen and {Zonca}, Andrea and {Astropy Project Contributors}},
        title = "{The Astropy Project: Sustaining and Growing a Community-oriented Open-source Project and the Latest Major Release (v5.0) of the Core Package}",
      journal = {\apj},
         year = 2022,
        month = aug,
       volume = {935},
       number = {2},
          eid = {167},
        pages = {167},
          doi = {10.3847/1538-4357/ac7c74},
archivePrefix = {arXiv},
       eprint = {2206.14220},
 primaryClass = {astro-ph.IM},
       adsurl = {https://ui.adsabs.harvard.edu/abs/2022ApJ...935..167A}
}

@Article{         numpy,
 title         = {Array programming with {NumPy}},
 author        = {Charles R. Harris and K. Jarrod Millman and St{\'{e}}fan J.
                 van der Walt and Ralf Gommers and Pauli Virtanen and David
                 Cournapeau and Eric Wieser and Julian Taylor and Sebastian
                 Berg and Nathaniel J. Smith and Robert Kern and Matti Picus
                 and Stephan Hoyer and Marten H. van Kerkwijk and Matthew
                 Brett and Allan Haldane and Jaime Fern{\'{a}}ndez del
                 R{\'{i}}o and Mark Wiebe and Pearu Peterson and Pierre
                 G{\'{e}}rard-Marchant and Kevin Sheppard and Tyler Reddy and
                 Warren Weckesser and Hameer Abbasi and Christoph Gohlke and
                 Travis E. Oliphant},
 year          = {2020},
 month         = sep,
 journal       = {Nature},
 volume        = {585},
 number        = {7825},
 pages         = {357--362},
 doi           = {10.1038/s41586-020-2649-2},
 publisher     = {Springer Science and Business Media {LLC}},
 url           = {https://doi.org/10.1038/s41586-020-2649-2}
}

@Article{matplotlib,
  Author    = {Hunter, J. D.},
  Title     = {Matplotlib: A 2D graphics environment},
  Journal   = {Computing in Science \& Engineering},
  Volume    = {9},
  Number    = {3},
  Pages     = {90--95},
  publisher = {IEEE COMPUTER SOC},
  doi       = {10.1109/MCSE.2007.55},
  year      = 2007
}

@ARTICLE{meszaros2015,
       author = {{M{\'e}sz{\'a}ros}, Szabolcs and {Martell}, Sarah L. and {Shetrone}, Matthew and {Lucatello}, Sara and {Troup}, Nicholas W. and {Bovy}, Jo and {Cunha}, Katia and {Garc{\'\i}a-Hern{\'a}ndez}, Domingo A. and {Overbeek}, Jamie C. and {Allende Prieto}, Carlos and {Beers}, Timothy C. and {Frinchaboy}, Peter M. and {Garc{\'\i}a P{\'e}rez}, Ana E. and {Hearty}, Fred R. and {Holtzman}, Jon and {Majewski}, Steven R. and {Nidever}, David L. and {Schiavon}, Ricardo P. and {Schneider}, Donald P. and {Sobeck}, Jennifer S. and {Smith}, Verne V. and {Zamora}, Olga and {Zasowski}, Gail},
        title = "{Exploring Anticorrelations and Light Element Variations in Northern Globular Clusters Observed by the APOGEE Survey}",
      journal = {\aj},
         year = 2015,
        month = may,
       volume = {149},
       number = {5},
          eid = {153},
        pages = {153},
          doi = {10.1088/0004-6256/149/5/153},
archivePrefix = {arXiv},
       eprint = {1501.05127},
 primaryClass = {astro-ph.SR},
       adsurl = {https://ui.adsabs.harvard.edu/abs/2015AJ....149..153M}
}

@ARTICLE{roediger2014,
       author = {{Roediger}, Joel C. and {Courteau}, St{\'e}phane and {Graves}, Genevieve and {Schiavon}, Ricardo P.},
        title = "{Constraining Stellar Population Models. I. Age, Metallicity and Abundance Pattern Compilation for Galactic Globular Clusters}",
      journal = {\apjs},
         year = 2014,
        month = jan,
       volume = {210},
       number = {1},
          eid = {10},
        pages = {10},
          doi = {10.1088/0067-0049/210/1/10},
archivePrefix = {arXiv},
       eprint = {1310.3275},
 primaryClass = {astro-ph.GA},
       adsurl = {https://ui.adsabs.harvard.edu/abs/2014ApJS..210...10R}
}

@ARTICLE{ivans1999,
       author = {{Ivans}, Inese I. and {Sneden}, Christopher and {Kraft}, Robert P. and {Suntzeff}, Nicholas B. and {Smith}, Verne V. and {Langer}, G. Edward and {Fulbright}, Jon P.},
        title = "{Star-to-Star Abundance Variations among Bright Giants in the Mildly Metal-poor Globular Cluster M4}",
      journal = {\aj},
         year = 1999,
        month = sep,
       volume = {118},
       number = {3},
        pages = {1273-1300},
          doi = {10.1086/301017},
archivePrefix = {arXiv},
       eprint = {astro-ph/9905370},
 primaryClass = {astro-ph},
       adsurl = {https://ui.adsabs.harvard.edu/abs/1999AJ....118.1273I}
}

@ARTICLE{carretta2015,
       author = {{Carretta}, E. and {Bragaglia}, A. and {Gratton}, R.~G. and {D'Orazi}, V. and {Lucatello}, S. and {Sollima}, A. and {Momany}, Y. and {Catanzaro}, G. and {Leone}, F.},
        title = "{The normal chemistry of multiple stellar populations in the dense globular cluster NGC 6093 (M 80)}",
      journal = {\aap},
         year = 2015,
        month = jun,
       volume = {578},
          eid = {A116},
        pages = {A116},
          doi = {10.1051/0004-6361/201525951},
archivePrefix = {arXiv},
       eprint = {1503.03074},
 primaryClass = {astro-ph.GA},
       adsurl = {https://ui.adsabs.harvard.edu/abs/2015A&A...578A.116C}
}

@ARTICLE{vandenberg2018,
       author = {{VandenBerg}, Don A. and {Denissenkov}, P.~A.},
        title = "{Constraints on the Distance Moduli, Helium and Metal Abundances, and Ages of Globular Clusters from their RR Lyrae and Non-variable Horizontal-branch Stars. III. M55 and NGC 6362}",
      journal = {\apj},
         year = 2018,
        month = jul,
       volume = {862},
       number = {1},
          eid = {72},
        pages = {72},
          doi = {10.3847/1538-4357/aaca9b},
archivePrefix = {arXiv},
       eprint = {1806.02916},
 primaryClass = {astro-ph.SR},
       adsurl = {https://ui.adsabs.harvard.edu/abs/2018ApJ...862...72V}
}

@ARTICLE{helmi2018,
       author = {{Helmi}, Amina and {Babusiaux}, Carine and {Koppelman}, Helmer H. and {Massari}, Davide and {Veljanoski}, Jovan and {Brown}, Anthony G.~A.},
        title = "{The merger that led to the formation of the Milky Way's inner stellar halo and thick disk}",
      journal = {\nat},
         year = 2018,
        month = oct,
       volume = {563},
       number = {7729},
        pages = {85-88},
          doi = {10.1038/s41586-018-0625-x},
archivePrefix = {arXiv},
       eprint = {1806.06038},
 primaryClass = {astro-ph.GA},
       adsurl = {https://ui.adsabs.harvard.edu/abs/2018Natur.563...85H}
}

@ARTICLE{Ratcliffe2023,
       author = {{Ratcliffe}, Bridget and {Minchev}, Ivan and {Anders}, Friedrich and {Khoperskov}, Sergey and {Guiglion}, Guillaume and {Buck}, Tobias and {Cunha}, Katia and {Queiroz}, Anna and {Nitschelm}, Christian and {Meszaros}, Szabolcs and {Steinmetz}, Matthias and {de Jong}, Roelof S. and {Nepal}, Samir and {Lane}, Richard R. and {Sobeck}, Jennifer},
        title = "{Unveiling the time evolution of chemical abundances across the Milky Way disc with APOGEE}",
      journal = {\mnras},
         year = 2023,
        month = oct,
       volume = {525},
       number = {2},
        pages = {2208-2228},
          doi = {10.1093/mnras/stad1573},
archivePrefix = {arXiv},
       eprint = {2305.13378},
 primaryClass = {astro-ph.GA},
       adsurl = {https://ui.adsabs.harvard.edu/abs/2023MNRAS.525.2208R}
}

@ARTICLE{Bensby2014,
       author = {{Bensby}, T. and {Feltzing}, S. and {Oey}, M.~S.},
        title = "{Exploring the Milky Way stellar disk. A detailed elemental abundance study of 714 F and G dwarf stars in the solar neighbourhood}",
      journal = {\aap},
         year = 2014,
        month = feb,
       volume = {562},
          eid = {A71},
        pages = {A71},
          doi = {10.1051/0004-6361/201322631},
archivePrefix = {arXiv},
       eprint = {1309.2631},
 primaryClass = {astro-ph.GA},
       adsurl = {https://ui.adsabs.harvard.edu/abs/2014A&A...562A..71B}
}

@ARTICLE{Carrera2019,
       author = {{Carrera}, R. and {Bragaglia}, A. and {Cantat-Gaudin}, T. and {Vallenari}, A. and {Balaguer-N{\'u}{\~n}ez}, L. and {Bossini}, D. and {Casamiquela}, L. and {Jordi}, C. and {Sordo}, R. and {Soubiran}, C.},
        title = "{Open clusters in APOGEE and GALAH. Combining Gaia and ground-based spectroscopic surveys}",
      journal = {\aap},
         year = 2019,
        month = mar,
       volume = {623},
          eid = {A80},
        pages = {A80},
          doi = {10.1051/0004-6361/201834546},
archivePrefix = {arXiv},
       eprint = {1901.09302},
 primaryClass = {astro-ph.GA},
       adsurl = {https://ui.adsabs.harvard.edu/abs/2019A&A...623A..80C}
}

@ARTICLE{Bochanski2010,
       author = {{Bochanski}, John J. and {Hawley}, Suzanne L. and {Covey}, Kevin R. and {West}, Andrew A. and {Reid}, I. Neill and {Golimowski}, David A. and {Ivezi{\'c}}, {\v{Z}}eljko},
        title = "{The Luminosity and Mass Functions of Low-mass Stars in the Galactic Disk. II. The Field}",
      journal = {\aj},
         year = 2010,
        month = jun,
       volume = {139},
       number = {6},
        pages = {2679-2699},
          doi = {10.1088/0004-6256/139/6/2679},
archivePrefix = {arXiv},
       eprint = {1004.4002},
 primaryClass = {astro-ph.SR},
       adsurl = {https://ui.adsabs.harvard.edu/abs/2010AJ....139.2679B}
}

@ARTICLE{Li2023,
       author = {{Li}, Jiadong and {Liu}, Chao and {Zhang}, Zhi-Yu and {Tian}, Hao and {Fu}, Xiaoting and {Li}, Jiao and {Yan}, Zhi-Qiang},
        title = "{Stellar initial mass function varies with metallicity and time}",
      journal = {\nat},
         year = 2023,
        month = jan,
       volume = {613},
       number = {7944},
        pages = {460-462},
          doi = {10.1038/s41586-022-05488-1},
archivePrefix = {arXiv},
       eprint = {2301.07029},
 primaryClass = {astro-ph.GA},
       adsurl = {https://ui.adsabs.harvard.edu/abs/2023Natur.613..460L}
}

@ARTICLE{Li2026,
       author = {{Li}, Jiadong and {Rix}, Hans-Walter and {Ting}, Yuan-Sen and {Wang}, Yu-Ting and {M{\'e}sz{\'a}ros}, Szabolcs and {Medan}, Ilija and {Liu}, Chao and {Yan}, Zhiqiang and {Smith}, Peter J. and {Qiu}, Dan and {Roman-Lopes}, Alexandre and {Green}, Gregory M. and {Horta}, Danny and {Way}, Zachary and {Matsuno}, Tadafumi and {Souza}, Stefano O. and {Fern{\'a}ndez-Trincado}, Jos{\'e} G.},
        title = "{Variations in the Milky Way's Stellar Mass Function at [Fe/H] < {\ensuremath{-}}1}",
      journal = {\apjl},
         year = 2026,
        month = feb,
       volume = {998},
       number = {2},
          eid = {L33},
        pages = {L33},
          doi = {10.3847/2041-8213/ae3d39},
archivePrefix = {arXiv},
       eprint = {2601.19522},
 primaryClass = {astro-ph.GA},
       adsurl = {https://ui.adsabs.harvard.edu/abs/2026ApJ...998L..33L}
}

@ARTICLE{Badenes2018,
       author = {{Badenes}, Carles and {Mazzola}, Christine and {Thompson}, Todd A. and {Covey}, Kevin and {Freeman}, Peter E. and {Walker}, Matthew G. and {Moe}, Maxwell and {Troup}, Nicholas and {Nidever}, David and {Allende Prieto}, Carlos and {Andrews}, Brett and {Barb{\'a}}, Rodolfo H. and {Beers}, Timothy C. and {Bovy}, Jo and {Carlberg}, Joleen K. and {De Lee}, Nathan and {Johnson}, Jennifer and {Lewis}, Hannah and {Majewski}, Steven R. and {Pinsonneault}, Marc and {Sobeck}, Jennifer and {Stassun}, Keivan G. and {Stringfellow}, Guy S. and {Zasowski}, Gail},
        title = "{Stellar Multiplicity Meets Stellar Evolution and Metallicity: The APOGEE View}",
      journal = {\apj},
         year = 2018,
        month = feb,
       volume = {854},
       number = {2},
          eid = {147},
        pages = {147},
          doi = {10.3847/1538-4357/aaa765},
archivePrefix = {arXiv},
       eprint = {1711.00660},
 primaryClass = {astro-ph.SR},
       adsurl = {https://ui.adsabs.harvard.edu/abs/2018ApJ...854..147B}
}

@ARTICLE{elb2018,
       author = {{El-Badry}, Kareem and {Rix}, Hans-Walter},
        title = "{Imprints of white dwarf recoil in the separation distribution of Gaia wide binaries}",
      journal = {\mnras},
         year = 2018,
        month = nov,
       volume = {480},
       number = {4},
        pages = {4884-4902},
          doi = {10.1093/mnras/sty2186},
archivePrefix = {arXiv},
       eprint = {1807.06011},
 primaryClass = {astro-ph.SR},
       adsurl = {https://ui.adsabs.harvard.edu/abs/2018MNRAS.480.4884E}
}

@ARTICLE{DESI2023_mws,
       author = {{Cooper}, Andrew P. and {Koposov}, Sergey E. and {Allende Prieto}, Carlos and {Manser}, Christopher J. and {Kizhuprakkat}, Namitha and {Myers}, Adam D. and {Dey}, Arjun and {G{\"a}nsicke}, Boris T. and {Li}, Ting S. and {Rockosi}, Constance and {Valluri}, Monica and {Najita}, Joan and {Deason}, Alis and {Raichoor}, Anand and {Wang}, M.-Y. and {Ting}, Y.-S. and {Kim}, Bokyoung and {Carrillo}, Andreia and {Wang}, Wenting and {Beraldo e Silva}, Leandro and {Han}, Jiwon Jesse and {Ding}, Jiani and {S{\'a}nchez-Conde}, Miguel and {Aguilar}, Jessica N. and {Ahlen}, Steven and {Bailey}, Stephen and {Belokurov}, Vasily and {Brooks}, David and {Cunha}, Katia and {Dawson}, Kyle and {de la Macorra}, Axel and {Doel}, Peter and {Eisenstein}, Daniel J. and {Fagrelius}, Parker and {Fanning}, Kevin and {Font-Ribera}, Andreu and {Forero-Romero}, Jaime E. and {Gazta{\~n}aga}, Enrique and {Gontcho a Gontcho}, Satya and {Guy}, Julien and {Honscheid}, Klaus and {Kehoe}, Robert and {Kisner}, Theodore and {Kremin}, Anthony and {Landriau}, Martin and {Levi}, Michael E. and {Martini}, Paul and {Meisner}, Aaron M. and {Miquel}, Ramon and {Moustakas}, John and {Nie}, Jundan J.~D. and {Palanque-Delabrouille}, Nathalie and {Percival}, Will J. and {Poppett}, Claire and {Prada}, Francisco and {Rehemtulla}, Nabeel and {Schlafly}, Edward and {Schlegel}, David and {Schubnell}, Michael and {Sharples}, Ray M. and {Tarl{\'e}}, Gregory and {Wechsler}, Risa H. and {Weinberg}, David H. and {Zhou}, Zhimin and {Zou}, Hu},
        title = "{Overview of the DESI Milky Way Survey}",
      journal = {\apj},
         year = 2023,
        month = apr,
       volume = {947},
       number = {1},
          eid = {37},
        pages = {37},
          doi = {10.3847/1538-4357/acb3c0},
archivePrefix = {arXiv},
       eprint = {2208.08514},
 primaryClass = {astro-ph.GA},
       adsurl = {https://ui.adsabs.harvard.edu/abs/2023ApJ...947...37C}
}

@ARTICLE{GALAH2025,
       author = {{Buder}, Sven and {Kos}, Janez and {Wang}, Xi Ella and {McKenzie}, Madeleine and {Howell}, Madeleine and {Martell}, Sarah and {Hayden}, Michael R. and {Zucker}, Daniel B. and {Nordlander}, Thomas and {Montet}, Benjamin and {Traven}, Gregor and {Bland-Hawthorn}, Joss and {de Silva}, Gayandhi M. and {Freeman}, Kenneth and {Lewis}, Geraint and {Lind}, Karin and {Sharma}, Sanjib and {Simpson}, Jeffrey D. and {Stello}, Dennis and {Zwitter}, Tomaz and {Amarsi}, Anish M. and {Armstrong}, Joseph J. and {Banks}, Kirsten and {Beavis}, Mark and {Beeson}, Kevin-Luke and {Chen}, Boquan and {Ciuc{\u{a}}}, Ioana and {da Costa}, Gary S. and {de Grijs}, Richard and {Martin}, Bailey and {Nataf}, David Moise and {Ness}, Melissa and {Rains}, Adam D. and {Scarr}, Tim and {Vogrin{\v{c}}i{\v{c}}}, Rok and {Wang}, Zixian Purmortal and {Wittenmyer}, Rob A. and {Xie}, Yi Anne and {The Galah Collaboration}},
        title = "{The GALAH survey: Data release 4}",
      journal = {\pasa},
         year = 2025,
        month = may,
       volume = {42},
          eid = {e051},
        pages = {e051},
          doi = {10.1017/pasa.2025.26},
archivePrefix = {arXiv},
       eprint = {2409.19858},
 primaryClass = {astro-ph.GA},
       adsurl = {https://ui.adsabs.harvard.edu/abs/2025PASA...42...51B}
}

@ARTICLE{Gaia2023GSPSpec,
       author = {{Recio-Blanco}, A. and {de Laverny}, P. and {Palicio}, P.~A. and {Kordopatis}, G. and {{\'A}lvarez}, M.~A. and {Schultheis}, M. and {Contursi}, G. and {Zhao}, H. and {Torralba Elipe}, G. and {Ordenovic}, C. and {Manteiga}, M. and {Dafonte}, C. and {Oreshina-Slezak}, I. and {Bijaoui}, A. and {Fr{\'e}mat}, Y. and {Seabroke}, G. and {Pailler}, F. and {Spitoni}, E. and {Poggio}, E. and {Creevey}, O.~L. and {Abreu Aramburu}, A. and {Accart}, S. and {Andrae}, R. and {Bailer-Jones}, C.~A.~L. and {Bellas-Velidis}, I. and {Brouillet}, N. and {Brugaletta}, E. and {Burlacu}, A. and {Carballo}, R. and {Casamiquela}, L. and {Chiavassa}, A. and {Cooper}, W.~J. and {Dapergolas}, A. and {Delchambre}, L. and {Dharmawardena}, T.~E. and {Drimmel}, R. and {Edvardsson}, B. and {Fouesneau}, M. and {Garabato}, D. and {Garc{\'\i}a-Lario}, P. and {Garc{\'\i}a-Torres}, M. and {Gavel}, A. and {Gomez}, A. and {Gonz{\'a}lez-Santamar{\'\i}a}, I. and {Hatzidimitriou}, D. and {Heiter}, U. and {Jean-Antoine Piccolo}, A. and {Kontizas}, M. and {Korn}, A.~J. and {Lanzafame}, A.~C. and {Lebreton}, Y. and {Le Fustec}, Y. and {Licata}, E.~L. and {Lindstr{\o}m}, H.~E.~P. and {Livanou}, E. and {Lobel}, A. and {Lorca}, A. and {Magdaleno Romeo}, A. and {Marocco}, F. and {Marshall}, D.~J. and {Mary}, N. and {Nicolas}, C. and {Pallas-Quintela}, L. and {Panem}, C. and {Pichon}, B. and {Riclet}, F. and {Robin}, C. and {Rybizki}, J. and {Santove{\~n}a}, R. and {Silvelo}, A. and {Smart}, R.~L. and {Sarro}, L.~M. and {Sordo}, R. and {Soubiran}, C. and {S{\"u}veges}, M. and {Ulla}, A. and {Vallenari}, A. and {Zorec}, J. and {Utrilla}, E. and {Bakker}, J.},
        title = "{Gaia Data Release 3. Analysis of RVS spectra using the General Stellar Parametriser from spectroscopy}",
      journal = {\aap},
         year = 2023,
        month = jun,
       volume = {674},
          eid = {A29},
        pages = {A29},
          doi = {10.1051/0004-6361/202243750},
archivePrefix = {arXiv},
       eprint = {2206.05541},
 primaryClass = {astro-ph.GA},
       adsurl = {https://ui.adsabs.harvard.edu/abs/2023A&A...674A..29R}
}

@dataset{LAMOST2026DR11,
       author = {{Luo}, A.-L. and {Zhao}, Y.-H. and {Zhao}, G. and {et al.}},
        title = "{VizieR Online Data Catalog: LAMOST DR11 catalogs (Luo+, 2026)}",
 howpublished = {VizieR On-line Data Catalog: V/162.  Originally published in: 2026RAA..in.prep..L},
         year = 2026,
        month = jan,
          eid = {V/162},
       adsurl = {https://ui.adsabs.harvard.edu/abs/2026yCat.5162....0L}
}

@ARTICLE{AllendePrieto2006FERRE,
       author = {{Allende Prieto}, Carlos and {Beers}, Timothy C. and {Wilhelm}, Ronald and {Newberg}, Heidi Jo and {Rockosi}, Constance M. and {Yanny}, Brian and {Lee}, Young Sun},
        title = "{A Spectroscopic Study of the Ancient Milky Way: F- and G-Type Stars in the Third Data Release of the Sloan Digital Sky Survey}",
      journal = {\apj},
         year = 2006,
        month = jan,
       volume = {636},
       number = {2},
        pages = {804-820},
          doi = {10.1086/498131},
archivePrefix = {arXiv},
       eprint = {astro-ph/0509812},
 primaryClass = {astro-ph},
       adsurl = {https://ui.adsabs.harvard.edu/abs/2006ApJ...636..804A}
}

@ARTICLE{Husser2013_PHOENIX,
       author = {{Husser}, T.-O. and {Wende-von Berg}, S. and {Dreizler}, S. and {Homeier}, D. and {Reiners}, A. and {Barman}, T. and {Hauschildt}, P.~H.},
        title = "{A new extensive library of PHOENIX stellar atmospheres and synthetic spectra}",
      journal = {\aap},
         year = 2013,
        month = may,
       volume = {553},
          eid = {A6},
        pages = {A6},
          doi = {10.1051/0004-6361/201219058},
archivePrefix = {arXiv},
       eprint = {1303.5632},
 primaryClass = {astro-ph.SR},
       adsurl = {https://ui.adsabs.harvard.edu/abs/2013A&A...553A...6H}
}

@ARTICLE{Ting2025,
       author = {{Ting}, Yuan-Sen},
        title = "{Why Machine Learning Models Systematically Underestimate Extreme Values}",
      journal = {The Open Journal of Astrophysics},
         year = 2025,
        month = jul,
       volume = {8},
          eid = {95},
        pages = {95},
          doi = {10.33232/001c.142224},
archivePrefix = {arXiv},
       eprint = {2412.05806},
 primaryClass = {astro-ph.IM},
       adsurl = {https://ui.adsabs.harvard.edu/abs/2025OJAp....8E..95T}
}

@ARTICLE{donor2018,
       author = {{Donor}, John and {Frinchaboy}, Peter M. and {Cunha}, Katia and {Thompson}, Benjamin and {O'Connell}, Julia and {Zasowski}, Gail and {Jackson}, Kelly M. and {Meyer McGrath}, Brianne and {Almeida}, Andr{\'e}s and {Bizyaev}, Dmitry and {Carrera}, Ricardo and {Garc{\'\i}a-Hern{\'a}ndez}, D.~A. and {Nitschelm}, Christian and {Pan}, Kaike and {Zamora}, Olga},
        title = "{The Open Cluster Chemical Abundances and Mapping Survey. II. Precision Cluster Abundances for APOGEE Using SDSS DR14}",
      journal = {\aj},
         year = 2018,
        month = oct,
       volume = {156},
       number = {4},
          eid = {142},
        pages = {142},
          doi = {10.3847/1538-3881/aad635},
archivePrefix = {arXiv},
       eprint = {1807.09791},
 primaryClass = {astro-ph.GA},
       adsurl = {https://ui.adsabs.harvard.edu/abs/2018AJ....156..142D}
}

@ARTICLE{lamostOCs2025,
       author = {{Yang}, Guochao and {Zhao}, Jingkun and {Yang}, Yong and {Liu}, Nian and {Luo}, Yangping and {Zhao}, Gang},
        title = "{Chemical Abundance Gradients in Open Clusters from the Gaia/LAMOST Sample}",
      journal = {\aj},
         year = 2025,
        month = apr,
       volume = {169},
       number = {4},
          eid = {214},
        pages = {214},
          doi = {10.3847/1538-3881/adba45},
       adsurl = {https://ui.adsabs.harvard.edu/abs/2025AJ....169..214Y}
}

@ARTICLE{Andrae2023_xgboost,
       author = {{Andrae}, Ren{\'e} and {Rix}, Hans-Walter and {Chandra}, Vedant},
        title = "{Robust Data-driven Metallicities for 175 Million Stars from Gaia XP Spectra}",
      journal = {\apjs},
         year = 2023,
        month = jul,
       volume = {267},
       number = {1},
          eid = {8},
        pages = {8},
          doi = {10.3847/1538-4365/acd53e},
archivePrefix = {arXiv},
       eprint = {2302.02611},
 primaryClass = {astro-ph.SR},
       adsurl = {https://ui.adsabs.harvard.edu/abs/2023ApJS..267....8A}
}

@ARTICLE{Casey2016,
       author = {{Casey}, Andrew R. and {Hogg}, David W. and {Ness}, Melissa and {Rix}, Hans-Walter and {Ho}, Anna Q Y and {Gilmore}, Gerry},
        title = "{The Cannon 2: A data-driven model of stellar spectra for detailed chemical abundance analyses}",
      journal = {arXiv e-prints},
         year = 2016,
        month = mar,
          eid = {arXiv:1603.03040},
        pages = {arXiv:1603.03040},
          doi = {10.48550/arXiv.1603.03040},
archivePrefix = {arXiv},
       eprint = {1603.03040},
 primaryClass = {astro-ph.SR},
       adsurl = {https://ui.adsabs.harvard.edu/abs/2016arXiv160303040C}
}

@ARTICLE{sdssdr17,
       author = {{Abdurro'uf} and {Accetta}, Katherine and {Aerts}, Conny and {Silva Aguirre}, V{\'\i}ctor and {Ahumada}, Romina and {Ajgaonkar}, Nikhil and {Filiz Ak}, N. and {Alam}, Shadab and {Allende Prieto}, Carlos and {Almeida}, Andr{\'e}s and {Anders}, Friedrich and {Anderson}, Scott F. and {Andrews}, Brett H. and {Anguiano}, Borja and {Aquino-Ort{\'\i}z}, Erik and {Arag{\'o}n-Salamanca}, Alfonso and {Argudo-Fern{\'a}ndez}, Maria and {Ata}, Metin and {Aubert}, Marie and {Avila-Reese}, Vladimir and {Badenes}, Carles and {Barb{\'a}}, Rodolfo H. and {Barger}, Kat and {Barrera-Ballesteros}, Jorge K. and {Beaton}, Rachael L. and {Beers}, Timothy C. and {Belfiore}, Francesco and {Bender}, Chad F. and {Bernardi}, Mariangela and {Bershady}, Matthew A. and {Beutler}, Florian and {Bidin}, Christian Moni and {Bird}, Jonathan C. and {Bizyaev}, Dmitry and {Blanc}, Guillermo A. and {Blanton}, Michael R. and {Boardman}, Nicholas Fraser and {Bolton}, Adam S. and {Boquien}, M{\'e}d{\'e}ric and {Borissova}, Jura and {Bovy}, Jo and {Brandt}, W.~N. and {Brown}, Jordan and {Brownstein}, Joel R. and {Brusa}, Marcella and {Buchner}, Johannes and {Bundy}, Kevin and {Burchett}, Joseph N. and {Bureau}, Martin and {Burgasser}, Adam and {Cabang}, Tuesday K. and {Campbell}, Stephanie and {Cappellari}, Michele and {Carlberg}, Joleen K. and {Wanderley}, F{\'a}bio Carneiro and {Carrera}, Ricardo and {Cash}, Jennifer and {Chen}, Yan-Ping and {Chen}, Wei-Huai and {Cherinka}, Brian and {Chiappini}, Cristina and {Choi}, Peter Doohyun and {Chojnowski}, S. Drew and {Chung}, Haeun and {Clerc}, Nicolas and {Cohen}, Roger E. and {Comerford}, Julia M. and {Comparat}, Johan and {da Costa}, Luiz and {Covey}, Kevin and {Crane}, Jeffrey D. and {Cruz-Gonzalez}, Irene and {Culhane}, Connor and {Cunha}, Katia and {Dai}, Y. Sophia and {Damke}, Guillermo and {Darling}, Jeremy and {Davidson}, Jr., James W. and {Davies}, Roger and {Dawson}, Kyle and {De Lee}, Nathan and {Diamond-Stanic}, Aleksandar M. and {Cano-D{\'\i}az}, Mariana and {S{\'a}nchez}, Helena Dom{\'\i}nguez and {Donor}, John and {Duckworth}, Chris and {Dwelly}, Tom and {Eisenstein}, Daniel J. and {Elsworth}, Yvonne P. and {Emsellem}, Eric and {Eracleous}, Mike and {Escoffier}, Stephanie and {Fan}, Xiaohui and {Farr}, Emily and {Feng}, Shuai and {Fern{\'a}ndez-Trincado}, Jos{\'e} G. and {Feuillet}, Diane and {Filipp}, Andreas and {Fillingham}, Sean P. and {Frinchaboy}, Peter M. and {Fromenteau}, Sebastien and {Galbany}, Llu{\'\i}s and {Garc{\'\i}a}, Rafael A. and {Garc{\'\i}a-Hern{\'a}ndez}, D.~A. and {Ge}, Junqiang and {Geisler}, Doug and {Gelfand}, Joseph and {G{\'e}ron}, Tobias and {Gibson}, Benjamin J. and {Goddy}, Julian and {Godoy-Rivera}, Diego and {Grabowski}, Kathleen and {Green}, Paul J. and {Greener}, Michael and {Grier}, Catherine J. and {Griffith}, Emily and {Guo}, Hong and {Guy}, Julien and {Hadjara}, Massinissa and {Harding}, Paul and {Hasselquist}, Sten and {Hayes}, Christian R. and {Hearty}, Fred and {Hern{\'a}ndez}, Jes{\'u}s and {Hill}, Lewis and {Hogg}, David W. and {Holtzman}, Jon A. and {Horta}, Danny and {Hsieh}, Bau-Ching and {Hsu}, Chin-Hao and {Hsu}, Yun-Hsin and {Huber}, Daniel and {Huertas-Company}, Marc and {Hutchinson}, Brian and {Hwang}, Ho Seong and {Ibarra-Medel}, H{\'e}ctor J. and {Chitham}, Jacob Ider and {Ilha}, Gabriele S. and {Imig}, Julie and {Jaekle}, Will and {Jayasinghe}, Tharindu and {Ji}, Xihan and {Johnson}, Jennifer A. and {Jones}, Amy and {J{\"o}nsson}, Henrik and {Katkov}, Ivan and {Khalatyan}, Dr., Arman and {Kinemuchi}, Karen and {Kisku}, Shobhit and {Knapen}, Johan H. and {Kneib}, Jean-Paul and {Kollmeier}, Juna A. and {Kong}, Miranda and {Kounkel}, Marina and {Kreckel}, Kathryn and {Krishnarao}, Dhanesh and {Lacerna}, Ivan and {Lane}, Richard R. and {Langgin}, Rachel and {Lavender}, Ramon and {Law}, David R. and {Lazarz}, Daniel and {Leung}, Henry W. and {Leung}, Ho-Hin and {Lewis}, Hannah M. and {Li}, Cheng and {Li}, Ran and {Lian}, Jianhui and {Liang}, Fu-Heng and {Lin}, Lihwai and {Lin}, Yen-Ting and {Lin}, Sicheng and {Lintott}, Chris and {Long}, Dan and {Longa-Pe{\~n}a}, Pen{\'e}lope and {L{\'o}pez-Cob{\'a}}, Carlos and {Lu}, Shengdong and {Lundgren}, Britt F. and {Luo}, Yuanze and {Mackereth}, J. Ted and {de la Macorra}, Axel and {Mahadevan}, Suvrath and {Majewski}, Steven R. and {Manchado}, Arturo and {Mandeville}, Travis and {Maraston}, Claudia and {Margalef-Bentabol}, Berta and {Masseron}, Thomas and {Masters}, Karen L. and {Mathur}, Savita and {McDermid}, Richard M. and {Mckay}, Myles and {Merloni}, Andrea and {Merrifield}, Michael and {Meszaros}, Szabolcs and {Miglio}, Andrea and {Di Mille}, Francesco and {Minniti}, Dante and {Minsley}, Rebecca and {Monachesi}, Antonela},
        title = "{The Seventeenth Data Release of the Sloan Digital Sky Surveys: Complete Release of MaNGA, MaStar, and APOGEE-2 Data}",
      journal = {\apjs},
         year = 2022,
        month = apr,
       volume = {259},
       number = {2},
          eid = {35},
        pages = {35},
          doi = {10.3847/1538-4365/ac4414},
archivePrefix = {arXiv},
       eprint = {2112.02026},
 primaryClass = {astro-ph.GA},
       adsurl = {https://ui.adsabs.harvard.edu/abs/2022ApJS..259...35A}
}

@ARTICLE{Behmard2025,
       author = {{Behmard}, Aida and {Ness}, Melissa K. and {Casey}, Andrew R. and {Angus}, Ruth and {Cunha}, Katia and {Souto}, Diogo and {Lu}, Yuxi(Lucy) and {Johnson}, Jennifer A.},
        title = "{A Data-driven M Dwarf Model and Detailed Abundances for {\ensuremath{\sim}}17,000 M Dwarfs in SDSS-V}",
      journal = {\apj},
         year = 2025,
        month = mar,
       volume = {982},
       number = {1},
          eid = {13},
        pages = {13},
          doi = {10.3847/1538-4357/adaf1f},
archivePrefix = {arXiv},
       eprint = {2501.14955},
 primaryClass = {astro-ph.SR},
       adsurl = {https://ui.adsabs.harvard.edu/abs/2025ApJ...982...13B}
}

@ARTICLE{jones2005,
       author = {{Jones}, Hugh R.~A. and {Pavlenko}, Yakiv and {Viti}, Serena and {Barber}, R.~J. and {Yakovina}, Larisa A. and {Pinfield}, David and {Tennyson}, Jonathan},
        title = "{Carbon monoxide in low-mass dwarf stars}",
      journal = {\mnras},
         year = 2005,
        month = mar,
       volume = {358},
       number = {1},
        pages = {105-112},
          doi = {10.1111/j.1365-2966.2005.08736.x},
archivePrefix = {arXiv},
       eprint = {astro-ph/0412387},
 primaryClass = {astro-ph},
       adsurl = {https://ui.adsabs.harvard.edu/abs/2005MNRAS.358..105J}
}

@ARTICLE{charams,
       author = {{ten Brummelaar}, T.~A. and {McAlister}, H.~A. and {Ridgway}, S.~T. and {Bagnuolo}, Jr., W.~G. and {Turner}, N.~H. and {Sturmann}, L. and {Sturmann}, J. and {Berger}, D.~H. and {Ogden}, C.~E. and {Cadman}, R. and {Hartkopf}, W.~I. and {Hopper}, C.~H. and {Shure}, M.~A.},
        title = "{First Results from the CHARA Array. II. A Description of the Instrument}",
      journal = {\apj},
         year = 2005,
        month = jul,
       volume = {628},
       number = {1},
        pages = {453-465},
          doi = {10.1086/430729},
archivePrefix = {arXiv},
       eprint = {astro-ph/0504082},
 primaryClass = {astro-ph},
       adsurl = {https://ui.adsabs.harvard.edu/abs/2005ApJ...628..453T}
}

@article{way2026,
doi = {10.3847/1538-3881/ae48ae},
url = {https://doi.org/10.3847/1538-3881/ae48ae},
year = {2026},
month = {mar},
publisher = {The American Astronomical Society},
volume = {171},
number = {4},
pages = {252},
author = {Way, Zachary and Lépine, Sébastien and Gagné, Jonathan and Medan, Ilija},
title = {21,864 Unresolved, Low-mass Binaries Identified via Their Overluminosity in Gaia Data Release 3 and a Catalog of 347,440 Systems within 100 pc of the Sun},
journal = {The Astronomical Journal}
}

@ARTICLE{NUTS,
       author = {{Hoffman}, Matthew D. and {Gelman}, Andrew},
        title = "{The No-U-Turn Sampler: Adaptively Setting Path Lengths in Hamiltonian Monte Carlo}",
      journal = {arXiv e-prints},
         year = 2011,
        month = nov,
          eid = {arXiv:1111.4246},
        pages = {arXiv:1111.4246},
          doi = {10.48550/arXiv.1111.4246},
archivePrefix = {arXiv},
       eprint = {1111.4246},
 primaryClass = {stat.CO},
       adsurl = {https://ui.adsabs.harvard.edu/abs/2011arXiv1111.4246H}
}
\bibliographystyle{aasjournalv7}

\appendix

\section{Comparing Hessian Covariances to MCMC Sampling}\label{app:mcmc}

\added{To assess the use of the covariances estimated from the inverse of the Hessian, we run a full MCMC on a relatively small set of stars that cover the full range of our parameter space. We restrict our test sample to solutions marked with \texttt{flag\_bad = False}. We do quantile binning in SNR, $T_\mathrm{eff}$, $\log g$ and $[\mathrm{Fe/H}]$ for 7 bins per dimension. We then randomly select 5 stars per bin. If there are less than 5 stars, we select all stars in that bin. In total, this results in 9,954 stars for this test.}

\added{For the MCMC, we use The No-U-Turn Sampler \citep[NUTS;][]{NUTS} consisting of four chains each with a warm-up of 750 draws and then 300 subsequent samples. For priors, we assume each parameter is a Gaussian with a mean equal to the inferred BOSS-CLAM parameters and standard deviation equal to the approximate scatter found during the testing stage (275 K, 0.3 dex, 0.15 dex and 0.11 dex for $T_\mathrm{eff}$, $\log g$, $[\mathrm{Fe/H}]$ and $[\alpha/\mathrm{M}]$, respectively; Figure \ref{fig:clam_results}).}

\begin{figure}
    \centering
    \includegraphics[width=0.49\textwidth]{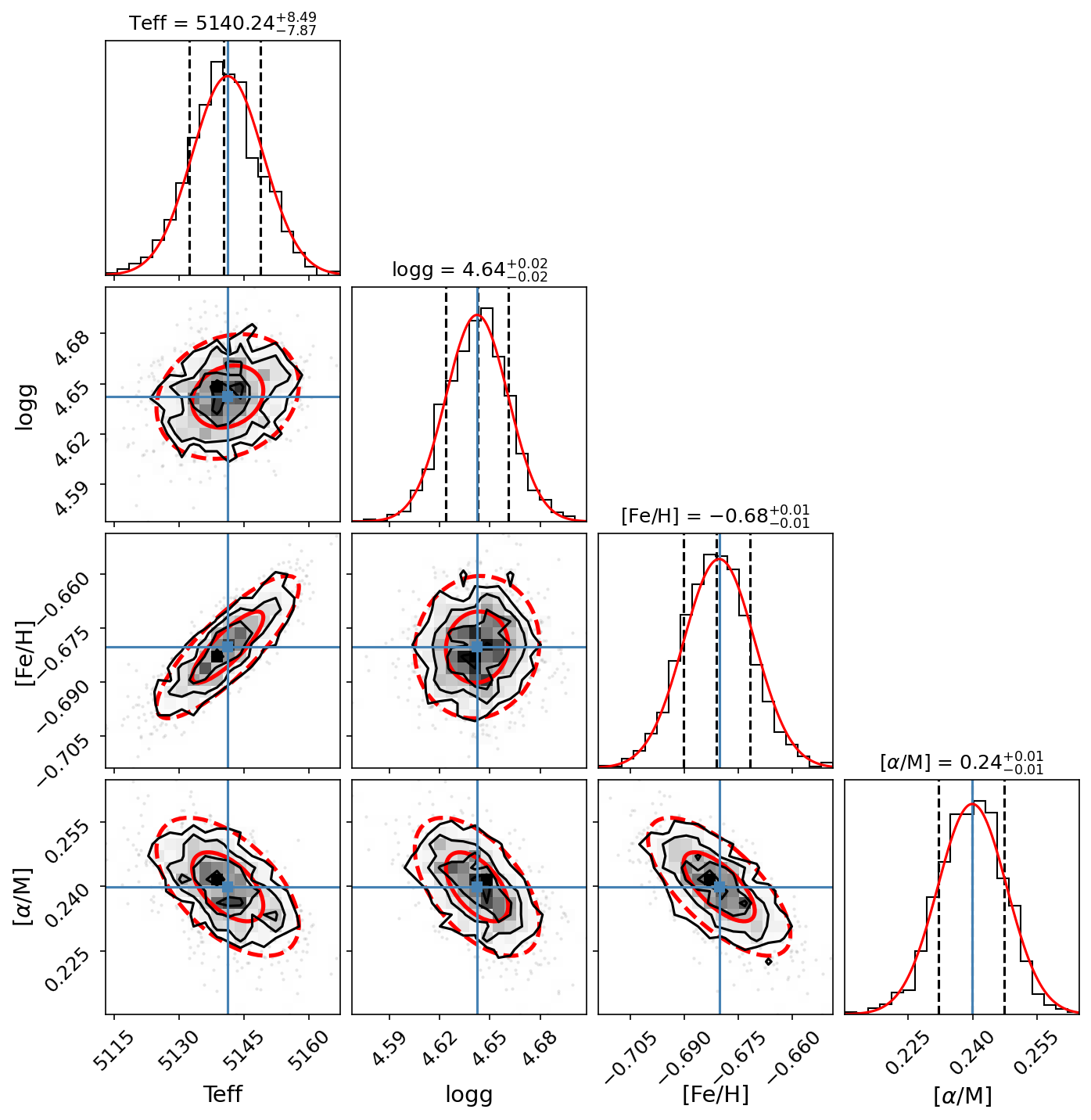}
    \includegraphics[width=0.49\textwidth]{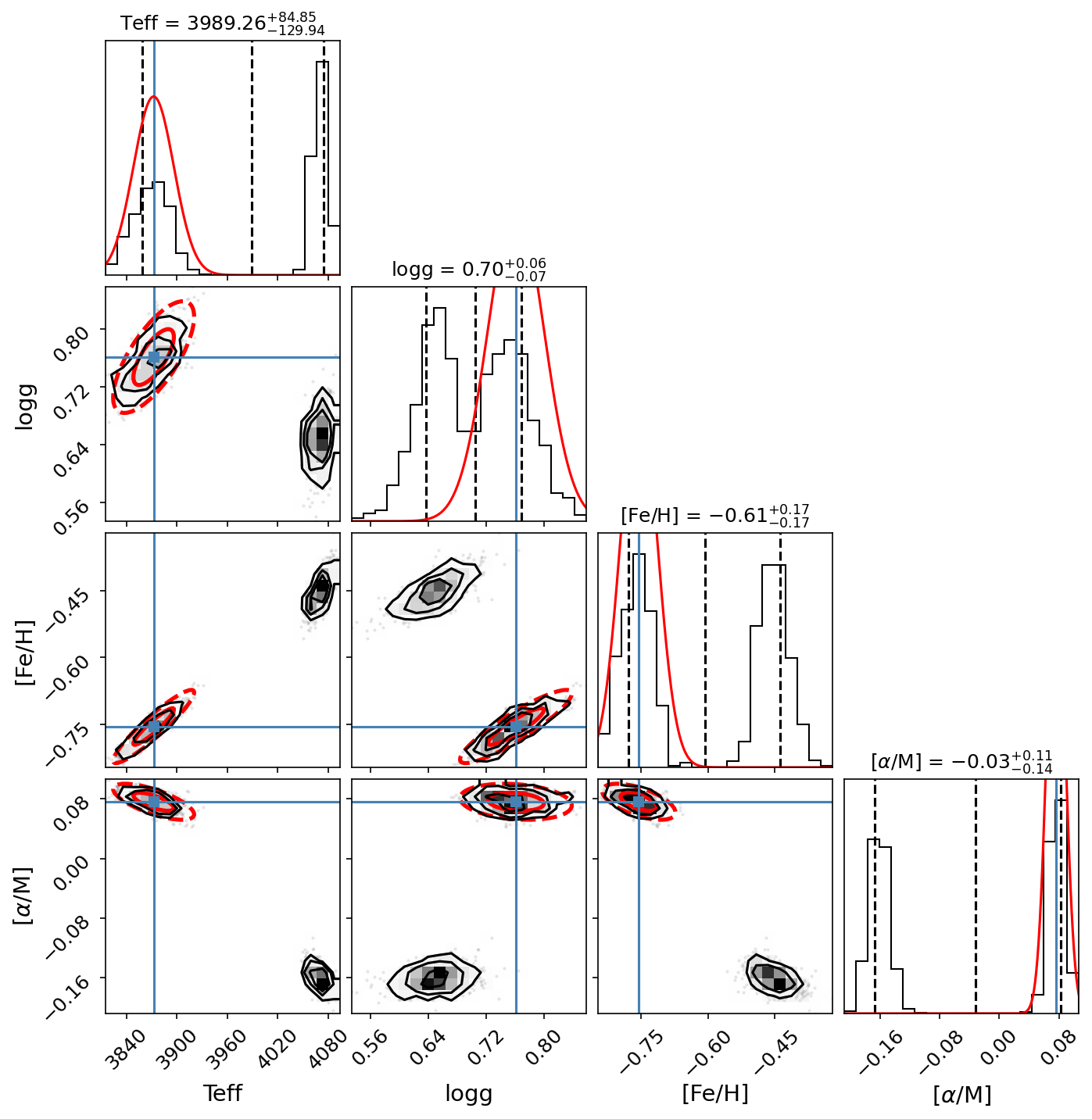}
    \caption{Examples of the resulting MCMC posterior (gray contour) compared to the estimate from the inverse of the Hessian (red lines). The left panel shows an example where the Hessian approximation well matches the MCMC posterior. The right panel demonstrates when this approximation usually fails; when the resulting posterior is bimodal.}
    \label{fig:mcmc_ex}
\end{figure}

\added{Generally, we find that the resulting MCMC posteriors are quite similar to the approximate variances and covariances released with this work. The left panel of Figure \ref{fig:mcmc_ex} shows a corner plot comparing the resulting MCMC posterior to the estimate from the inverse of the Hessian. Not only does the overall variances of the individual parameters match, but so do the covariances in all cases. There are some rare cases where these seem to greatly differ from one another based on this test. This is when the resulting posterior is bimodal (right panel, Figure \ref{fig:mcmc_ex}).}

\added{To quantify the average differences between the MCMC and Hessian approximations, Table \ref{tab:hessian_mcmc_comparison} summarizes the percent differences in the marginal, 1D uncertainty:
\begin{equation}
    \sigma_{\boldsymbol{\ell}} = 100\times\left(\frac{\sigma_{\rm Hessian}}{\sigma_{\rm MCMC}} - 1\right)
\end{equation}
and the Pearson correlation coefficient between parameter pairs:
\begin{equation}
    \rho(\boldsymbol{\ell}_1, \boldsymbol{\ell}_2) = 100\times\frac{(\rho_{\rm Hessian} - \rho_{\rm MCMC})}{|\tilde{\rho}_{\rm MCMC}|}
\end{equation}
where  $\tilde{\rho}_{\rm MCMC}$ is the pair's median correlation across the test sample. Here, we find that the median difference between the MCMC derived variance and covariance is within a few percent of the Hessian approximation in all cases. Additionally, we find that 95\% of stars in our test cases have variances and covariances typically within $\sim 20\%$, though some covariances can be as high as $\sim 35\%$. Overall though, this demonstrates that in the majority of the cases, the Hessian approximation is adequate for estimating the statistical errors on the stellar parameters from BOSS-CLAM.}

\begin{table}
\centering
\caption{Hessian vs.\ MCMC posterior uncertainties for the stratified validation subsample. Top: percent difference in the marginal (1D) uncertainty, $100\times(\sigma_{\rm Hessian}/\sigma_{\rm MCMC} - 1)$. Bottom: percent difference in the Pearson correlation coefficient between parameter pairs, $100\times(\rho_{\rm Hessian} - \rho_{\rm MCMC})/|\tilde{\rho}_{\rm MCMC}|$, where $\tilde{\rho}_{\rm MCMC}$ is that pair's median correlation across the test sample. The final column gives the smallest symmetric interval about zero containing 95\% of the sample.}
\label{tab:hessian_mcmc_comparison}
\begin{tabular}{lccc}
\hline
\hline
Quantity & Median diff. & Scatter & 95\% within \\
\hline
$\sigma_{\rm Teff}$ & +2.4\% & 14.2\% & 19.5\% \\
$\sigma_{\log g}$ & +2.0\% & 17.9\% & 19.0\% \\
$\sigma_{\rm [Fe/H]}$ & +2.7\% & 18.2\% & 21.9\% \\
$\sigma_{[\alpha/\rm M]}$ & +2.3\% & 24.8\% & 20.6\% \\
\hline
$\rho(\mathrm{Teff}, \log g)$ & +0.7\% & 14.0\% & 17.3\% \\
$\rho(\mathrm{Teff}, \mathrm{[Fe/H]})$ & +0.5\% & 11.8\% & 12.5\% \\
$\rho(\mathrm{Teff}, [\alpha/\mathrm{M}])$ & -1.1\% & 23.0\% & 25.0\% \\
$\rho(\log g, \mathrm{[Fe/H]})$ & +1.6\% & 24.2\% & 32.3\% \\
$\rho(\log g, [\alpha/\mathrm{M}])$ & -2.1\% & 28.6\% & 35.1\% \\
$\rho(\mathrm{[Fe/H]}, [\alpha/\mathrm{M}])$ & -0.3\% & 17.7\% & 14.4\% \\
\hline
\hline
\end{tabular}
\end{table}

\end{document}